\documentclass[11pt,a4paper]{article}

\pdfoutput=1
\usepackage{jcappub}
\usepackage{graphicx}
\usepackage{multirow}
\usepackage{hyperref}
\usepackage{amsmath,amssymb}
\usepackage{graphicx}
\usepackage{bm}
\usepackage{latexsym}
\usepackage{epsfig}
\usepackage{psfrag}
\usepackage{ulem}
\usepackage{color}
\usepackage[dvipsnames]{xcolor}
\usepackage{subfigure}
\usepackage[utf8]{inputenc}
\usepackage[a4paper, total={7in, 9in}]{geometry}

\makeatletter
\gdef\@fpheader{}
\makeatother

\newcommand{\be}{\begin{equation}}
\newcommand{\ee}{\end{equation}}
\newcommand{\bea}{\begin{eqnarray}}
\newcommand{\eea}{\end{eqnarray}}

\def\f{\frac}

\def\l{\left}
\def\r{\right}
\def\d{{\mathrm{d}}}

\def\ps{\mathcal{P}_{_{\mathrm{S}}}}
\def\pt{\mathcal{P}_{_{\mathrm{T}}}}
\def\ns{n_{_{\mathrm{S}}}}

\allowdisplaybreaks

\begin{document}

%%%%%%%%%%%%%%%%%%%%%%%%%%%%%%%%%%%%%%%%%%%%%%%%%%%%%%%%%%%%%%%%%%%%%%%%%%%%%%%

\title{Non-thermal moduli production during preheating in $\alpha$-attractor inflation models}
\author[a]{Khursid Alam,}
\author[b]{Mar Bastero-Gil,}
\author[a]{Koushik Dutta,}
\author[c]{and H.~V.~Ragavendra} 
\affiliation[a]{Department of Physical Sciences, Indian Institute of Science 
Education and Research Kolkata, Mohanpur, Nadia 741246, India}
\affiliation[b]{Departamento de Física Teórica y del Cosmos, Universidad de
Granada, Granada-18071, Spain}
\affiliation[c]{Raman Research Institute, C. V. Raman Avenue, Sadashivanagar, Bengaluru 560080, India}
\emailAdd{ka20rs114@iiserkol.ac.in}
\emailAdd{mbg@ugr.es}
\emailAdd{koushik@iiserkol.ac.in}
\emailAdd{ragavendra@rrimail.rri.res.in}
%\date{May 2022}

%%%%%%%%%%%%%%%%%%%%%%%%%%%%%%%%%%%%%%%%%%%%%%%%%%%%%%%%%%%%%%%%%%%%%%%%%%%%%%%

\abstract{Production of gravitationally coupled light moduli fields must be suppressed in the early universe so that its decay products do not alter Big Bang Nucleosynthesis (BBN) predictions for light elements. On the other hand, the moduli quanta can be copiously produced non-thermally during preheating after the end of inflation. In this work, we study the production of moduli in the $\alpha$-attractor inflationary model through parametric resonances. For our case, where the inflationary potential at its minimum is quartic, the inflaton field self-resonates, and subsequently induces a large production of moduli particles. We find that this production is suppressed for small values of $\alpha$. Combining semi-analytical estimation and numerical lattice simulations, we infer the parametric dependence on $\alpha$ and learn that $\alpha$ needs to be $\lesssim 10^{-8}\,m_{\rm Pl}^2$ to be consistent with BBN for ${\cal O}(1)$ coupling between inflaton and moduli. This in turn predicts an upper bound on the energy scale of inflation and on
the reheating temperature.}

\maketitle

%%%%%%%%%%%%%%%%%%%%%%%%%%%%%%%%%%%%%%%%%%%%%%%%%%%%%%%%%%%%%%%%%%%%%%%%%%%%%%%

\section{Introduction}
The paradigm of cosmic inflation driven by a scalar field is the most well-accepted mechanism to produce coherent super-Hubble perturbations that seed cosmic structures. In addition, the setup allows the production of a thermal bath of particles of the Standard Model (SM) and beyond the Standard Model (BSM) by converting inflation energy density via the mechanism of reheating. It is important that the idea of inflation be embedded in the physics of BSM as conceiving inflaton being any of the SM particles is difficult \cite{Baumann:2014nda}. 

Among several models of BSM physics, inflation is commonly considered in the context of Supergravity (SUGRA) or more generally in the context of String Theory. There are several well-motivated constructions where inflaton is identified with a BSM field. But, in SUGRA/String Theory, there are many other scalar fields that are gauge singlets and gravitationally coupled with the visible sector. These two facts lead to these fields having a large decay time with $\Gamma \sim m_{\sigma}^3/{m_{\rm Pl}}^2$, where $m_{\sigma}$ is the mass of the field dubbed as moduli and $m_{\rm Pl}$ is the Planck mass. These moduli fields get mass by supersymmetry breaking. In the context of String Theory, the vevs of the moduli fields fix several low-energy observables including gauge coupling constants, and in certain cases, it also fixes the volume of the compactified internal space \cite{Binetruy:2006ad, Cicoli:2023opf}.

First of all, it is crucial that the moduli are stabilized at fixed vevs and stabilization must hold during the cosmic history of the Universe. This issue is dependent on the initial field configurations of the fields, the form of the potential (which usually has finite barrier heights), and the content of the Universe at the end of inflation \cite{Brustein:1992nk, Buchmuller:2004xr, Buchmuller:2004tz, Alam:2022rtt}. The existence of moduli fields can alter cosmological dynamics in several interesting ways. Typically, the homogeneous zero-momentum mode of the field is shifted away from its post-inflationary global minimum \cite{Dine:1995uk, Cicoli:2016olq}, and the field starts to oscillate coherently when the Hubble parameter becomes smaller than the mass of the moduli. This leads to an epoch of early matter domination (EMD) \cite{Allahverdi:2020bys}. But the EMD epoch must end before BBN and the Universe must be reheated with a temperature above MeV energy, the lower bound being of the order of a few MeV \cite{Hasegawa:2019jsa,Kawasaki:2000en}. Generically, it requires the mass of the moduli to be heavier than $100$ TeV. Therefore, for light moduli masses, the universe may not be reheated with enough high temperature for successful BBN, and this is called the Cosmological Moduli Problem (CMP) \cite{Coughlan:1983ci, Banks:1993en, deCarlos:1993wie}. At the same time, inflationary predictions are sensitive to the post-inflationary history of the universe. The moduli-induced EMD epoch alters the predictions for inflationary observable \cite{Dutta:2014tya, Das:2015uwa, Allahverdi:2018iod}.

In addition to the background field, as mentioned, the fluctuations of the moduli fields and their quanta also play important roles in cosmology.  As noted above, if a modulus field is lighter than a few TeV in the range of $100$ GeV to $10$ TeV, it will decay during/after the time of BBN, and the decay products will alter the successful BBN predictions. Therefore, its abundance must be suppressed even if it does not dominate the energy budget: $(n_{\sigma}/s) \lesssim 10^{-12}$, where $n_\sigma$ is the number density of the modulus field, and $s$ is the entropy density \cite{Kawasaki:1994af, Kawasaki:2004qu, Kawasaki:2004yh, Ellis:1984er}. On the other hand, in the case of heavy moduli, the field will decay before BBN, and its decay products may thermalize. 
The moduli quanta can be produced in several ways. The moduli can be generated through thermal scatterings in the thermal bath produced during reheating. In this case, its abundance is determined by the thermal distribution. This abundance puts an upper limit on the reheating temperature \cite{Ellis:1982yb, Ellis:1984eq,Nanopoulos:1983up}. These fields can also be produced solely through gravitational interactions~\cite{Chung:1998zb, Giudice:1999yt}. In this work, we will consider moduli productions due to parametric resonances when the field is coupled with the inflaton field. 

The issue of non-thermal moduli production due to non-perturbative preheating processes was first discussed in~\cite{Giudice:2001ep}. It was concluded that for wide varieties of inflation models, and for generic couplings between the moduli and in the inflaton, the abundance is unacceptably high. In the work of ~\cite{Hagihara:2018uix}, it was pointed out that even for a modulus mass heavier than the scale of inflation, the modulus field can be displaced from its global minimum due to preheating effects, and that eventually leads to a matter-dominated phase from the oscillating field. 

In our work, we will consider $\alpha$-attractor inflation that has a parameter $\alpha$ which allows reducing the tensor amplitudes arbitrarily low. One important feature of this kind of model is that the dynamics of the inflaton field at the end of inflation are mostly independent of the physics during inflation. More specifically, even though the inflationary potential is exponentially flat, the potential at the time of preheating can be approximated as monomials. We will consider linear moduli coupling with the inflaton potential, and in the limit of low mass moduli, we will assume moduli mass is zero. We will numerically calculate the number density of the moduli and will show how $n_{\sigma}/s$ gets suppressed with the smaller values of $\alpha$. Using a semi-analytical approach \cite{Giudice:2001ep}, we will show how parametric dependence on $\alpha$ matches with the numerical estimates. We determine values of $\alpha$ for which the production is within the observational bound from BBN \cite{Kawasaki:1994af, Kawasaki:2004qu}. We find out that to suppress non-thermal light moduli production, the associated tensor-to-scalar ratio needs to be unobservably small in these kinds of inflation models. 

At this moment, we would like to emphasize the approach of this work. From several observations, it is becoming clear that the observable during inflation in terms of properties of the fluctuations are becoming inadequate to constrain inflation models. On the other hand, the physics of reheating of producing a thermal bath is still poorly understood. Therefore it is imperative to study post-inflationary physics more carefully to constraint inflation models \cite{Dai:2014jja, Bhattacharya:2017ysa, Maity:2018qhi,Drewes:2023bbs}. In this work, we attempt to constrain $\alpha$-attractor inflation model demanding that during preheating it does not produce many moduli particles. 

In other words, the current work deals with the moduli quanta production in $\alpha$-attractor inflation models through the non-perturbative preheating mechanism at the end of inflation where self-resonance of the inflaton field sources the moduli fluctuations. The final goal is to calculate $n_{\sigma}/s$ such that we can compare it with the bounds set by BBN. In the end, we explore non-trivial dependence on $\alpha$ on the moduli abundance. The broad idea of the work is to explore the constraints on inflation models from post-inflationary physics. The above-mentioned specific goal of calculating $n_{\sigma}/s$ can be achieved either semi-analytically or purely numerically. As we discuss in the manuscript, both have their limitations: 

(i) The semi-analytical process is limited due to the inherent non-linearities of the problem but is good at extracting parametric dependence. Even this process will require some inputs from numerical simulations, and thus {\it semi}-analytical.  

(ii) On the other hand, numerical estimates for the spectrum of the fields are reliable for all values of $\alpha$, but extraction of $n_{\sigma}/s$ from the spectrum is limited by numerical precision (not numerical error) due to subtraction of vacuum fluctuations contributions.

The work is organized as follows. In the next section, we discuss the main aspects of $\alpha$-attractor inflation model and its nature of coupling with the moduli that we use. In Sec.~3, we do the semi-analytical estimate of the abundance as a function of $\alpha$. In Sec.~4, we show the results of full numerical lattice simulations to tally our results with the previous section. In Sec.~5, we use these results to determine the observational bound on $\alpha$ and discuss the associated implications for reheating temperature and the tensor-to-scalar ratio. Finally, we conclude in Sec.~6. We have added a few Appendices to show some technical details.

As to the notations used in this work, we denote physical time with $t$ and conformal time with $\tau$. We denote Planck mass as $m_{_{\rm Pl}}=1.22\times 10^{19}$ GeV. An overdot represents a derivative with respect to physical time whereas an overprime denotes a derivative with respect to conformal time.

%%%%%%%%%%%%%%%%%%%%%%%%%%%%%%%%%%%%%%%%%%%%%%%%%%%%%%%%%%%%%%%%%%%%%%%%%%%%%%%

\section{$\alpha$-attractor potential and coupling with moduli field}
\label{attractor_potential_and_moduli_coupling}
In this section, we shall discuss the form and behavior of the base potential of 
interest that drives inflation. As mentioned earlier, we shall work with the 
$\alpha$-attractor class of models of 
inflation~\cite{Kallosh:2013daa,Kallosh:2013hoa,Kallosh:2013yoa,Kallosh:2022ggf}. 
The generic form of such a potential is given by
\begin{eqnarray}
V(\phi) &=& V_0\,\tanh^p\left(\frac{\phi}{\sqrt{6\alpha}}\right)\,,
\label{eq:Vphi}
\end{eqnarray}
where $V_0$ sets the overall energy scale. The power of the monomial $p$ is 
assumed to take positive, even integral values. The parameter $\alpha$ defines 
the range of the field excursion. 
The slow-roll driven accelerated expansion occurs over the range where $\phi \gg \sqrt{6\alpha}$, while reheating takes place during the oscillation of the field at the minimum, with $\phi < \sqrt{6\alpha}$.

%%%%%%%%%%%%%%%%%%%%%%%%%%%%%%%%%%%%%%%%%%%%%%%%%%%%%%%%%%%%%%%%%%%%%%%%%%%%%%%

Given the potential, one may solve the Klein-Gordon equation governing the evolution of the homogeneous part of the field $\phi_0(t)$,
\begin{equation}
\ddot \phi_0  + 3H\dot \phi_0 + \f{\partial V}{\partial \phi_0} = 0\,,
\end{equation}
where $H= d \ln a/d t$ is the Hubble parameter, $a$ the scale factor and an overdot denotes derivative with respect to cosmic time. Under the slow roll approximation described by 
$\ddot \phi_{0} \ll 3H\dot \phi_{0}$ and $\dot \phi_{0}^2 \ll V(\phi_{0})$, one may recast the 
above equation in terms of the no. of e-folds $N$ as 
\begin{equation}
  \f{\rm d \phi_{0}}{{\rm d} N} \simeq -\f{1}{3 H^2}\f{\partial V}{\partial \phi_{0}} \simeq
  -\f{4p}{\sqrt{6\alpha}}\f{m^2_{\rm Pl}}{8\pi}
  \exp{\left(-\f{2\phi_{0}}{\sqrt{6\alpha}}\right)} \,, \label{dphiN}
\end{equation}
where in the second equality we have used $ 3 m_{\rm Pl} ^2 H^2/8\pi \simeq V(\phi_{0})\simeq V_0$. 
Note that the parameter $\alpha$ is expressed in units of $m^2_{\rm Pl}$. However, 
the relevant dimensionless combination that appears here and in the rest of 
calculations is $8\pi\alpha/m^2_{\rm Pl}$.
Hence, we define a dimensionless parameter $\Tilde{\alpha}=8\pi\alpha/m^2_{\rm pl}$, 
in terms of which we shall express the quantities of interest. We shall later vary
this parameter over a range of values and observe its effects on these quantities.

Solving the above equation, we obtain the field value in terms of the model parameters $p$ and $\tilde\alpha$ as
\begin{equation}
\phi_{0}(N) \simeq \f{1}{4}\sqrt{\f{3\Tilde{\alpha}}{\pi}}
\ln\left(1 + \frac{4p\,N}{3\Tilde{\alpha}}\right)\,m_{\rm Pl}
\simeq \f{1}{4}\sqrt{\f{3\Tilde{\alpha}}{\pi}}
\ln\left(\frac{4pN}{3\Tilde{\alpha}}\right)\,,
\label{phiN}
\end{equation}
where $N=\ln(a_{end}/a)$ is the number of e-folds counted from the end of inflation and we have set $\phi_{\rm end} = \phi_0(N=0) \ll m_{\rm Pl}$. The second expression is valid for large values of $N$, typically $50$---$60$, during the early stages of inflation. 

The power spectra of scalar and tensor perturbations are given by
\begin{eqnarray}
\ps(k) &=& \left( \frac{H}{\dot \phi_{0}} \right)^2 \left( \frac{H}{2 \pi} \right)^2 \simeq \f{32}{9\,m_{\rm pl}^4}\f{V_0}{\Tilde{\alpha}}\,N_k^2 \,, \\
\pt(k) &=& 8 \left( \sqrt{\frac{2}{\pi}}\frac{H}{m_{\rm Pl}} \right)^2 \simeq \frac{128}{3\,m_{\rm Pl}^4}\,V_0,
\label{eq:power_spectra}
\end{eqnarray}
where $N_k= \ln(a_{end}/a_k)$ is e-fold at which the mode $k$ crosses the Hubble radius i.e. $k=a_k H$, and we have used  the slow-roll approximations 
Eqs.~\eqref{dphiN} and \eqref{phiN}, and the fact that $\phi \gg \sqrt{6\alpha}$ in arriving at these estimates.  The 
corresponding scalar spectral index $\ns$ and the tensor-to-scalar ratio $r$ can be readily obtained as
\begin{eqnarray}
\ns-1 &\equiv& \f{\d \ln \ps(k)}{\d \ln k} \simeq -\frac{2}{N_k}, \\
r &\equiv& \f{\pt(k)}{\ps(k)} \simeq \f{12\,\Tilde{\alpha}}{N_k^2}.
\label{eq:r}
\end{eqnarray}
These are well-known results of $\alpha$-attractor class of models~\cite{Kallosh:2022ggf,Bhattacharya:2022akq}.
The scalar spectral index at the pivot scale $k_\ast$ turns out to be $\ns \simeq 0.96$ 
if $N_{k_\ast} \simeq 50$, i.e., if $k_\ast$ exits the Hubble radius around 
$50$ e-folds before the end of inflation. Notice, that the spectral index is 
independent of $\Tilde{\alpha}$. However, and importantly, the tensor-to-scalar ratio 
is directly proportional to $\Tilde{\alpha}$. This implies that for a low value of $r$, 
as is being progressively favoured by data on large scales, the value of $\tilde\alpha$ has to be proportionately small.
The current constraint on $r$ around $k=5\times 10^{-2}\,{\rm Mpc}^{-1}$ is
$r \leq 3.6\times10^{-2}$~\cite{BICEPKeck:2022mhb}. 
For the value of $N_k\simeq 50$, we get $r \lesssim 4.8\times 10^{-3}\,\Tilde{\alpha}$.
Hence, one can infer that the typical range of values is 
$\Tilde{\alpha} \lesssim {\cal O}(1)$.

At this stage of discussion, we should mention the constraint on the 
parameters of the model due to the combination with which they appear in the 
expressions of the power spectra. Note that the scalar power spectrum is
proportional to the ratio~$V_0/\tilde\alpha$ in Planck units. Therefore, in order to achieve the COBE normalized value of $\ps(k) \simeq 2.1\times 10^{-9}$ at the pivot scale $k_\ast=0.05\,{\rm Mpc}^{-1}$~\cite{Akrami:2018odb,Aghanim:2018eyx}, we require
\begin{equation}
\frac{V_0}{\Tilde{\alpha}} \simeq 2.36 \times 10^{-13} m^4_{\rm pl}\,
\left(\frac{50}{N_{k_\ast}}\right)^2\,.
\label{eq:V0byalpha}
\end{equation}
In our subsequent analyses, whenever we vary the parameter $\Tilde{\alpha}$, 
we shall ensure that $V_0$ is also varied such that the ratio of~$V_0/\Tilde{\alpha}$ is kept constant and $\ps(k_\ast)$ is consistent with its well 
constrained value.
This in turn means that the smaller the value of $\Tilde{\alpha}$, the smaller the
value of $V_0$. Hence, if we reduce $\Tilde{\alpha}$, we implicitly decrease the 
energy scale of inflation. Further, this diminishes the amplitude of tensor 
power as well.
%%%%%%%%%%%%%%%%%%%%%%%%%%%%%%%%%%%%%%%%%%%%%%%%%%%%%%%%%%%%%%%%%%%%%%%%%%%%%%%

Having revised the behavior of potential and its implications for the power
spectra of perturbations, we turn to its behavior close to the end of inflation.
The range of excursion for the field $\phi$ close to the minimum of the potential 
is given by the condition $\phi < \sqrt{6\alpha}$. Around this region, the 
potential may be approximated as
\begin{equation}
V(\phi) \simeq V_0\l(\f{\phi}{\sqrt{6\alpha}}\r)^p \,.
\end{equation}
In such a monomial shape of the potential, the mass of inflaton is simply
given by 
\begin{equation}
m^2_\phi = \f{V_0}{6\alpha}p(p-1)\l(\f{\phi}{\sqrt{6\alpha}}\r)^{p-2} \,.
\label{eq:effmass}
\end{equation}
It is around this region that we shall study the effects of parametric 
resonance in the preheating stage of the field evolution.
Notice that for the case $p=2$ the potential reduces to 
a quadratic form $V(\phi) = m_\phi^2\phi^2/2$ with $m^2_\phi=V_0/(3 \alpha)$, 
whereas for the case $p=4$ we have the well 
known quartic form $ V(\phi)=\lambda\phi^4/4$ with $\lambda= V_0/(9 \alpha^2)$. 

%%%%%%%%%%%%%%%%%%%%%%%%%%%%%%%%%%%%%%%%%%%%%%%%%%%%%%%%%%%%%%%%%%%%%%%%%%%%%%%

To understand the dynamics of preheating and production of moduli during this
stage, we introduce the potential governing the moduli field $\sigma$.
From here on, we shall restrict to the case of $p=4$ to examine moduli 
production in this scenario. We introduce the term involving $\sigma$ in 
the potential that is linearly coupled to the inflaton as follows
\begin{eqnarray}\label{pot}
V(\phi, \sigma) 
&=& V_0\left(\frac{\phi}{\sqrt{6\alpha}}\right)^4\,
\bigg[1 + c\frac{\sigma}{m_{\rm pl}}\bigg]\,.
\end{eqnarray}
Such linear couplings have been discussed in the literature for different 
models~\cite{Giudice:2001ep,Hagihara:2018uix}.
Evidently, the strength of coupling is determined by the parameter $c$, and from the naturalness argument we will set $c=1$ throughout our analysis.
Note that by reducing $c$, the moduli abundance can always be suppressed. On the other hand, by making $c$ relatively small, it is difficult to analyze the dependence on the inflation parameter $\tilde\alpha$ as the resonance particle productions for moduli (not for the inflaton) become very small. To avoid that situation, we have kept $c = 1$ with the understanding that the production reduces as we reduce $c$. The broader goal of the article is to constrain the inflation model by studying post-inflationary physics. In that sense, analyzing with $c \sim \mathcal{O}(1)$ is suitable for us to find the dependence on inflation parameter $\tilde\alpha$. The effects of coupling between the modulus sector and the $\alpha$-attractor inflation sector have been discussed in \cite{Roest:2016lrb}\footnote{In the context of supergravity, the coupling between the modulus and the inflaton sector has been analyzed in \cite{Brax:2005jv, Davis:2008sa, Antusch:2008pn, Antusch:2009ty}. As the K\"ahler potential is usually an expansion of the field, with an appropriate choice of K\"ahler potential for the modulus field, we can get a coupling with the inflaton sector that is linear in the modulus field.}.

Further, we do not introduce any mass term for the moduli field or higher order terms in $\sigma$. For our purpose, mass dependence on the moduli abundance is negligible. First of all, the production of heavy moduli quanta is not problematic in the context of BBN as they decay much earlier without affecting the BBN. The only way, the heavy moduli quanta can affect cosmology is via its coherent oscillations leading to an early matter-dominated epoch \cite{Dutta:2014tya, Das:2015uwa}. Our current manuscript does not attempt to explore that. Secondly, the moduli mass that can affect BBN predictions should be of the order of $1$-$10$ TeVs, and this mass can be treated massless for all effective purposes as all other scales including the inflaton mass are much heavier than the moduli masses. In our numerical simulations, we have checked explicitly that adding moduli mass of this order does not change results in any significant manner.

In this minimal setup, we will concentrate on varying the defining parameter of this class of models, $\alpha$, and study the implications for the behavior of $\phi$ and $\sigma$ as they oscillate around the minimum of the potential. 

At this stage, we should remark on the initial condition of the field that
shall be used for the subsequent semi-analytical and numerical calculations.
We choose the initial value of the inflaton field for studying preheating, $\phi_0(0)$, at the
point when inflation terminates and the field begins its oscillations at the
minimum of the potential. This value is determined by setting the first potential
slow roll parameter $\epsilon_V$ to unity. We know that for the given potential
\be
\epsilon_V = \frac{m^2_{\rm pl}}{16\pi}
\left(\frac{\partial V}{\partial \phi}\frac{1}{V}\right)^2 \simeq  \frac{1}{\Tilde{\alpha}}
\left[ 8 \left(\frac{\sqrt{\alpha}}{\phi}\right)^2 - \frac{16}{9} 
+ {\cal O}\left(\f{\phi}{\sqrt{\alpha}}\right)^2
\right] \,.
\ee
Note that at the end of inflation, around the minimum of the potential, 
$\phi \ll \sqrt{6\alpha}$. Hence, we may truncate the above series up to just
the first two terms and determine $\phi_0(0)$ by setting $\epsilon_V=1$. Thus we obtain
\begin{eqnarray}
\phi_0(0) &\simeq & 0.42\,\sqrt{\Tilde{\alpha}}\,m_{\rm Pl}\,.
\label{eq:init_field_value}
\end{eqnarray}
%%%%%%%%%%%%%%%%%%%%%%%%%%%%%%%%%%%%%%%%%%%%%%%%%%%%%%%%%%%%%%%%%%%%%%%%%%%%%%%
  Therefore, we will always start our numerical simulations in the positive curvature region of the potential, where the potential can be well approximated by the first term in the expansion.  However it should be mentioned that as the field approaches the end of inflation, signaled by the condition $\epsilon_H= -\dot H/H^2 \simeq \epsilon_V \simeq 1$, the negative curvature of the potential increases and this leads to a tachyonic instability \cite{Rubio:2019ypq,Karam:2021sno,Tomberg:2021bll,Koivunen:2022mem}. The transition from $|\eta_V|= |\partial^2_\phi V/(3 H^2)| \simeq 1$ to $\epsilon_V \simeq 1 $ lasts roughly 2 e-folds, independent of $\alpha$, and sub-horizon inflaton fluctuations within a certain range of k-modes may be amplified during this period, before the field starts oscillating, with the enhancement being larger for smaller values of $\tilde \alpha$. This will drain energy from the inflaton oscillations, and for values of $\tilde \alpha \lesssim 10^{-5}$, it has been argued that it can lead to the total fragmentation of the inflaton field \cite{Tomberg:2021bll}, i.e., the background inflaton energy density being fully transferred into its fluctuations. However, one may expect that backreaction and re-scattering effects during the tachyonic instability affect the exponential growth of the fluctuations and prevent the full conversion of background energy into fluctuations. In this work, we do not attempt to model this period of tachyonic instability. We implicitly assume that the growth of the fluctuations somehow ``smoothens'' out the transition, still allowing an oscillating phase but only in the positive curvature region of the potential.  

%%%%%%%%%%%%%%%%%%%%%%%%%%%%%%%%%%%%%%%%%%%%%%%%%%%%%%%%%%%%%%%%%%%%%%%%%%%%%%%
\section{Estimation of moduli production: semi-analytical approach}
\label{emi-analytical approach}
Before discussing the results of the numerical analysis of preheating, 
we first derive a semi-analytical expression for the moduli abundance arising due to the potential of interest.
Such an estimate has been earlier obtained for a similar potential in Ref.~\cite{Giudice:2001ep}. Here we shall follow the same prescription highlighting
only the steps where we have done specific modifications as per our model of interest. We want to extract the parametric dependence of the abundance on $\alpha$. 

The equation of the motion of the inflaton field and the moduli field in FLRW spacetime for the potential of interest Eq.~\eqref{pot} are
\begin{align}
\label{EOM}
&\ddot{\phi}+ 3H\dot{\phi}- \frac{\nabla^2\phi}{a^2}+\lambda\phi^3\left(1+c\frac{\sigma}{m_{\rm Pl}}\right)=0, \\
&\ddot{\sigma}+ 3H\dot{\sigma}- \frac{\nabla^2\sigma}{a^2}+\frac{\lambda\phi^4}{4 m_{\rm Pl}}=0. \label{EOM1}
\end{align}
We will use the notation $\phi_{0}(t)$ for the background field, and $\phi(\bm x, t)$ for the full space-time dependent field, such that $\phi_{0}(t)=\langle\phi\rangle$,  where $\langle\cdots \rangle$ denotes spatial average. The equation for the inflaton and moduli are written in conformal time $d\tau=\frac{dt}{a(t)}$. We rescale the field variables $\varphi(\bm x,\tau)=a(t)\phi(\bm x, t)$, and $\overline{\sigma}(\bm x,\tau)=a(t)\sigma(\bm x, t)$ as,
\begin{align}
\label{EOM_conf}
    &\varphi''-\bm\nabla^2\varphi-\frac{a''}{a}\varphi+\lambda\varphi^3\simeq 0 ,\\
    &\overline{\sigma}''-\bm\nabla^2\overline{\sigma}-\frac{a''}{a}\overline{\sigma}=-c\frac{V_0}{36\alpha^2 m_{\rm pl}}\frac{\varphi^4}{a(\tau)}, \label{EOM_conf1}
\end{align}
where the overprime denotes the derivative with respect to the conformal time. Here, we have neglected the last term in Eq.~\eqref{EOM} because inflaton energy dominates over the moduli.

To account for the effect of homogeneous and inhomogeneous components of the inflaton field, we can write $\varphi(\bm x, \tau)$ as,
\begin{equation}\label{inf_fluc}
    \varphi(\bm x, \tau)=\varphi_{0}(\tau) + \delta\varphi(\bm x, \tau),
\end{equation}
where $\varphi_{0}(\tau)$ is the homogeneous background part, and $\delta\varphi(\bm x, \tau)$ is the fluctuating component of the inflaton field. Now, the background equation of the inflaton and the fluctuations of the inflaton field up to linear order in $\delta\varphi(\bm x, \tau)$ are,
\begin{align}
\label{EOM_conf_re}
    &\varphi_{0}''+\lambda\varphi_{0}^3= 0 ,\\
    &\delta\varphi''-\bm\nabla^2\delta\varphi+\lambda\varphi_{0}^2\delta\varphi= 0, \label{EOM_conf_eq}
\end{align}
where the background behaves with a radiation-like equation of state of $\omega = 1/3$. The equation for the fluctuations in Fourier space is,
\begin{equation}\label{inflaton_mode_equ}
    \delta\varphi''_{k}+(k^2+\lambda\varphi^2_{0})\delta\varphi_{k}=0,
\end{equation}
which is a harmonic oscillator with a time-dependent frequency. This leads to parametric resonance and inflaton fluctuations behaving like $\delta\varphi_{k} \sim e^{\mu\tau}$ where $\mu$ is called the Floquet exponent. We have calculated $\mu$ numerically, the details of which are explained in App.~\ref{app:floq}. We find the value of $\mu \simeq 0.03\,\phi_{0}(0)\sqrt{{V_0}/{9\alpha^2}}$ where $\phi_{0}(0)$ is the value of the inflaton field at the end of inflation. Note that $\mu$ from this expression,
along with our earlier arguments in Eqs.~~\eqref{eq:V0byalpha} and~\eqref{eq:init_field_value}, turns out to be independent of $\alpha$\footnote{Note that this closely follows the corresponding 
estimation of Ref.~\cite{Giudice:2001ep}, but the calculation is 
performed in conformal units in our case as opposed to physical 
units in the reference. This makes an important difference in our 
further analysis.}.

Solving Eq.~\eqref{EOM_conf1} using Green's function method, we can write the complete solution for $\overline{\sigma}(x)$ as
\begin{equation}
  \overline{\sigma}(x)=\overline{\sigma}_{ws}(x)-\frac{c V_{0}}{36\alpha^2 m_{\rm pl}}\int d^4x' G_{\rm ret}[x- x']\frac{\varphi^4(x')}{a(\tau')},
\label{eq:mod-soln}
\end{equation}
where $x=({\bm x}, \tau)$ represents the observer coordinate, and $x'=({\bm x'}, \tau')$ represents the source coordinate.
The first term $\overline{\sigma}_{ws}(x)$ is the homogeneous solution (without the source term) of the 
equation of motion in Eq.~\eqref{EOM_conf1} , and the second term is the particular solution capturing the contribution of the source term. The function $G_{\rm ret}[x-x']$ is the retarded Green's function, given by
\begin{equation}
   G_{\rm ret}[ x-x']=\theta(\tau-\tau')\int\frac{d^3k}{(2\pi)^3}\frac{\sin{\overline{\omega}_{k}(\tau-\tau')}}{\overline{\omega}_{k}}e^{i\bm k\cdot(\bm x-\bm x')}.
\end{equation}
Here, $\overline{\omega}_{k}=a(t)\omega_{k}$ where $\omega_{k} = k/a(t)$ is the physical frequency of oscillation of the moduli field.
On substituting $\varphi(x)$ in Eq.~\eqref{eq:mod-soln} and retaining
terms upto linear order in $\delta\varphi(x)$ we get
\begin{align}\label{sol1}
\overline{\sigma}(x)=\overline{\sigma}_{ws}(x)-\frac{c V_{0}}{36\alpha^2 m_{\rm pl}}&\int d^3\bm x' d\tau' G_{\rm ret}[x - x']\varphi_{0}^4(\tau') \nonumber \\ 
&-\frac{c V_{0}}{9\alpha^2 m_{\rm pl}}\int d^3\bm x' d\tau' G_{\rm ret}[x -x']\varphi_{0}^3(\tau')\delta\varphi(x')\,.
\end{align}
We shall identify each of the three terms in Eq. \eqref{sol1} as
\begin{equation}\label{eq:sigma_soln}
    \overline{\sigma}(x)=\overline{\sigma}_{ws}(x)+ \overline{\sigma}_1(x)+\delta\overline{\sigma}(x). 
\end{equation}
Notice that the second term, $\overline{\sigma}_{1}(x)$, 
is independent of $\delta\varphi$ and shall not alter the existing space dependence of $\overline{\sigma}$. Hence it can be treated in the same manner
as the homogeneous solution $\overline{\sigma}_{ws}(x)$. The term that captures the
effect due to the fluctuations in $\varphi$, and hence the resonance, is the third term $\delta \overline{\sigma}(x)$.

Here on we shall focus on explicitly computing $\delta \overline{\sigma}(x)$, and in particular their modes in Fourier space $\delta \overline{\sigma}_k(\tau)$:
\begin{equation}\label{var}
    \delta\overline{\sigma}(x)=\int \frac{d^3k}{(2\pi)^3}\delta \overline{\sigma}_{k}(\tau)e^{i\bm k \cdot \bm x}\,.
\end{equation}
Using the form of $G_{\rm ret}[ x - x']$ in Eq.~\eqref{sol1} these are given by
\begin{equation}\label{eq:time_dep_mode}
    \delta \overline{\sigma}_{k}(\tau)=-\frac{c V_{0}}{9\alpha^2 m_{\rm pl}}\int_{0}^{\tau}d\tau' \frac{\sin[\overline{\omega}_{k}(\tau-\tau')]}{\overline{\omega}_{k}}\varphi_{0}^3(\tau')
    \delta\varphi_{k}(\tau').
\end{equation}
Again, we work at linear order and therefore the dominant contribution to the mode function $\delta\overline{\sigma}_k$ is linearly proportional to the inflaton fluctuations $\delta\varphi_{k}$. The subsequent higher order contributions to $\delta\overline{\sigma}_k$ shall be proportional to $\varphi_0^2\delta \varphi^2$ and $\varphi_0\delta\varphi^3$. From the numerical analyses, one can show that the ratio of 
$\langle \delta \varphi^2\rangle/\varphi_0^2$ is always less than unity, which 
supports our approximation for the semi-analytical estimation of the moduli abundance. 

The variance of the moduli from Eq.~\eqref{var} is given by:
\begin{equation}
    \label{var_sq}\langle\delta\overline{\sigma}^2(\tau)\rangle=\int\frac{d^3k}{(2\pi)^3}|\delta \overline{\sigma}_{k}(\tau)|^2,
\end{equation}
where we have taken the spatial average on the RHS, and $\delta \overline{\sigma}_k(\tau)$ is given by Eq. \eqref{eq:time_dep_mode}. We are interested in the variance up to the time when the resonance ends.
After completion of resonance at $t_r$, quantities such as mode amplitudes, number densities, and energy densities will be only red-shifted due to the expansion of the universe. During the resonance, $\delta\varphi_{k}$ changes exponentially in the integrand of Eq. \eqref{eq:time_dep_mode}, but other quantities do not change appreciably in comparison. So we perform the integration only over the exponential behavior in time. Next to integrate over the modes in Fourier space, we will focus solely on the resonance mode ($k_{r}$) as it gives the dominant contribution to the variance. Therefore, at the end of the resonance phase $\tau_{r}$, the average moduli particle production can be analytically approximated as
\begin{equation}\label{mode_variance}
   \langle \delta\overline{\sigma}^2(\tau_{r})\rangle \simeq
   \frac{c^2}{m^2_{\rm pl}} \frac{V^2_0}{81\,\alpha^4}
   \left(\frac{\varphi^3_{0}(\tau_{r})}{a(\tau_{r})}\right)^2\frac{
   \langle\delta\varphi^2(\tau_{r})\rangle}{2\overline{\omega}_{k_{r}}^2(\tau_{r})\mu^2} \,, 
\end{equation}
where $k_r$ is the mode maximally produced during the resonance, and the variance of the inflaton field is given by 
\begin{equation}\label{variance}
    \langle\delta\varphi^2(\tau)\rangle=\int\frac{d^3k}{(2\pi)^3}|\delta\varphi_{k}(\tau)|^2 \,.
\end{equation}
We also know that at the resonance, the moduli field starts to oscillate with the same frequency as the inflaton, $\overline{\omega}^2_{k_{r}} \simeq 3\left(V_{0}/9\alpha^2\right)\varphi^2_{0}(t)$, where $\varphi_{0}(t)=a(t)\phi_0(t)$ is the amplitude of the oscillating inflaton field \cite{Giudice:2001ep}\footnote{This statement will be numerically confirmed in Fig. \ref{fig:var_nk_alpha} in Sec. 4 where we see that particle production for both inflaton and moduli fields happen at same momentum values. More details can be found in Appendix A.}. Finally, substituting the expressions of $\overline{\omega}_{k_{\rm r}}$ and $\mu$ in terms of  the model parameters $V_0$ and $\alpha$, we obtain for the variance of the moduli field: 
\begin{equation}\label{eq:variance}
   \langle\delta\sigma^2(t_{r})\rangle \simeq 2\times 10^{2}\,c^2 \, \left(\frac{\phi_{0}(0)}{a(t_{r})}\right)^2\frac{\langle\delta\phi^2(t_r)\rangle}{m^2_{p}}\,.
\end{equation}
Notice that we have transformed the variances of the inflaton and moduli into the physical coordinates from conformal coordinates. 
Also, we should note that the above expression differs from the
corresponding expression of Ref.~\cite{Giudice:2001ep}.
The variance of $\delta \sigma$ depends linearly on the variance of $\delta \phi$ apart from the amplitude of the oscillations of the background component.
 
Now, we can estimate the comoving number density of the moduli field at the end of the resonance as 
\be
n_{\sigma}a^3(t_r) \simeq   a^3(t_r)\omega_{k_{r}}(t_{r})\langle \delta\sigma^2(t_{r})\rangle 
    \simeq  5.71\times 10^{-13}\,c^2
    \left(\frac{\ps(k_\ast)}{2.1\times 10^{-9}} \right)^{1/2}
    \left(\frac{\Tilde{\alpha}}{1/3}\right)\frac{\langle\delta\phi^2(t_{r})\rangle}{1.5\times 10^{-8}m_{\rm Pl}^2}\,\, m^3_{\rm Pl} \,,
    \label{nsigma}
\ee
where $n_{\sigma}$ is the physical number density. We should recall that we maintain the ratio of $V_0/\alpha$ a constant such that the scalar power spectrum is suitably normalized over CMB scales. So, from the Eq.~\eqref{nsigma}, we see that the comoving number density of the moduli is proportional to $\Tilde \alpha$ because $\phi_{0}(0)\propto{\Tilde \alpha}^{1/2}$ [cf. Eq.~\eqref{eq:init_field_value}]. 

To compute the required ratio of $n_\sigma/s$, we need the estimate of the entropy density $s$. The comoving energy density of the system is given by
\be
\rho(t_r) a^4(t_r) \simeq  \frac{V_0}{36\alpha^2}\phi_0^4(0)\simeq  4.44\times 10^{-14}
 \left( \frac{\ps(k_\ast)}{2.1\times 10^{-9}} \right)
\left(\frac{\Tilde{\alpha}}{1/3}\right)\,m_{_{\rm Pl}}^4 \,,
\ee
where $\rho$ is the physical energy density of the system. We can readily see that the comoving energy density depends linearly on $\Tilde \alpha$. 
Hence the comoving entropy density, when the initial energy density of
inflaton has been completely transferred to decay products, is given by 
\begin{equation}
s\,a^3(t_r) \sim \, a^3(t_{r}) \rho^{3/4} 
 \sim 9.68\times10^{-11}
 \left( \frac{\ps(k_\ast)}{2.1\times 10^{-9}} \right)^{3/4}
 \left(\frac{\Tilde{\alpha}}{1/3}\right)^{3/4} m^3_{p} \,,
\end{equation}
where once again retaining the ratio of $V_0/\alpha$ constant gives us the dependence of comoving entropy density on $\Tilde \alpha$ to be $\Tilde \alpha^{3/4}$. Therefore the ratio of the comoving number density of the moduli to comoving entropy density is 
\begin{eqnarray}
\frac{n_\sigma}{s}
 & \simeq 5.90\times10^{-3} c^2 
 \left(\displaystyle\frac{\ps(k_\ast)}{2.1\times 10^{-9}}\right)^{-1/4}
 \left(\displaystyle\frac{\Tilde{\alpha}}{1/3}\right)^{1/4}\left(\displaystyle\frac{\langle\delta\phi^2(t_r)\rangle}{1.5\times 10^{-8}\,m_{\rm pl}^2}\right)\,.
\label{abundance}
\end{eqnarray}
From this expression, we may expect that $n_\sigma/s$ shall scale as 
${\Tilde \alpha}^{1/4}$ apart from any implicit dependence on $\Tilde \alpha$ arising from $\langle\delta \phi^2(t_r)\rangle$. Hence, in order to obtain this, and  the complete dependence of the moduli abundance on
$\Tilde \alpha$, we now turn to numerics.
Besides, we can see that $n_\sigma/s \propto c^2$ and so obviously
the abundance can be suppressed by decreasing the coupling strength $c$. The dependence of the moduli abundance on the coupling constant $c$ has been explored in more detail in Appendix D. We recover the expected dependence on $c$ numerically.
%%%%%%%%%%%%%%%%%%%%%%%%%%%%%%%%%%%%%%%%%%%%%%%%%%%%%%%%%%%%%%%%%%%%%%%%%%%%%%%
\section{Estimation of moduli production: numerical approach}
To examine the exact behavior of the inflaton and moduli fields during preheating
and arrive at the estimate of moduli abundance arising out of parametric resonance, 
we perform a lattice simulation of our system using the publicly available numerical 
package designed for this purpose, called $\texttt{Latticeeasy}$~\cite{Felder:2000hq}. This package allows us to evolve
a system of fields described by potentials along with canonical kinetic terms and couplings of our choice.

The system of equations of $\phi$ and $\sigma$ that is solved in the lattice is 
given in Eqs.~\eqref{EOM} and~\eqref{EOM1}.
Evidently, the equations are coupled through the terms arising from the potential and Hubble parameter.
The system is coded in suitably rescaled variables for efficient evolution in
the lattice. For a discussion about the program variables in which the system is
evolved in the lattice, refer to App.~\ref{app:prog_var}.
Our objective is to compute the observationally relevant quantity $n_\sigma/s$,
using the semi-analytical expression obtained earlier as well as directly from
the outputs of the lattice simulations.

\begin{figure}
\centering
\includegraphics[width=6in,height=4in]{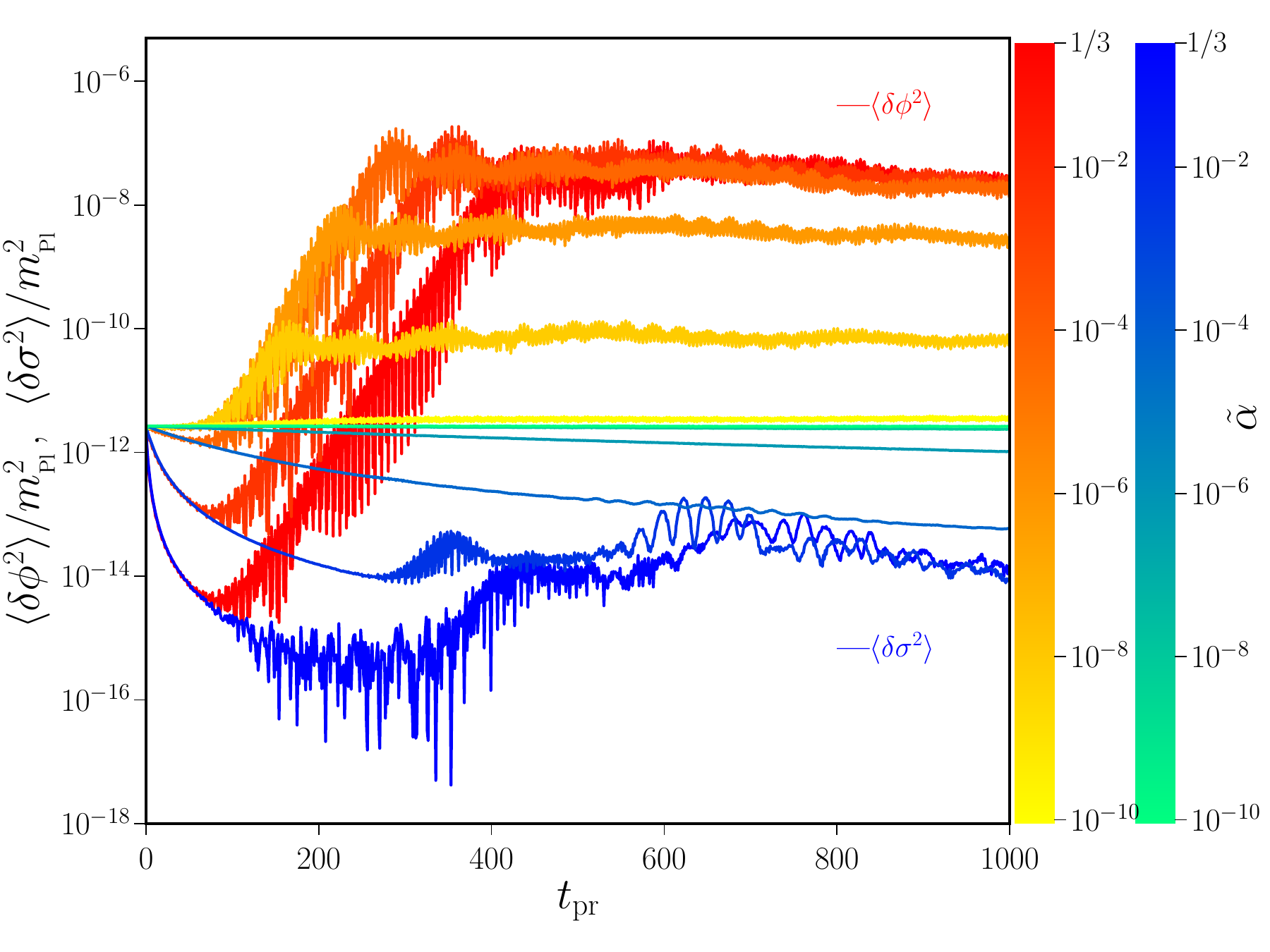}
\vskip -0.1in
\caption{Behavior of variances of $\phi$ (in shades of red to yellow) and 
$\sigma$ (in blue to cyan) obtained numerically, for different 
values of $\Tilde{\alpha}$ in log scale as indicated in the color bar.}
\label{fig:var_vs_alpha}
\end{figure}

We use a three-dimensional lattice of size $32$ in program units with $128$ points 
along each dimension\footnote{We ran several numerical simulations with the following combinations $N^3 = 128^3, N^3 = 64^3$ and  $L^3 = 32^3, L^3 = 60^3$, where $N$ is the lattice points, and $L$ is the lattice size. We have found that the results remained robust without any appreciable changes. In the draft, we have quoted results for $N^3 = 128^3$ and $L^3 =  32^3$.}. 
We perform the simulation for a range of our model parameter 
$\tilde \alpha$. We evolve each run of the lattice up to a time when we capture 
the complete resonant dynamics of the system and the variances of the fields 
settle to asymptotic values.
Having run a simulation for a given set of model parameters, we obtain the 
quantities of interest, such as the variance of the fields, energy densities
, and the spectra of particles produced. We repeat the exercise for different 
values of $\Tilde \alpha$ to observe the effect of this parameter on these 
quantities. We begin with the canonical value of $\Tilde{\alpha}=1/3$ and 
decrease it in logarithmic steps down to $10^{-9}$\,.
The corresponding behaviors of variances of $\phi$ and $\sigma$ are 
presented in Fig.~\ref{fig:var_vs_alpha}.

The behavior of variances contains the tell-tale signature of resonance. 
For the canonical value of $\Tilde{\alpha} = 1/3$, we see that  
$\langle \delta \phi^2 \rangle$ starts with a small initial value 
of around $10^{-12}\,m^2_{\rm Pl}$. 
This value can be understood from the initial conditions imposed on modes of the 
fluctuations $\delta\phi_k(0)\simeq 1/\sqrt{2\,\omega_k(0)}$. 
For a discussion about the imposition of this Bunch-Davies initial condition on modes, refer to Appendix~\ref{app:ini_con}.
On performing inverse Fourier transform of $\langle\vert \delta \phi_k\vert^2\rangle$, 
one can show that the initial value in real space to be $\langle\delta\phi^2(0)\rangle \simeq 50\,m_\phi^2/4$, 
which for our class of models turns out to be 
$\langle \delta\phi^2(0)\rangle \simeq 1.5\times10^{-12}\,m^2_{\rm Pl}$. We should 
note that the quantities involved in the imposition of this initial condition, namely 
$\omega_k$ and $m_\phi$, do not have any explicit dependence on $\tilde\alpha$ 
[cf.~Eq.~\eqref{eq:char-mass}]. Hence, the value is the same across the range of 
$\tilde\alpha$.
At a specific time, denoted as $t_{\rm r}$, the variance $\langle\delta\phi^2(t)\rangle$ rises exponentially for a brief 
while and settles to a very large value of about $\langle \delta \phi^2 \rangle \simeq 1.5 \times 10^{-8}\,m_{\rm Pl}^2$, 
with persistent oscillations throughout the evolution.
As we decrease the value of $\Tilde \alpha$, we see that $t_{\rm r}$ decreases, 
i.e. resonance occurs at earlier times. Also, there is a decrease in the 
amplitude to which $\langle \delta \phi^2 \rangle$ rises and settles.
A similar effect happens in the behavior of $\langle \delta \sigma^2 \rangle$.
The quantity starts with the same initial value as
$\langle \delta \phi^2 \rangle$ and redshifts along with it.
At a certain time after $t_{\rm r}$, it rises as $\sigma$ resonates due to
its coupling with $\phi$.
The asymptotic value at which $\langle \delta \sigma^2 \rangle$ settles is evidently lesser than that of $\langle \delta \phi^2 \rangle$.
This behavior occurs earlier and is less pronounced as we decrease $\Tilde \alpha$. Below $\Tilde{\alpha} = 10^{-9}$, we cannot distinguish the asymptotic
values of variances from their initial values as the quantities just remain
roughly constant throughout the time. 
We should note that the redshifting of the variances is also suppressed as we
decrease $\Tilde \alpha$. This is due to the decrease in energy density ($\rho \sim \Tilde \alpha$) resulting in a decrease of the  Hubble parameter and hence slower variation of the scale factor.
These results essentially convey that the phenomenon of resonance is suppressed
with the decrease of $\Tilde \alpha$. These effects shall be much clearer as we examine the spectra of particle and energy densities in the next subsection.

\begin{figure}
\centering
\includegraphics[width=6in,height=4in]{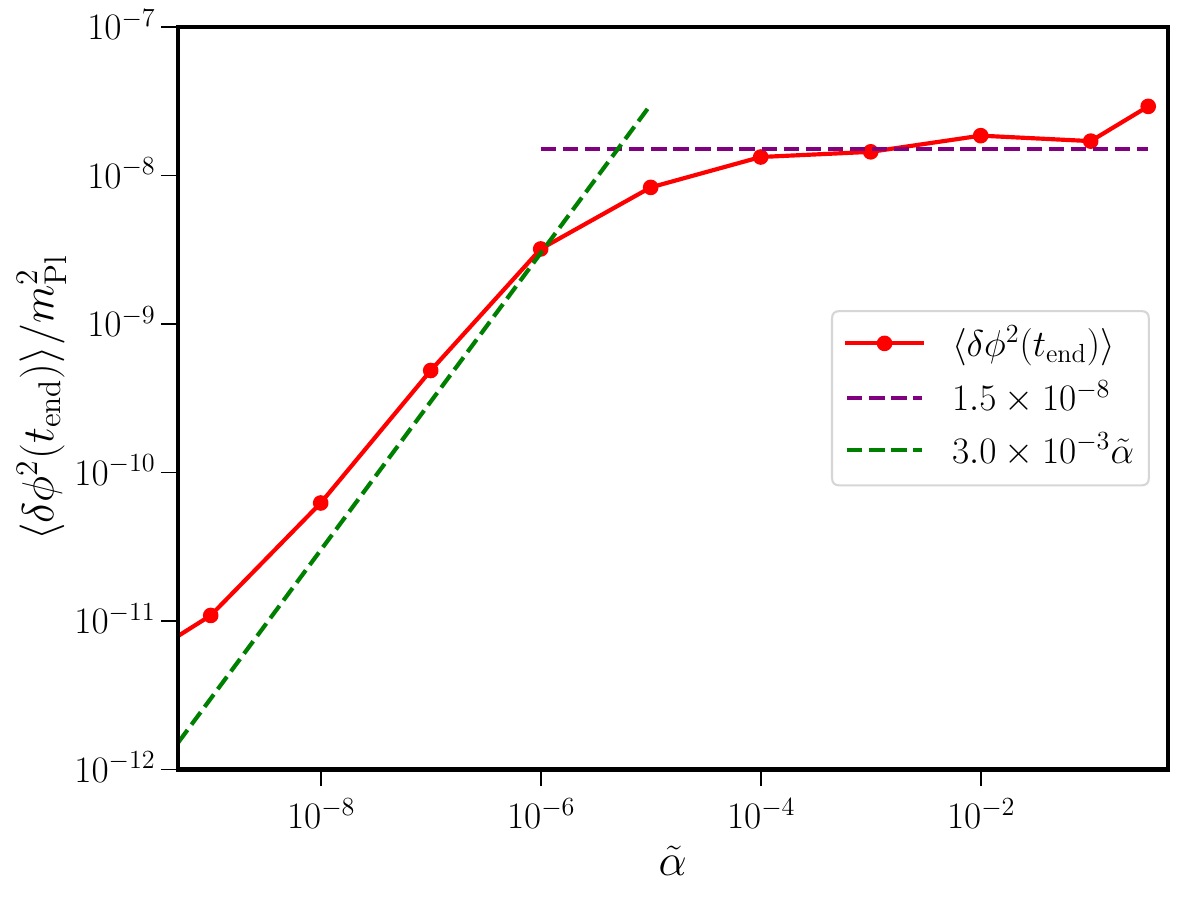}
\vskip -0.1in
\caption{The asymptotic values of $\langle \delta \phi^2(t_{\rm end}) \rangle$ is presented across the range of $\Tilde{\alpha}$.
This asymptotic value is a reliable measure of the $\langle\delta\phi^2(t_r)\rangle$ at the end of resonance, as it is taken well
past the rapid transient behavior.
This is the value that we shall be using in our semi-analytical analysis.
We find that the behavior of $\langle \delta \phi^2(t_{\rm end}) \rangle$ can be 
broadly approximated with a broken power-law of a constant value over 
$\Tilde{\alpha} > 10^{-5}$ and a form  $\sim \Tilde{\alpha}$ over 
$\Tilde{\alpha} < 10^{-5}$.}
\label{fig:var_end_vs_alpha}
\end{figure}

For the purpose of semi-analytical estimation of the abundance of moduli,
we require the behavior of $\langle \delta \phi^2 (t_r)\rangle$ as a function of
$\Tilde \alpha$ from the outputs of this numerical exercise [cf.~Eq.~\eqref{abundance}]. 
This behavior is depicted in Fig.~\ref{fig:var_end_vs_alpha}, where we have 
plotted the asymptotic values of $\langle \delta \phi^2(t_{\rm end}) \rangle$  for different values of $\Tilde{\alpha}$. 
To quantify the trend observed, we approximate the behavior of $\langle \delta 
\phi^2(t_{\rm end}) \rangle$ across $\Tilde \alpha$ using a broken power law as 
\begin{eqnarray}\langle \delta \phi^2(t_{\rm end}) \rangle &=& 
\begin{cases}
1.5\times 10^{-8}\,m^2_{\rm Pl}\,, & \text{for $\Tilde{\alpha} > 10^{-5}\,,$ }\\
3.0\times 10^{-3}\,\Tilde{\alpha}\,m^2_{\rm Pl}\,, & \text{for $\Tilde{\alpha} < 10^{-5}$}\,.
\end{cases}
\label{eq-broken-law}
\end{eqnarray}
This asymptotic value is a reliable measure of the value of the inflaton variance 
$\langle\delta\phi^2(t_r)\rangle$ at the end of resonance that is required in 
the semi-analytical approach of our analysis.
Having understood the complete dependence of $n_\sigma/s$ on $\tilde\alpha$
using Eq.~\eqref{eq-broken-law} in Eq.~\eqref{abundance}, we shall now proceed
to compare this estimate against the result from exact numerical simulations. We note that there is nothing special about $\tilde \alpha \sim 10^{-5}$. It is just due to the simplified form of the fitting function we have used for Fig.~\ref{fig:var_end_vs_alpha}. Instead of fitting with a broken power law with two linear slopes, if we have shown results of extracting $\langle \delta \phi^2 (t_r)\rangle$ via a polynomial fit, we would have gotten best-fit numbers for the coefficients of the polynomial. 
%%%%%%%%%%%%%%%%%%%%%%%%%%%%%%%%%%%%%%%%%%%%%%%%%%%%%%%%%%%%%%%%%%%%%%%%%%%%%%%
%%%%%%%%%%%%%%%%%%%%%%%%%%%%%%%%%%%%%%%%%%%%%%%%%%%%%%%%%%%%%%%%%%%%%%%%%%%%%%%
%%%%%%%%%%%%%%%%%%%%%%%%%%%%%%%%%%%%%%%%%%%%%%%%%%%%%%%%%%%%%%%%%%%%%%%%%%%%%%%

\subsection{Computing number density}
The spectrum of number density and energy density associated with the
fluctuations in $\phi$ and $\sigma$ fields are obtained as outputs of lattice
simulations.
The comoving number density of particles produced for a given field $f$ is 
computed from the spectra of $n_k$ and $\omega_k$ that are obtained as outputs, 
as follows~\cite{Dufaux:2006ee,Felder:2006cc,Dufaux:2007pt}
\begin{eqnarray}
\label{eq:nk_integ}
n_f &=& \frac{4\pi\,B^3}{N^3}
\int_{k_{\rm min}}^{k_{\rm max}}{\rm d}k_{\rm pr} 
k_{\rm pr}^2\,n_{k_{\rm pr}}\,b_{k_{\rm pr}}\,, 
\end{eqnarray}
where the subscript `pr' denotes quantities in program variables,
$b_{k_{\rm pr}}$ is the bin count corresponding to each $k_{\rm pr}$, $B$ is the
conversion factor and $N$ is the number of lattice points used in a simulation
[cf. App.~\ref{app:prog_var}].
The values of $k_{\rm min}$ and $k_{\rm max}$ are taken such that they cover 
sufficient range of wavenumbers around $k=m_\phi$, where we expect resonance to 
occur [cf. App.~\ref{app:prog_var} for details].
The choice of these values do not affect our conclusions as long as they span at 
least a decade on either side of $k=m_\phi$\,.

\begin{figure}
\centering
\includegraphics[width=3.47in]{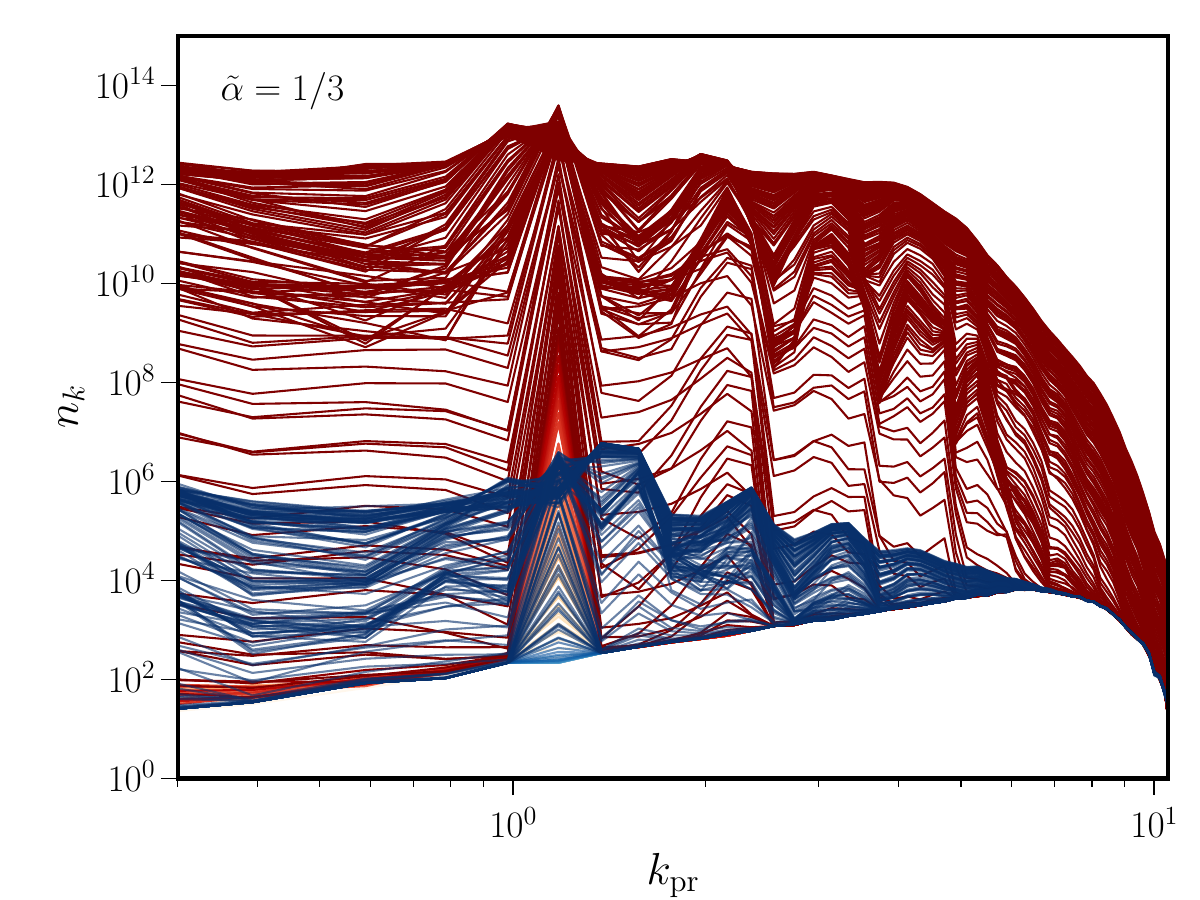}
\includegraphics[width=3.47in]{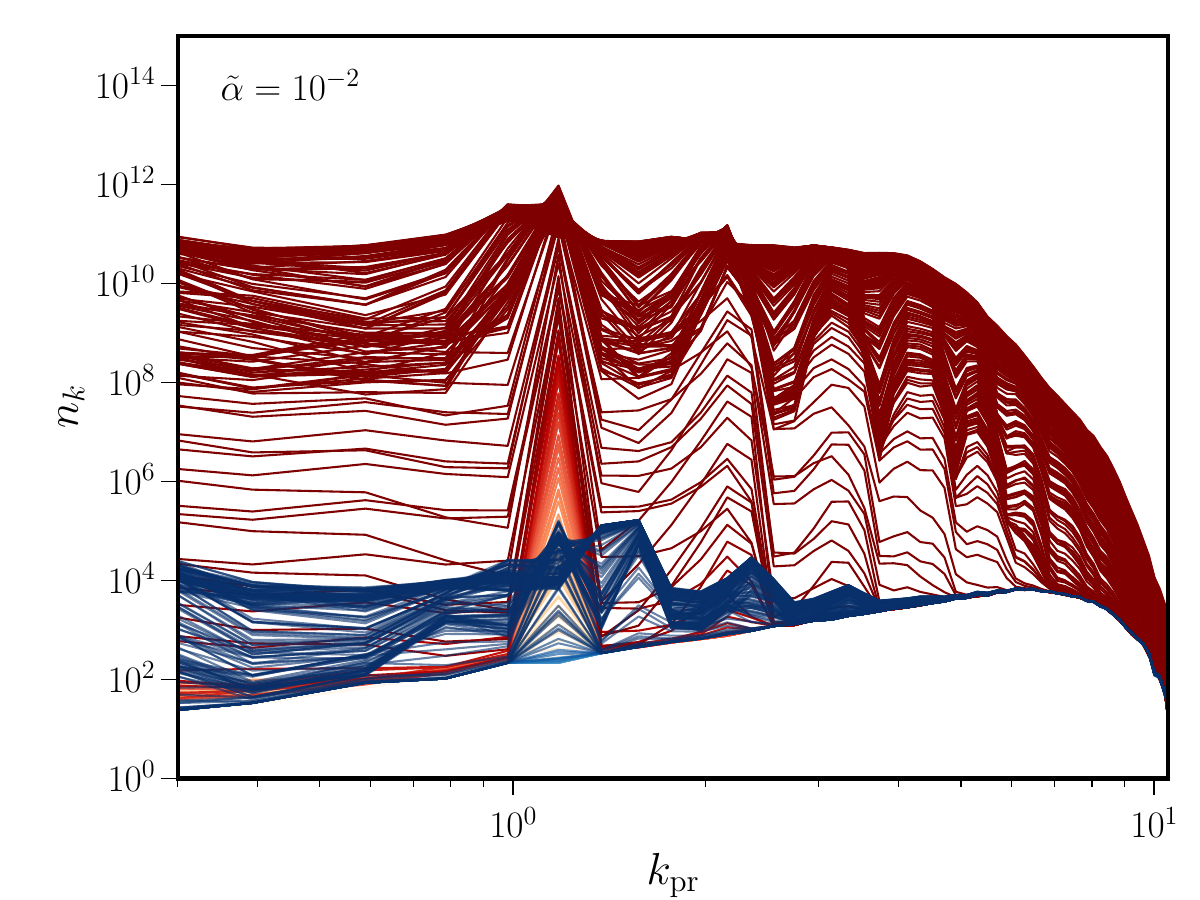}\\
\includegraphics[width=3.47in]{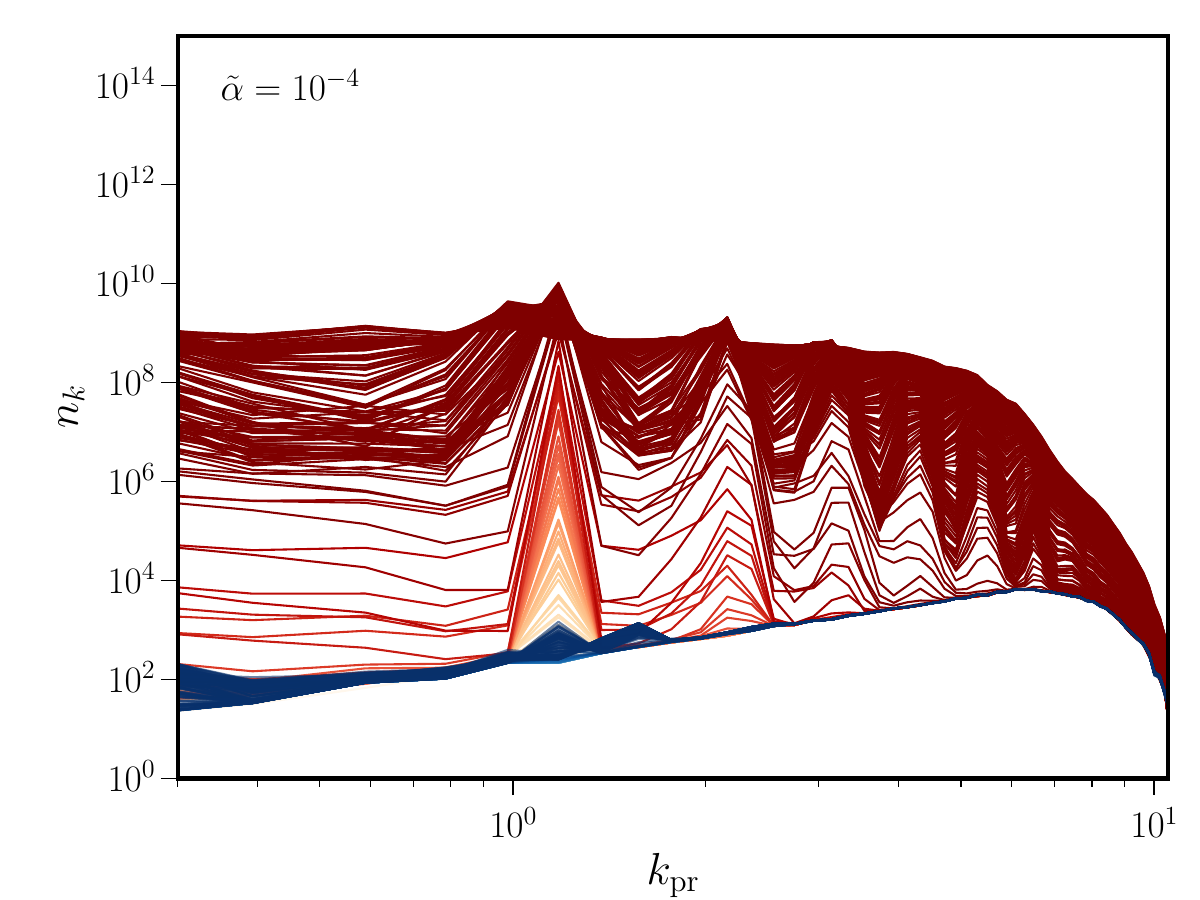}
\includegraphics[width=3.47in]{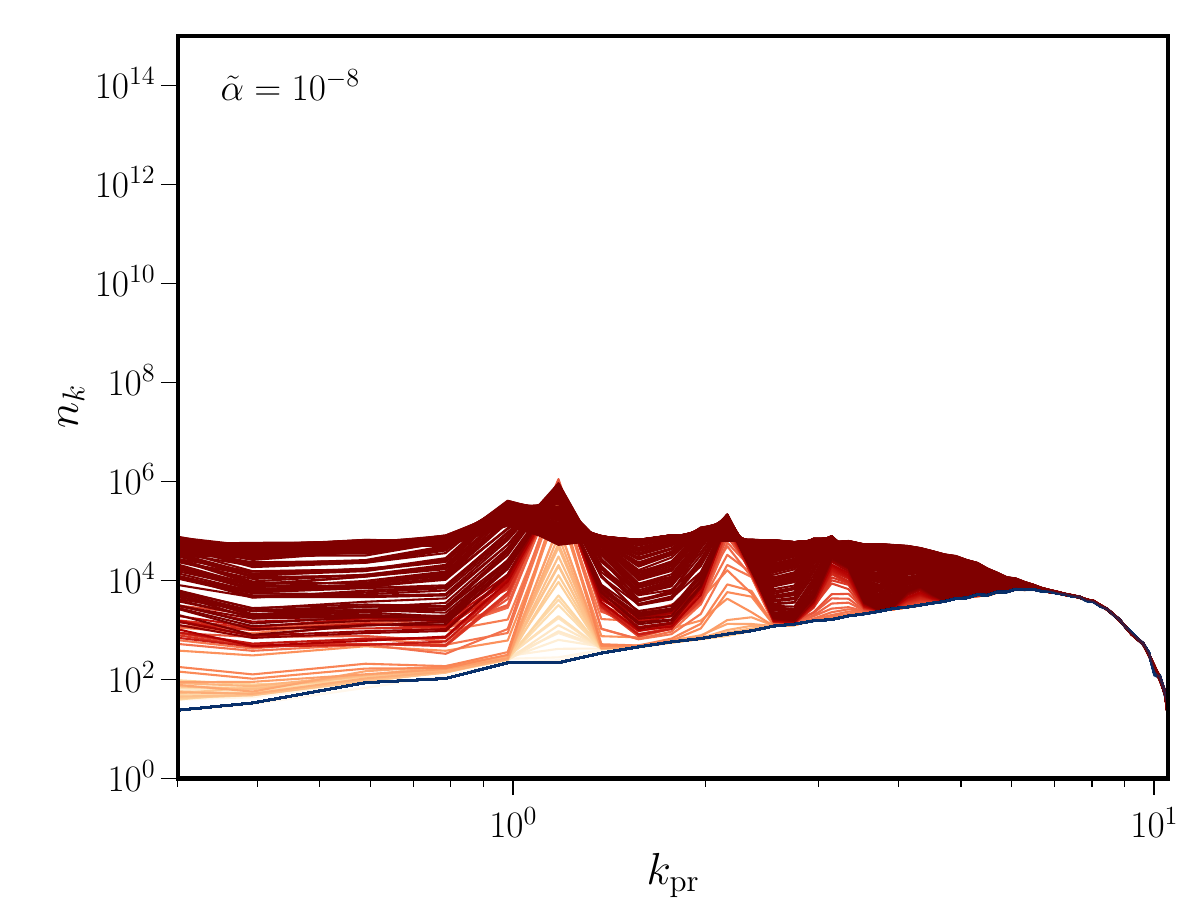}
\vskip -0.1in
\caption{Behavior of spectrum of particle number densities $n_k$ of $\phi$ 
(in red) and $\sigma$ (in blue) are presented for the different values of
$\Tilde{\alpha}$. 
The evolution of the spectrum is represented as shades of the spectra, with
lighter shades of a given spectrum corresponding to early times and darker 
shades to late times. 
Evidently, the resonances in $\phi$ and $\sigma$ are suppressed as $\tilde\alpha$ 
is decreased. Particularly, for values of $\Tilde{\alpha} < 10^{-6}$, there 
seems to be effectively no production of $\sigma$ particles over the spectrum 
of initial background fluctuation.}
\label{fig:var_nk_alpha}
\end{figure}

Before proceeding to compute these quantities, let us examine the spectra 
obtained for different values of $\Tilde{\alpha}$ in the model of interest. 
We present the spectra of number density $n_k$ for $\phi$ and $\sigma$ in 
Fig.~\ref{fig:var_nk_alpha}. The resonant particle production is seen occurring 
at specific wave numbers initially and gradually getting distributed across the
range as time progresses. The enhancement in $n_k$ for any given $\Tilde \alpha$ is 
particularly concentrated around $k_{\rm pr}\simeq1$ i.e. at $k \simeq m_\phi$, 
at initial times as expected.
%%%%%%%%%%%%%%%%%%%%%%%%%%%%%%%%%%%%%%%%%%%%%%%%%%%%%%%%%%%%%%%%%%%%%%%%%%%%%%%
%%%%%%%%%%%%%%%%%%%%%%%%%%%%%%%%%%%%%%%%%%%%%%%%%%%%%%%%%%%%%%%%%%%%%%%%%%%%%%%
This enhancement occurs at the same $k$ for both $\phi$ and $\sigma$, though
lesser in magnitude for $\sigma$, as can be understood by the driving of 
$\sigma$ by self resonance of $\phi$. 
The effect of variation in $\Tilde{\alpha}$ is seen as suppression in the production
of particles with a decrease in its value. We infer that the amplitude of 
enhancement in $n_k$ for both $\phi$ and $\sigma$ fields are significantly
reduced as we lower the value of $\Tilde{\alpha}$ thereby indicating that the 
self-resonance of $\phi$ and the subsequent driving of particle production in 
$\sigma$ is inhibited by reduction of $\Tilde{\alpha}$.

Upon integrating the spectrum of particles using Eq.~\eqref{eq:nk_integ}, we 
obtain the total number density of particles. We should mention here that the spectrum contains the initial background fluctuations that need to be
subtracted to arrive at the spectrum corresponding to resonant particles alone.
Hence, we have to perform numerical subtraction between the final and initial
spectrum of particles to extract the spectrum of resonant particles.

Let us briefly comment on the process of estimating the number density of 
particles produced solely due to resonance. From the output of the lattice 
simulations, we obtain the spectrum of the total number density of particles $n_k$, for 
the fields involved. We see that there always exists a background spectrum of particles, say $n_k^0$, at the initial time of 
evolution. After resonance, particularly for large values of $\tilde\alpha$, the spectrum 
is highly dominated by particles produced due to resonance at discrete increments of 
wavenumbers beginning with $k_{\rm pr}=1$.
The background spectrum $n_k^0$ can be understood to be the spectrum corresponding
to the solution of moduli field denoted earlier as $\overline{\sigma}_{ws}$ and $\overline{\sigma}_1$ whereas 
the spectrum of resonant particles corresponds to the solution denoted as 
$\delta\overline{\sigma}$ induced by $\delta \phi$ [cf.~Eq.~\eqref{eq:sigma_soln}].
For large $\tilde\alpha$, the spectrum of resonant particles can be essentially taken to 
be the final spectrum of particles and so the number density of resonant particles 
is the same as the total number density at the end of evolution.
However, for smaller values of $\tilde\alpha$, the resonance is less pronounced and the final
spectrum $n_k$ is closer to the initial background spectrum $n_k^0$. In these cases, 
one has to carefully subtract $n_k^0$ from $n_k$ to extract the number density of  
resonant particles.

\begin{figure}[t]
\includegraphics[width=3.47in]{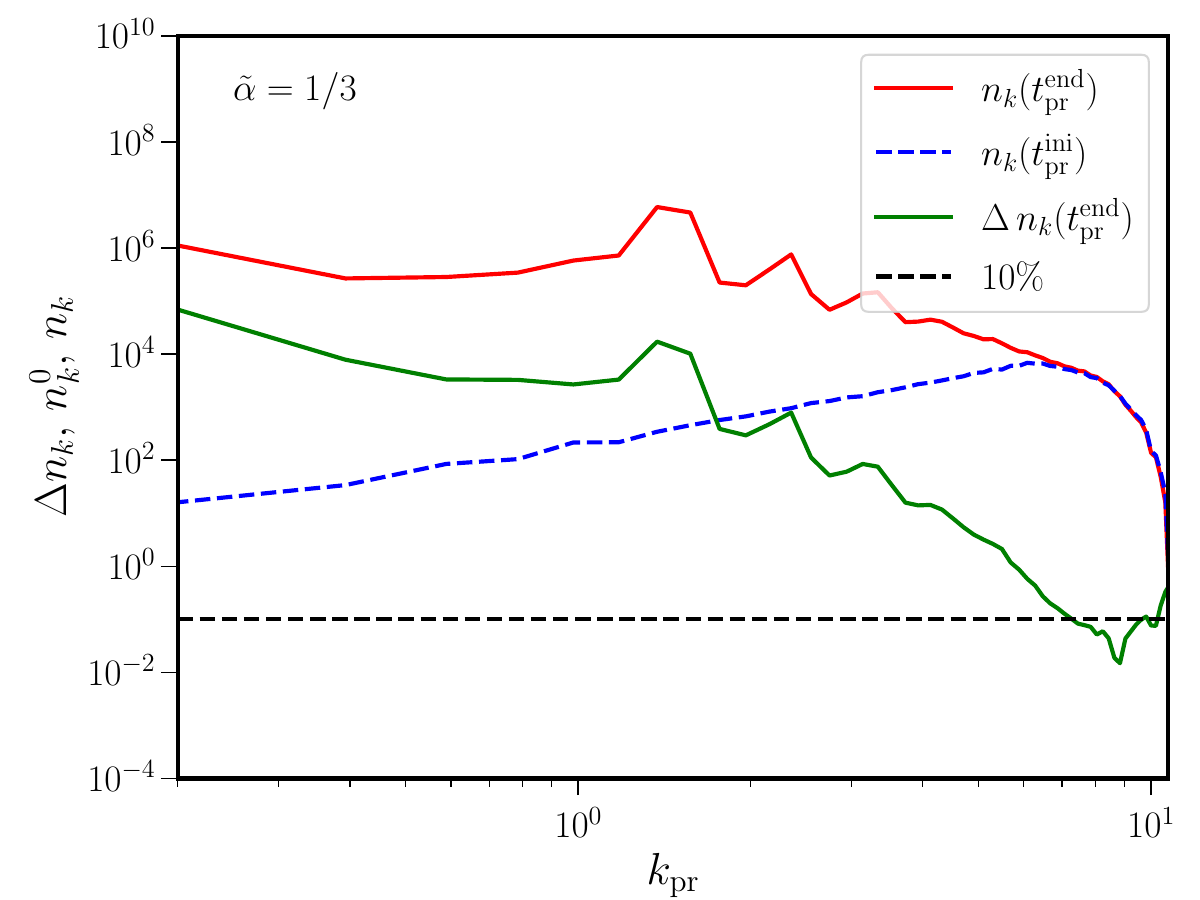}
\includegraphics[width=3.47in]{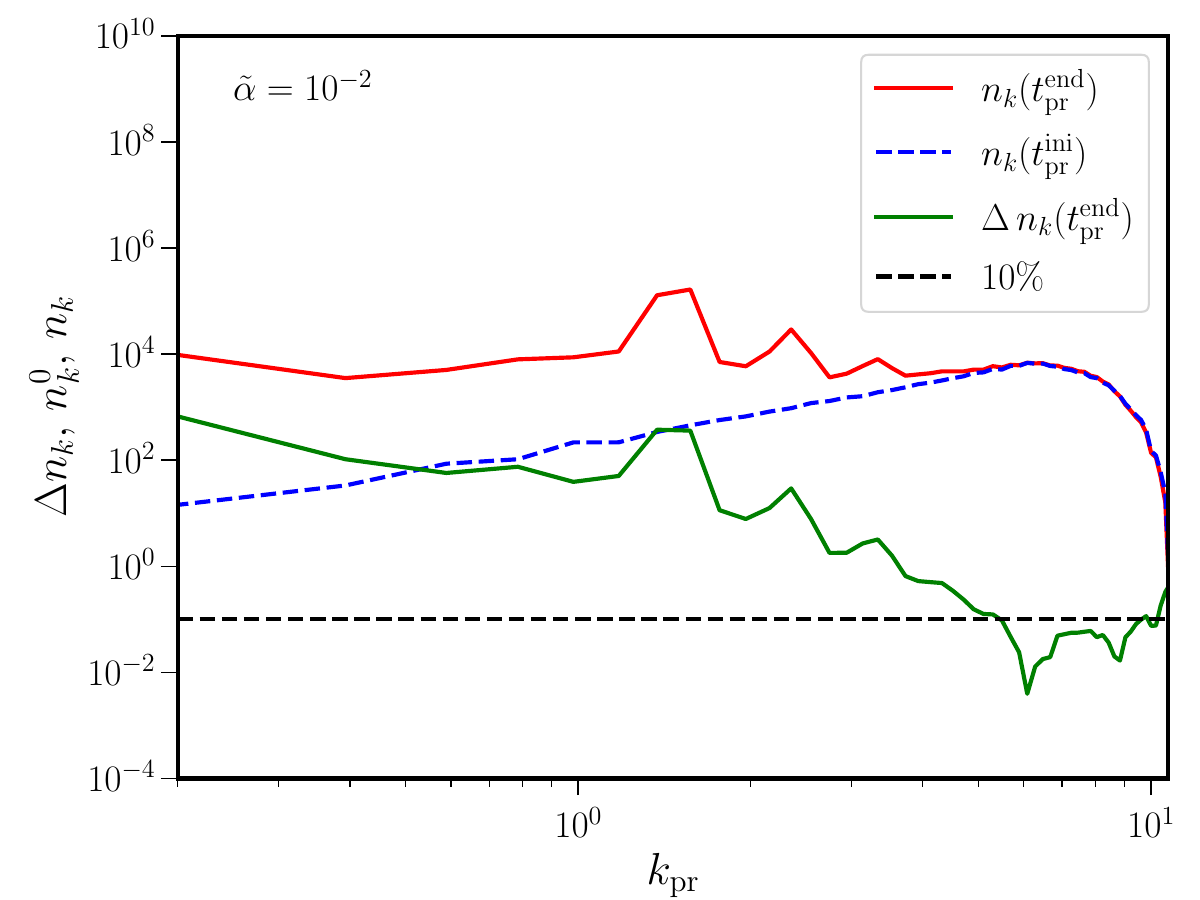}\\
\includegraphics[width=3.47in]{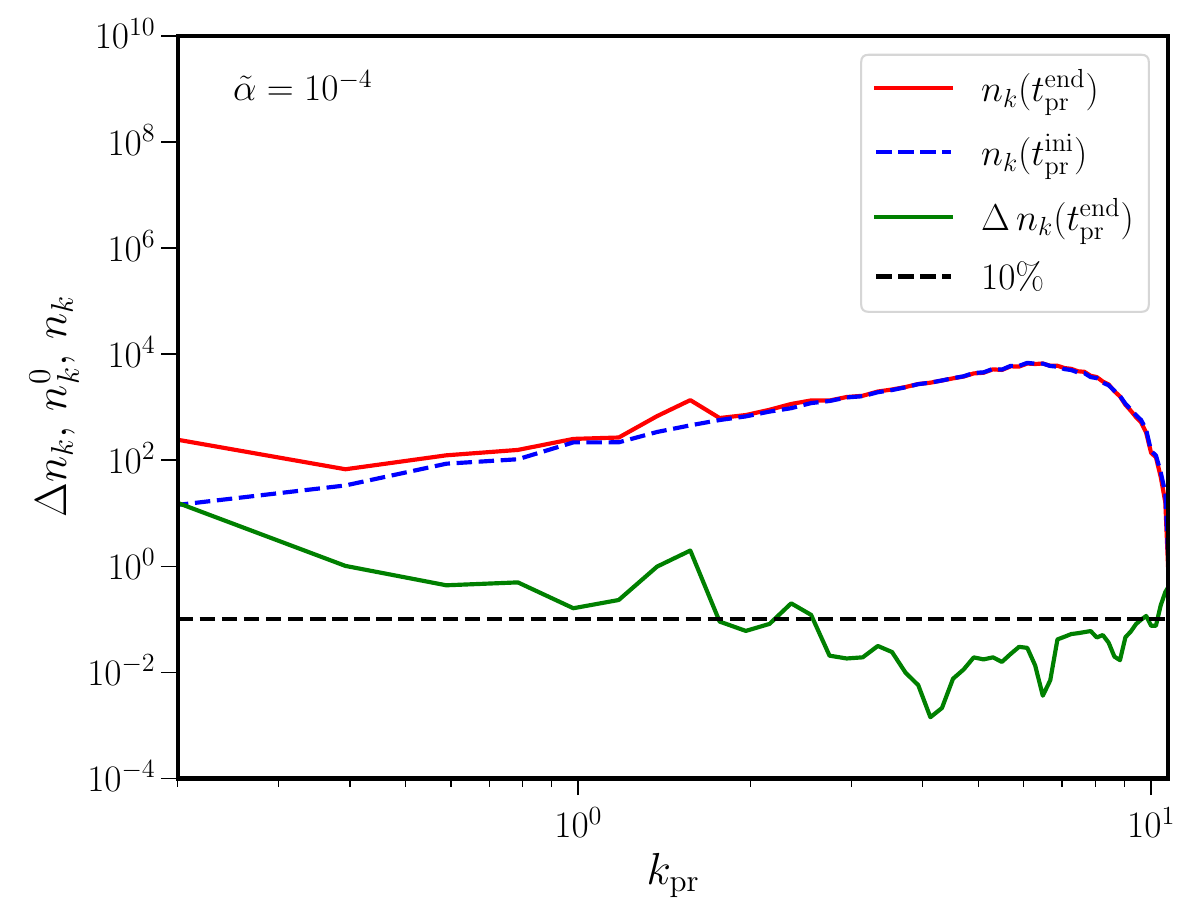}
\includegraphics[width=3.47in]{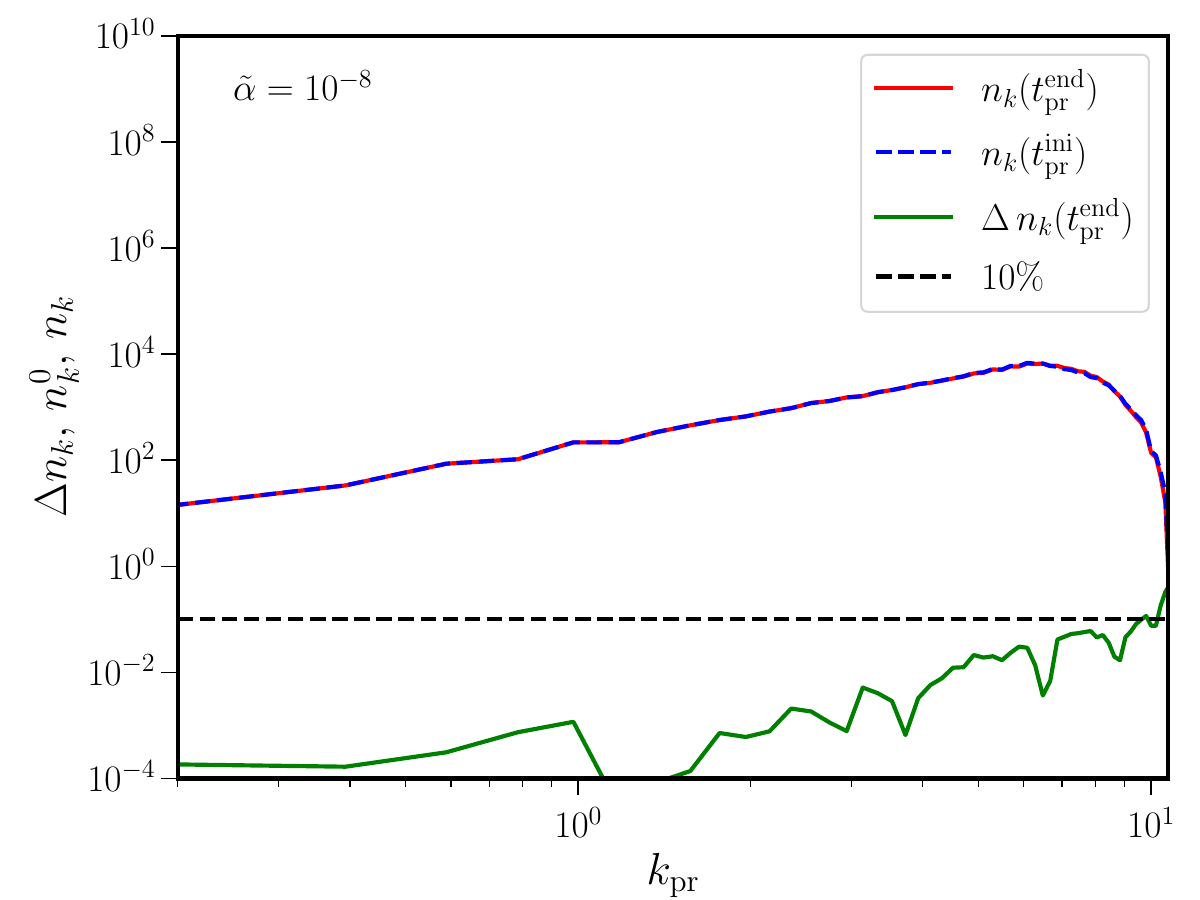}
\vskip -0.1in
\caption{The behavior of the relative difference of the spectrum of number 
density $\Delta n_k \equiv \vert n_k-n_k^0 \vert/n_k^0$ is plotted for the 
moduli field across different values of $\Tilde{\alpha}$ (in green). Also plotted 
are the final spectrum of number density $n_k$ (in red) and the offset spectrum 
$n_k^0$ (in dashed blue). The line corresponding to $10\%$ is plotted 
(in dashed black) as a suggestive value below which the relative difference can 
be treated effectively as zero.}
\label{fig:rel_nk_alpha}
\end{figure}

In our analysis, we have performed this subtraction of $n_k-n_k^0$ to arrive at 
the number density of the resonant particles as illustrated in Fig.~\ref{fig:rel_nk_alpha}. 
We can readily infer that the spectrum of relative difference in the number density, 
i.e. $\Delta n_k \equiv \vert n_k - n_k^0 \vert/n_k^0$ tends to small values as 
$\tilde\alpha$ is decreased.
Such a subtraction for the cases of small values of $\tilde\alpha$ means taking the 
numerical difference of two comparable values and hence is limited by the numerical
precision of the setup, below which we cannot make reliable estimates. We have taken  
$10\%$ as a threshold value so that $\Delta n_k$ smaller than this value may be treated 
as zero. Though the relative difference is small, the difference itself may be 
non-negligible. Upon integration, these non-negligible differences lead to 
considerable values of number densities. These number densities when converted
to physical variables with suitable dimensions turn out to have significant 
magnitudes and hence affect further computations.
Therefore, we should keep track of the effect of this error due to numerical
precision throughout the rest of the calculations especially for small values of
$\tilde\alpha$. 
We shall further comment on this number density of moduli, with the background 
spectrum subtracted when we utilize it in the computation of $n_\sigma/s$ in subsection \ref{nsigmas}.
%%%%%%%%%%%%%%%%%%%%%%%%%%%%%%%%%%%%%%%%%%%%%%%%%%%%%%%%%%%%%%%%%%%%%%%%%%%%%%%
%%%%%%%%%%%%%%%%%%%%%%%%%%%%%%%%%%%%%%%%%%%%%%%%%%%%%%%%%%%%%%%%%%%%%%%%%%%%%%%

\subsection{Estimation of $\rho_{\rm total}$}
\begin{figure}[t]
\includegraphics[width=3.47in]{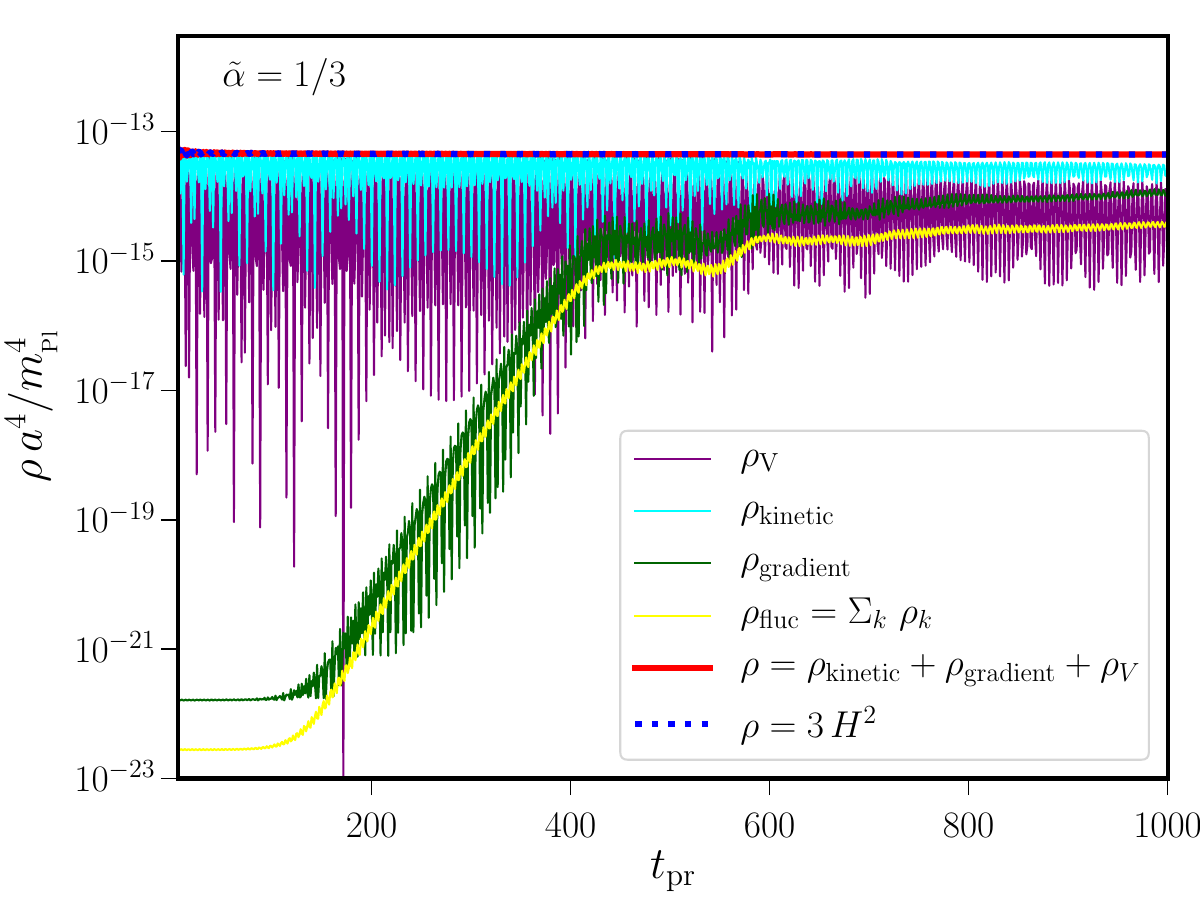}
\includegraphics[width=3.47in]{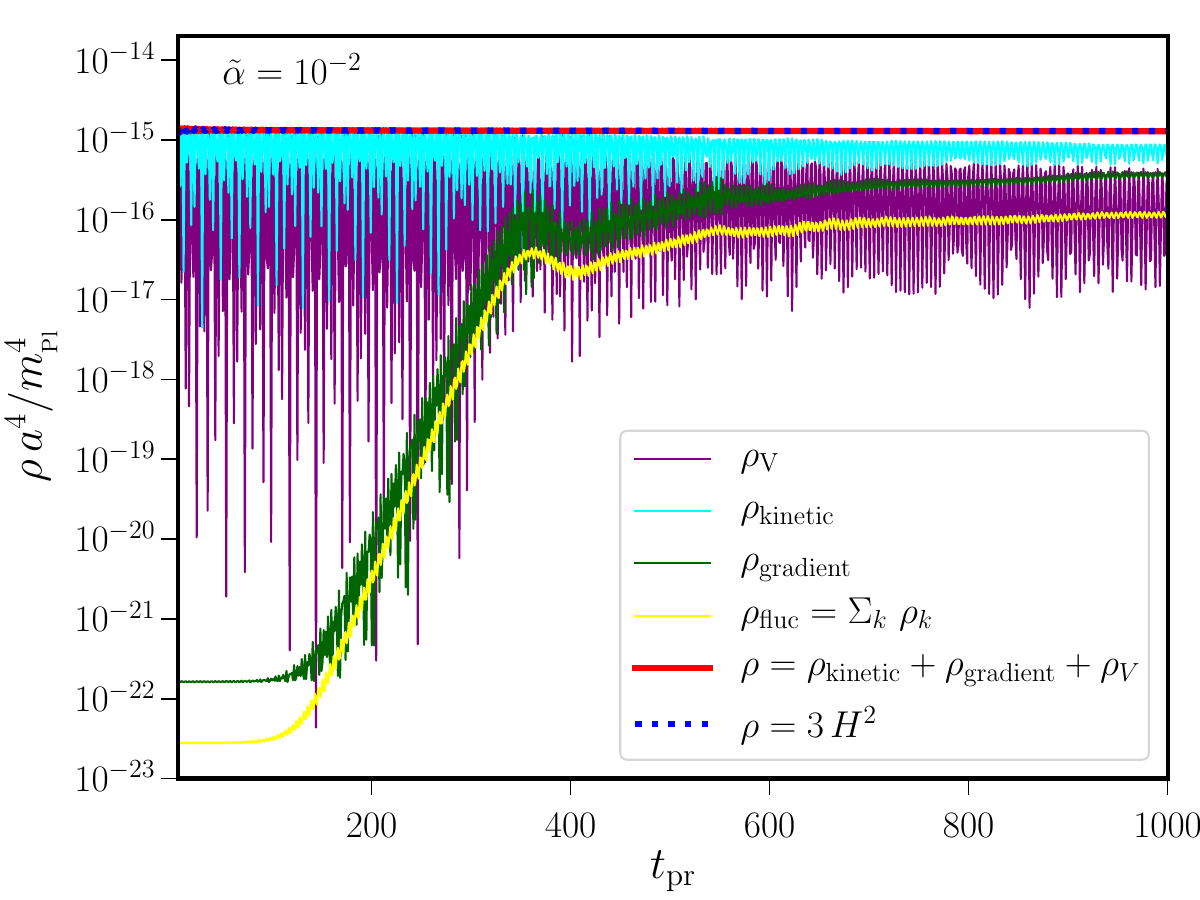}\\
\includegraphics[width=3.47in]{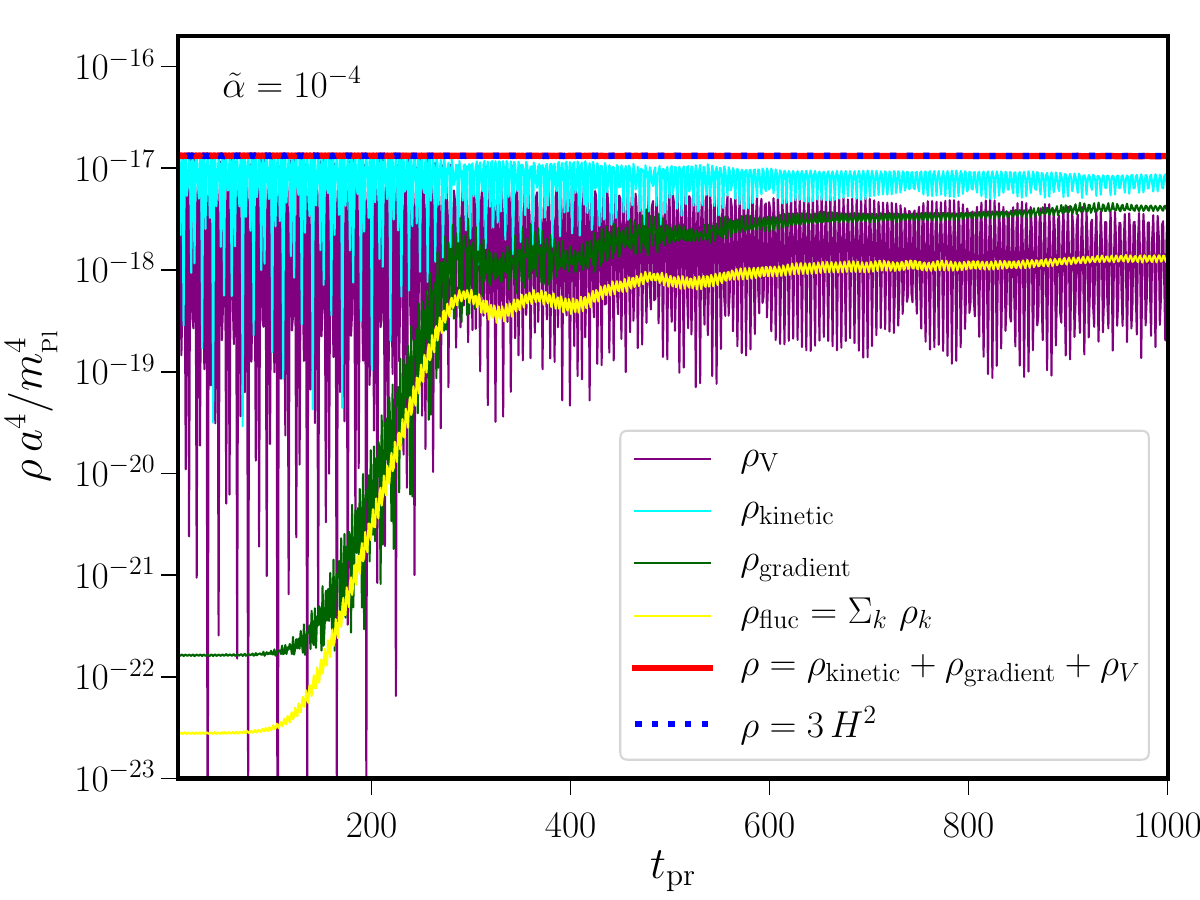}
\includegraphics[width=3.47in]{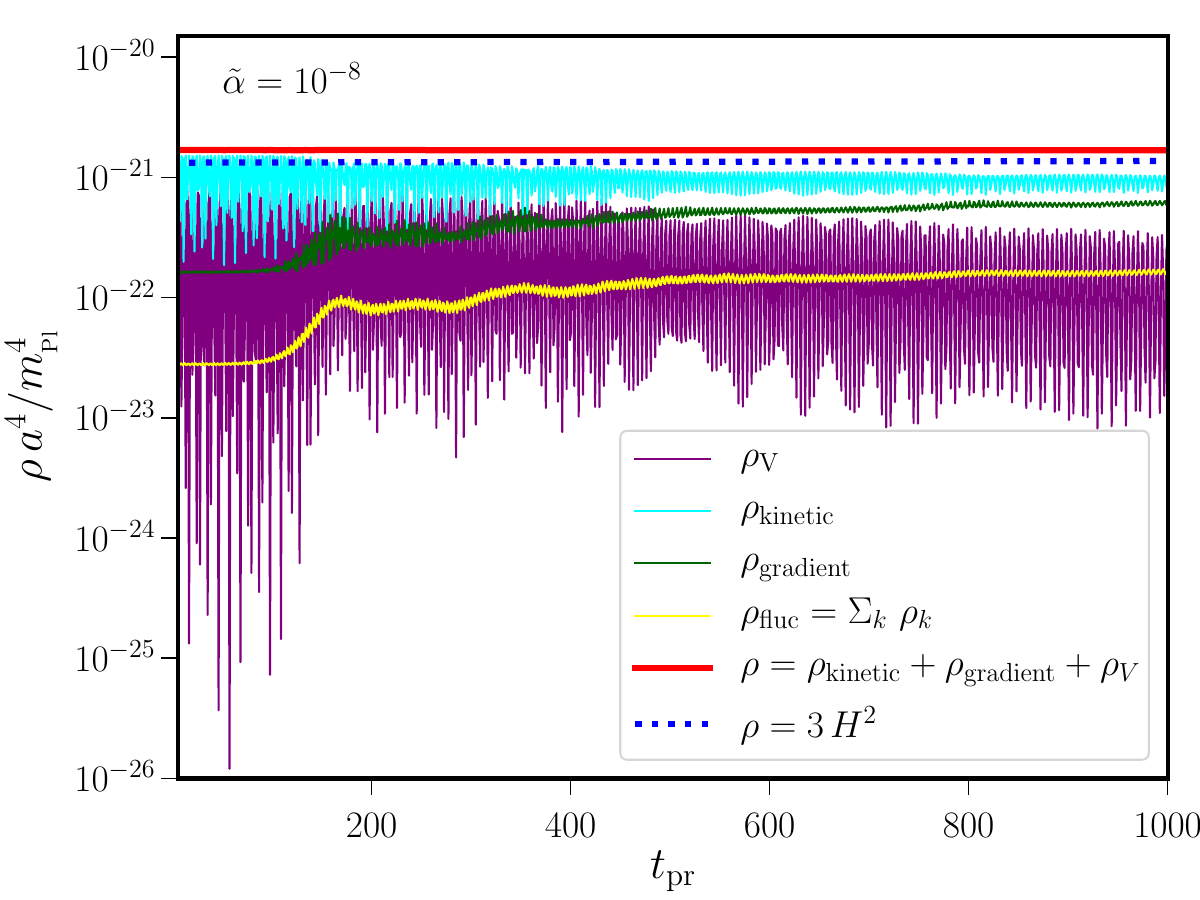}
\vskip -0.1in
\caption{The evolution of the comoving energy density $\rho\,a^4$ obtained from 
different methods is plotted across different values of $\Tilde{\alpha}$ (in red). 
The different components of the total energy are plotted separately for clarity, 
namely, the potential energy (in purple), total kinetic energy (in cyan), gradient 
energy (in green). The energy associated with the fluctuations obtained from summing 
the spectrum is also plotted (in yellow) and it closely follows the gradient energy.
Besides, the estimate of energy density using the Friedmann equation $3H^2m^2_{\rm Pl}$ is 
plotted (in dashed blue) to validate the estimate of total energy of the system.}
\label{fig:rho_alpha}
\end{figure}

We are interested in calculating the moduli abundance at the end of resonance time ($t_{r}$). To estimate the moduli abundance, we have to calculate the number density of the moduli field ($n_{\sigma}$) and the total entropy density ($s$) at the end of the resonance time. In the previous section, we discussed how to calculate the number density of the moduli field in detail. To calculate the entropy density, we need to know the total energy density of the system because there is a direct relation between the total energy density($\rho_{\rm total}$) and the entropy density($s\sim \rho_{\rm total}^{3/4}$). So, in this section, we will discuss the total energy density of the system in detail. Using the output files of the $\texttt{Latticeeasy}$~\cite{Felder:2000hq}, we can calculate the total energy density of the system in two ways. One method adds all energy components in real space, and another using the Friedmann equation. To calculate the total energy density using the first method, we have to know how the various parts of energy evolve with time. The various components of energy such as kinetic energy density, gradient energy density, and potential energy density are 
\begin{align}\label{component_of_energy}
\rho_{\rm kinetic}&=\frac{1}{2}\dot{\phi}^2+\frac{1}{2}\dot{\sigma}^2\,,\\ 
\rho_{\rm gradient}&=\frac{1}{2a^2}(\nabla\phi)^2+\frac{1}{2a^2}(\nabla\sigma)^2\,, \\
\rho_{V}&= V_{0}\left(\frac{\phi}{\sqrt{6\alpha}}\right)^4\left[1+c\frac{\sigma}{m_{\rm pl}}\right]\,.
\end{align}
So the total energy density ($\rho_{\rm total}$) of the system in the configuration space is,
\begin{equation}\label{total_conf_energy}
    \rho_{\rm total}=\rho_{\rm kinetic}+\rho_{\rm gradient}+\rho_{V}.
\end{equation}
On the other hand, the total energy density computed using the Friedmann equation is
\begin{equation}\label{Friedmann_equation}
H^2\,=\,\frac{8\pi}{3m^2_{_{\rm Pl}}}\,\,\rho_{\rm total} .
\end{equation}
In Fig.~\ref{fig:rho_alpha}, we have plotted the time evolution of the energy density of the various components, such as kinetic energy density, gradient energy density, and potential energy density with different colours such as purple, cyan, and green, respectively. The red and dashed blue line plots represent the total energy density of the system, which is calculated using the Eqs.~\eqref{total_conf_energy}, and \eqref{Friedmann_equation} respectively, and the total energy density calculated from both ways agree well.
%%%%%%%%%%%%%%%%%%%%%%%%%%%%%%%%%%%%%%%%%%%%%%%%%%%%%%%%%%%%%%%%%%%%%%%%%%%%%%%
%%%%%%%%%%%%%%%%%%%%%%%%%%%%%%%%%%%%%%%%%%%%%%%%%%%%%%%%%%%%%%%%%%%%%%%%%%%%%%%

From the Eq.~\eqref{EOM_conf_eq}, the particle production due to resonance is not possible without the presence of the gradient term. So the total gradient energy density in the configuration space should follow the total energy density of the particle production due to resonance. We will check it using the output files of the $\texttt{Latticeeasy}$~\cite{Felder:2000hq}. Now, we define the particle production due to resonance for the inflaton and the moduli,
\begin{align}
    \rho^{\rm fluc}_{\rm \phi}&=\frac{1}{(2\pi)^3a^4}\int^{k_{\max}}_{k_{\rm min}}d^3k~\omega_{k}^{\phi}n_{k}^{\phi}\,,\\
    \rho^{\rm fluc}_{\rm \sigma}&=\frac{1}{(2\pi)^3a^4}\int^{k_{\max}}_{k_{\rm min}}d^3k~\omega_{k}^{\sigma}n_{k}^{\sigma},
\end{align}
where $\omega^{\phi}_{k}$($\omega_{k}^{\sigma}$) is the frequency of the inflaton(moduli) for a particular mode $k$, and $n_{k}^{\phi}$($n_{k}^{\sigma}$) is the number density of the inflaton(moduli) for a particular mode $k$ which is shown in Fig.~\ref{fig:var_nk_alpha}. So, the total energy density of the particle due to resonance is $\rho_{\rm fluc}=\rho_{\rm fluc}^{\rm \phi}+\rho_{\rm fluc}^{\sigma}$. In Fig.~\ref{fig:rho_alpha}, the yellow line represents the total energy density of the particles due to resonance and the green line represents the total gradient energy density. From Fig.~\ref{fig:rho_alpha}, we see that the particle production due to resonance follows the gradient energy density across all values of $\Tilde{\alpha}$. It also shows that the total energy density of the particle production due to resonance and the total gradient energy density decreases as $\Tilde{\alpha}$ decreases.
%%%%%%%%%%%%%%%%%%%%%%%%%%%%%%%%%%%%%%%%%%%%%%%%%%%%%%%%%%%%%%%%%%%%%%%%%%%%%%%

\subsection{Computing $n_\sigma/s$}
\label{nsigmas}
Having obtained the numerical estimates of number density for the moduli field 
$n_\sigma$ and the total energy density $\rho$, we proceed to compute the ratio 
of observational relevance, $n_\sigma/s$. We shall obtain this numerically and 
compare against the analytically expected trend as given in 
Eq.~\eqref{abundance}. Once again, as commented earlier, the ratio is to be 
computed using the number density corresponding to the particles produced due to 
the resonance of the moduli fields, apart from the particles corresponding to the 
background vacuum fluctuations that is already present for any given field.
%%%%%%%%%%%%%%%%%%%%%%%%%%%%%%%%%%%%%%%%%%%%%%%%%%%%%%%%%%%%%%%%%%%%%%%%%%%%%%%
\begin{figure}[!t]
\includegraphics[width=3.47in]{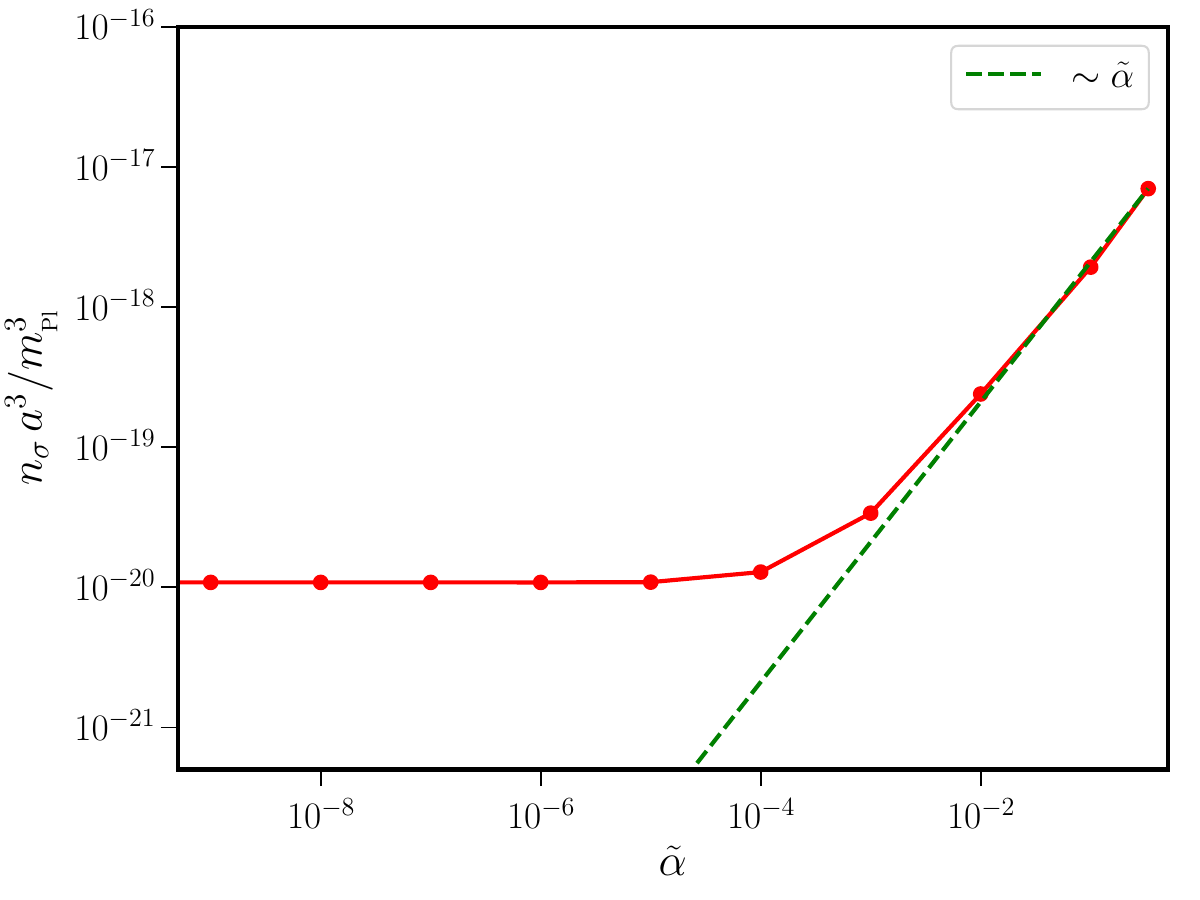}
\includegraphics[width=3.47in]{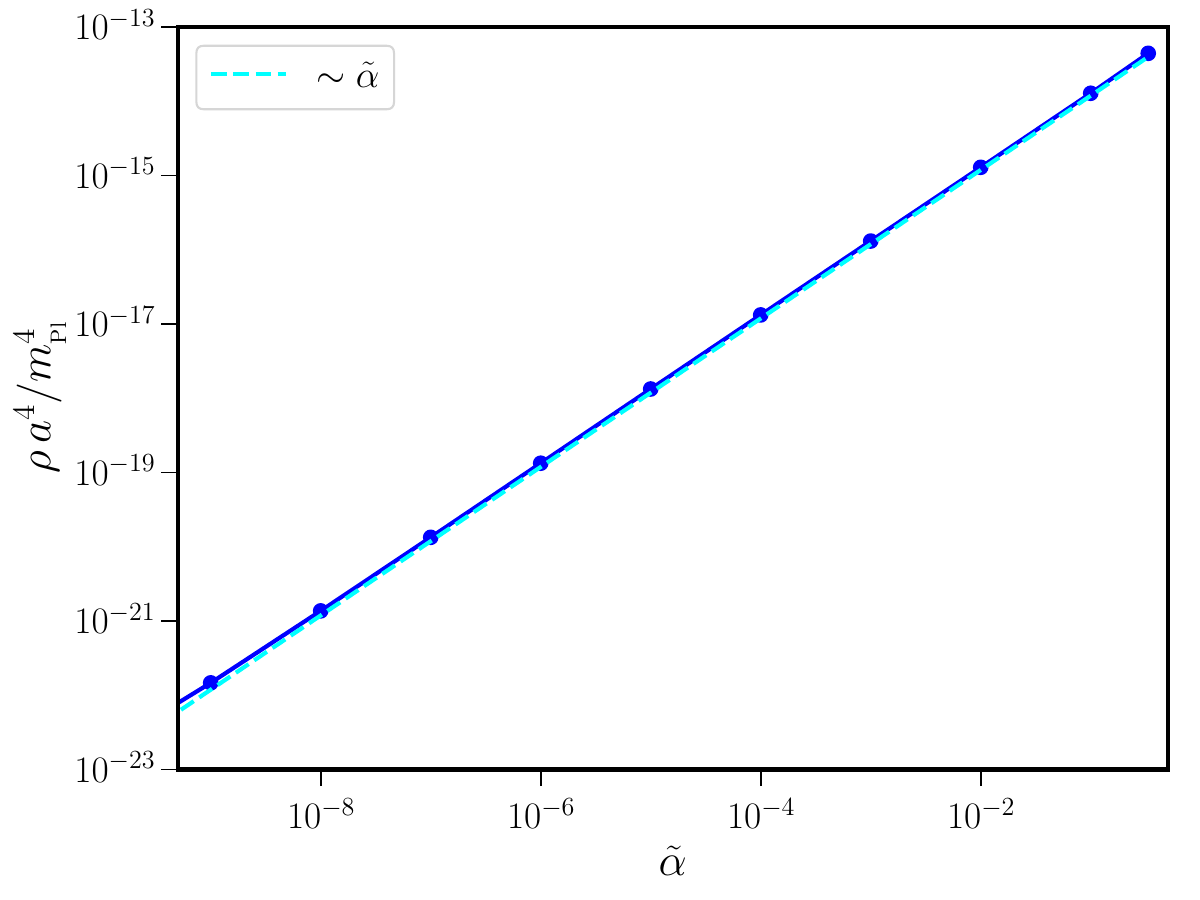}
\vskip -0.1in
\caption{The behavior of the comoving number density of moduli $n^{\sigma}a^3$ 
and the total comoving energy density of the system $\rho\,a^4$ are plotted at 
their asymptotic values against $\Tilde{\alpha}$.
Note that $n_\sigma$ plotted corresponds to just the resonant particles 
produced, excluding the number density due to background spectrum as
explained in the main text.  This comoving number density $n_\sigma$ behaves as 
$\sim \Tilde{\alpha}$ over large values of $\Tilde{\alpha}$ down to $10^{-4}$ but 
over smaller values, as errors dominate, it flattens to a constant value of $10^{-20}\,m^3_{_{\rm Pl}}$.
On the other hand, the overall scaling of $\rho$ is found to be $\Tilde{\alpha}$, 
which is as expected and explained in the main text.}
\label{fig:n_rho_alpha}
\end{figure}
%%%%%%%%%%%%%%%%%%%%%%%%%%%%%%%%%%%%%%%%%%%%%%%%%%%%%%%%%%%%%%%%%%%%%%%%%%%%%%%

In Fig.~\ref{fig:n_rho_alpha}, we present the comoving number density of moduli 
field $n_\sigma\,a^3$ and the total comoving energy density $\rho\,a^4$ at 
their asymptotic values, across the range of $\tilde\alpha$. We see that as 
$\tilde\alpha$ decreases, $n_\sigma\,a^3$ initially scales as $\tilde\alpha$ over 
$\tilde\alpha > 10^{-4}$. But for $\tilde\alpha \lesssim 10^{-4}$, 
it settles to a constant value of about $10^{-20}\,m^3_{\rm Pl}$. The behavior 
over the range of $\tilde\alpha > 10^{-4}$ is dominated by resonant 
particles whereas over $\tilde\alpha < 10^{-4}$ the settling is 
due to the unavoidable loss of numerical precision arising from the difference 
$n_k - n_k^0$. 
Hence we see the difference in the scaling with respect to $\Tilde \alpha$. 
This behavior due to numerical error is discussed in App.~\ref{app:num_err}.
On the other hand, $\rho\,a^4$ is proportional to $\Tilde \alpha$ throughout
the whole range and it involves no errors due to numerical difference.

At this stage, we shall compare the numerical results against the semi-analytical 
trends obtained earlier. The behaviors of the comoving densities across the range of 
$\tilde\alpha$ are as expected from the semi-analytical expressions that we have 
obtained. But at the value of $\tilde\alpha=1/3$ we see that the amplitudes do not 
match exactly. While the semi-analytical estimate gives the comoving number density 
to be $5.71\times 10^{-13}\,m_{_{\rm Pl}}^3$, the numerical simulation gives a 
value of $7.02\times 10^{-18}\,m_{_{\rm Pl}}^3$\,. This difference can be 
attributed to the approximations involved in arriving at the analytical expressions. 
To correct for this discrepancy arising due to approximations and match the 
analytical value with the numerical result at $\tilde \alpha = 1/3$, we introduce a correction factor to our expression of $n_\sigma$, denoted as ${\cal F}$.

Recall that the expression of $n_\sigma$ in terms of $\tilde\alpha$ and 
$\langle\delta\phi^2\rangle$ was obtained in Eq.~\eqref{nsigma}.
Introducing $\cal F$, we can write this expression as
\begin{eqnarray}
n_\sigma\,a(t_{\rm r})^3 &\simeq & 5.71\times 10^{-13}{\cal F} \,c^2\,
    \left(\frac{\ps(k_\ast)}{2.1\times 10^{-9}}\right)^{1/2}
    \left(\frac{\tilde\alpha}{1/3}\right)
    \left(\frac{\langle\delta\phi^2(t_{r})\rangle}{10^{-8}m^2_{\rm Pl}}\right)\,
    m^3_{\rm Pl}\,.
    \label{eq:nsigma-final}
\end{eqnarray}
We determine ${\cal F}=1.23\times 10^{-5}$ which lets us correct the
analytical estimate and match it against the numerics. In Fig.~\ref{fig:n_rho_alpha}, we have plotted the quantities accounting for the correction factor ${\cal F}$.

Further, utilizing the broken power law behavior of 
$\langle \delta \phi^2 \rangle$ as observed from numerics we may get the complete behavior of $n_\sigma$ as
\begin{eqnarray}
n_{\sigma}\,a^3(t_{\rm r}) &\simeq& 5.71 \times 10^{-13} {\cal F}
c^2
\left(\frac{\ps(k_\ast)}{2.1\times 10^{-9}}\right)^{1/2} \,m^3_{\rm Pl}
\begin{cases}
\displaystyle\f{\tilde\alpha}{1/3}\,
\left(\frac{\langle \delta \phi^2(t_r) \rangle}
{1.5\times 10^{-8}\,m^2_{\rm Pl}}\right)\,, & 
\text{for $\tilde\alpha > 10^{-5}$}\,, \\
\left(\displaystyle\f{\tilde\alpha}{1/3}\right)^2\,
\displaystyle\left(\frac{\langle \delta \phi^2(t_r) \rangle}
{3.0\times 10^{-3}\,m^2_{\rm Pl}}\right)\,, & 
\text{for $\tilde\alpha < 10^{-5}$}\,. \\
\end{cases}
\end{eqnarray}
This behavior closely matches the numerically obtained comoving number 
density over $\tilde\alpha > 10^{-4}$.
However, for smaller values of $\tilde\alpha$, the numerical value of $n_\sigma$ is determined by the numerical error and it is independent of $\tilde\alpha$ as explained in App.~\ref{app:num_err} and observed in Fig.~\ref{fig:n_rho_alpha}.  
Therefore, having verified the semi-analytical estimate over $\tilde\alpha > 10^{-4}$, we continue to rely on it over $\tilde\alpha \lesssim 10^{-4}$, where
numerics are limited by precision and use the trend to draw conclusions about
the observable quantities.

On the other hand, we obtain the entropy density $s$ from the estimate of total 
energy density $\rho$, assuming that all the energy of the system shall be 
converted to relativistic particles and they eventually thermalize. So, we assume $s=\rho^{3/4}$, as mentioned earlier, 
and we may expect the dependence of entropy density on $\tilde\alpha$ to be 
$s \sim \tilde\alpha^{3/4}$.
Hence, the ratio of interest can be written as
\begin{eqnarray}
\frac{n_{\sigma}}{s} &\simeq& 7.28 \times 10^{-8}\,c^2
\left(\frac{\ps(k_\ast)}{2.1\times 10^{-9}}\right)^{-1/4}
\left(\f{\tilde\alpha}{1/3}\right)^{1/4}\,\left(\frac{\langle \delta \phi^2(t_r) \rangle}
{1.5\times 10^{-8}\,m^2_{\rm Pl}}\right)\,. 
\label{eq:nbys-full}
\end{eqnarray}
With the broken power law approximation for $\langle \delta \phi^2 (t_r) \rangle$ we can write the explicit scaling to be
\begin{eqnarray}
\frac{n_{\sigma}}{s} &\simeq& 7.28 \times 10^{-8}\,c^2
\left(\frac{\ps(k_\ast)}{2.1\times 10^{-9}}\right)^{-1/4}
\begin{cases}\displaystyle\left(\f{\tilde\alpha}{1/3}\right)^{1/4}\,
\displaystyle\left(\frac{\langle \delta \phi^2(t_r) \rangle}
{1.5\times 10^{-8}\,m^2_{\rm Pl}}\right)\,, & 
\text{for $\tilde\alpha > 10^{-5}$}\,, \\
\displaystyle\left(\f{\tilde\alpha}{1/3}\right)^{5/4}\,
\displaystyle\left(\frac{\langle \delta \phi^2 (t_r)\rangle}
{3.0\times 10^{-3}\,m^2_{\rm Pl}}\right)\,, & 
\text{for $\tilde\alpha < 10^{-5}$}\,.\\
\end{cases}
\label{eq:nbys-final}
\end{eqnarray}
However, due to numerical errors, the numerical scaling over $\tilde\alpha<10^{-4}$
shall be (cf.~App.~\ref{app:num_err})
\begin{eqnarray}
\frac{n_{\sigma}}{s} &\simeq& 5\times 10^{-11}\,c^2
\left(\frac{\ps(k_\ast)}{2.1\times 10^{-9}}\right)^{-1/4}
\left(\f{\tilde\alpha}{1/3}\right)^{-3/4}\,\left(\frac{\langle \delta \phi^2 (t_r)\rangle}
{1.5\times 10^{-8}\,m^2_{\rm Pl}}\right)\,.~
%\text{for~}\tilde\alpha < 10^{-4}\,.
\label{eq:nbys-err}
\end{eqnarray}

\begin{figure}
\centering
\includegraphics[width=6in,height=4in]{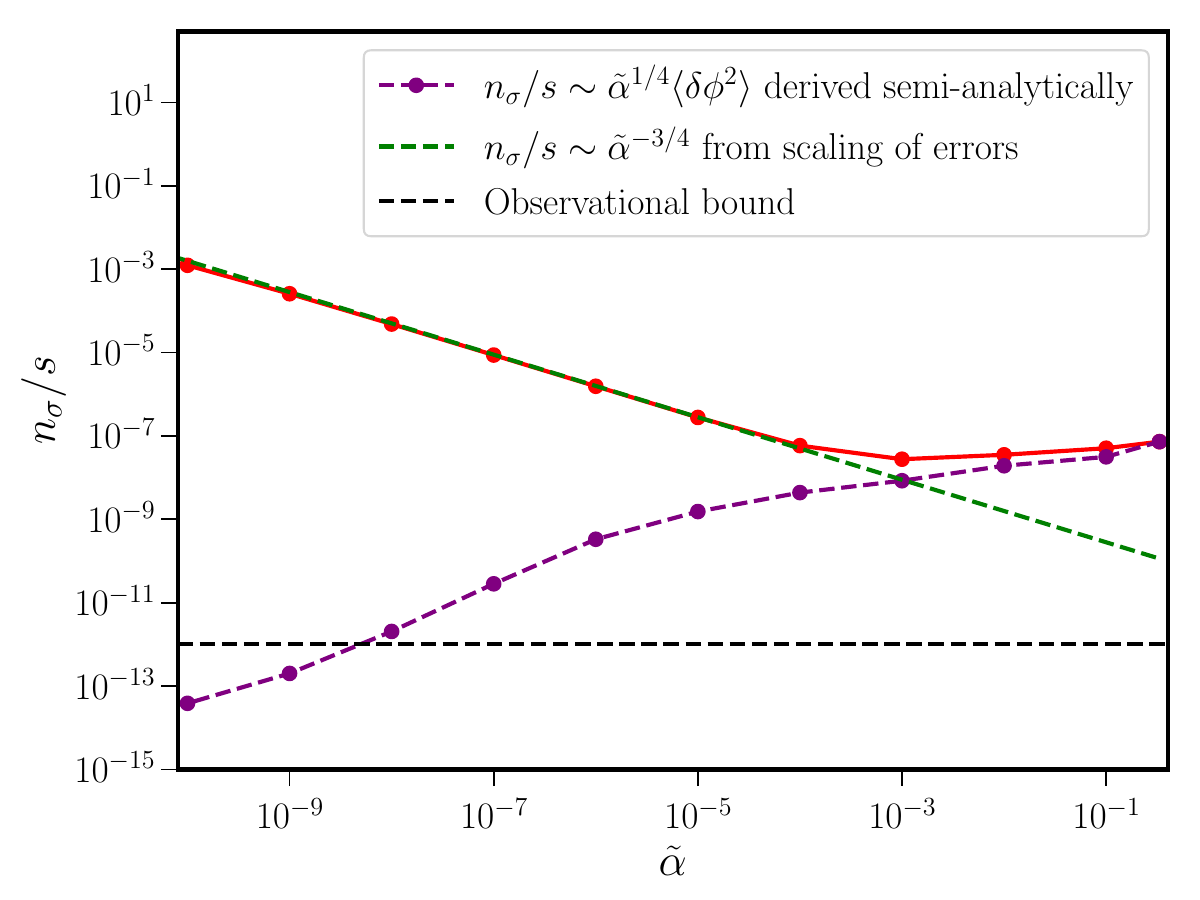}
\caption{The ratio $n_\sigma/s$ evaluated at the end of evolution is presented 
as a function of $\tilde\alpha$. Besides, we should remember that we set the coupling 
strength $c=1$ to arrive at these results. We see that over large values of 
$\Tilde{\alpha} > 10^{-4}$ it closely follows the semi-analytical trend as given in 
Eq.~\eqref{eq:nbys-full}. We should note that the semi-analytical behavior is plotted 
by substituting the numerically evaluated values of $\langle \delta \phi^2 \rangle$ 
in Eq.~\eqref{eq:nbys-full}. 
The difference between the analytical estimate from the exact numerics is due to
various approximations made in the derivation have been corrected using a
factor of ${\cal F}$ as explained in the main text.
Over smaller values of $\Tilde{\alpha} < 10^{-4}$, the numerical behavior scales 
as $\Tilde{\alpha}^{-3/4}$ as expected from the propagation of numerical 
errors~(cf.~App.~\ref{app:num_err}).
Evidently, the semi-analytical trend over this range is much smaller than the 
observed error-dominated values.
However, this trend can be utilized to determine the value of $\Tilde{\alpha}$ 
at which we suppress the abundance below the observational bound of 
$n_\sigma/s < 10^{-12}$.}
\label{fig:nbys_alpha}
\end{figure}

The behavior of $n_\sigma/s$ is presented in Fig.~\ref{fig:nbys_alpha}. This 
quantity matches the expected semi-analytical result of 
$n_\sigma/s \propto \tilde\alpha^{1/4}$ over $\tilde\alpha > 10^{-4}$ 
 given in Eq.~\eqref{eq:nbys-final}. 
Over $\tilde\alpha < 10^{-4}$, with errors dominating the estimate of 
$n_\sigma$, we see $n_\sigma/s$ growing as $\tilde\alpha^{-3/4}$. 
However, as the semi-analytical estimate has been verified over large values
of $\tilde\alpha$ we can continue to rely on it over smaller values where the 
numerics become dominated by errors. 
We may extend the semi-analytical trend for 
$\tilde\alpha < 10^{-4}$ and utilize the broken power-law to 
conclude that the actual $n_\sigma/s \sim \tilde\alpha^{5/4}$ over this regime.
We should clarify that we have used Eq.~\eqref{eq:nbys-full} to 
plot the semi-analytical trend in Fig.~\ref{fig:nbys_alpha} where we
substitute the exact values of $\langle \delta \phi^2 \rangle$ obtained
from the simulations. This is more reliable than the broken power-law
assumption as it is closer to the exact numerics and does not rely on 
any approximation of the variation with respect to $\tilde\alpha$. 
%%%%%%%%%%%%%%%%%%%%%%%%%%%%%%%%%%%%%%%%%%%%%%%%%%%%%%%%%%%%%%%%%%%%%%%%%%%%%%%
%%%%%%%%%%%%%%%%%%%%%%%%%%%%%%%%%%%%%%%%%%%%%%%%%%%%%%%%%%%%%%%%%%%%%%%%%%%%%%%

Thus we have obtained an expression of moduli abundance and its dependence on 
$\tilde\alpha$ in the class of models of interest. 
However, a few caveats have to be reiterated in this derivation. 
Firstly, the semi-analytical expression relies on the assumption that the dominant
contribution to $n_\sigma$ comes from the resonant mode and mainly at the time
of resonance. So the integrations over wavenumbers and time are performed only
around the corresponding ranges.
Secondly, in our numerical analysis, we extract the number density of particles 
corresponding to resonance by subtracting the initial background spectra of particles.
This invariably leads to numerical errors which eventually dominate the estimates
as resonance is suppressed by smaller values of $\tilde\alpha$. 
Hence, we compare and match the analytical estimate with numerical results over
the regime of $\tilde\alpha$ where errors are sub-dominant and then extrapolate 
the verified analytical trend over the regime where errors start dominating. This
semi-analytical approach enables us to study the system and connect to observational bounds as shall be discussed in the next section.
Lastly, the lattice simulation allows us to explore the range of $\tilde\alpha$ 
as small as $\tilde\alpha = 10^{-10}$\,. Beyond this range, the code does not 
support the stable evolution of the system. However, we believe the range we have 
explored is sufficient to understand the necessary physical implications 
regarding the abundance of resonant moduli production.

At the end of this section, and before moving forward to looking at the observational bounds, it is good to remind ourselves of the procedures that we have followed to calculate the moduli abundance.  In the case of the semi-analytic estimation of moduli production in Sec. 3, the calculations progress in the following ways: $n_{\sigma}/s \propto \langle \delta \sigma ^2 \rangle \propto c^2 \tilde\alpha^{1/4}\langle \delta \phi^2\rangle$. In deriving these expressions, a few approximations are made. But, the derivation brings out one important result, i.e the explicit $\tilde\alpha$ dependence. In addition, there is an implicit $\tilde\alpha$ dependence on $\langle \delta \phi^2\rangle$, and this dependence is borrowed from numerical calculations (Sec. 4) for all values of $\tilde\alpha$. Note that there are no numerical issues in calculating {\it variances} for smaller values of $\tilde\alpha$. In Ref.~\cite{Giudice:2001ep}, this was done for $\tilde\alpha \sim {\mathcal O}(1)$  without any direct estimation of abundance from numerical simulations. For our case, the additional complexity is the implicit $\tilde\alpha$ dependence on $\langle \delta \phi^2\rangle$ that we find out in Fig. 2. We do not expect this semi-analytic calculation scheme to be a good estimate (in terms of numerical values) for number density (even for $\tilde\alpha \sim 1$) due to the inherent non-linearities of the system and associated approximations. But, the process allows us to find out parametric dependence on $\tilde\alpha$. This parametric dependence is crucial to find out $n_{\sigma}/s$ for smaller values of $\tilde\alpha$.

We did an explicit calculation of $n_\sigma/s$ in Sec.~4. Note that this was not done in \cite{Giudice:2001ep}. In the calculation of $n_\sigma$, we need to subtract the vacuum contributions, and as $\tilde\alpha$ becomes progressively smaller, resonance particle production becomes comparable to the vacuum contributions, and a meaningful subtraction can not be done just due to the numerical precision. For the calculations of entropy density $s$, the issue does not arise as it is a sum of several energy densities. As noted above, the strength of the semi-analytical formula is in its parametric dependence on $\tilde\alpha$. Note that for $\tilde\alpha \gtrsim 10^{-4}$, we reconfirm this parametric dependence from numerical analysis.

%%%%%%%%%%%%%%%%%%%%%%%%%%%%%%%%%%%%%%%%%%%%%%%%%%%%%%%%%%%%%%%%%%%%%%%%%%%%%%%

\section{Observational bound}

From our numerical exercises, we find that the ratio $n_\sigma/s$ scales as 
$\tilde \alpha^{1/4}$ over large $\tilde \alpha$ and the semi-analytical estimate 
suggests that it decreases as $\Tilde{\alpha}^{5/4}$ over smaller values of 
$\alpha$. Though this trend could not be verified numerically below 
$\Tilde{\alpha} \simeq 10^{-4}$, 
we can rely on the semi-analytical estimate beyond this value. Besides, we know 
of the observational bound on the ratio to be 
$n_\sigma/s \lesssim 10^{-12}$ in the moduli mass range of $100$ GeV to $10$ TeV~\cite{Kawasaki:2004qu, Kawasaki:1994af}.
To translate this bound on the value of $\Tilde{\alpha}$, we consider the reliable 
semi-analytical trend and extrapolate for values smaller than 
$\Tilde{\alpha} \simeq 10^{-4}$. Making use of Eq.~\eqref{eq:nbys-final}, this means
\begin{eqnarray}
\frac{n_\sigma}{s} &\simeq& 5.36\times 10^{-9} 
c^2 \left(\frac{\Tilde{\alpha}}{10^{-5}}\right)^{5/4} < 10^{-12}\,.
\end{eqnarray}
The semi-analytical trend plotted in Fig.~\ref{fig:nbys_alpha} illustrates this behavior and hence we can infer that the observationally viable range 
of $\Tilde{\alpha}$ is
\begin{equation}
\Tilde{\alpha} < 10^{-8}\,c^{-8/5}.
\end{equation}
Recall that we have set $c=1$ in our analysis.
This bound is crucial in itself in that it informs us about the steep shape
of the potential around the minimum. Further, this also gives an idea about the 
field excursion allowed for the inflaton field during its evolution.

For such a bound on $\Tilde{\alpha}$ and the ratio of $V_0/\alpha$ already determined
by the CMB spectrum, we may arrive at a corresponding bound on the energy scale of
inflation and hence the reheating temperature. Since 
$V_0/\tilde\alpha \simeq 2.63 \times 10^{-13}\,m^4_{\rm Pl}$, the bound on $V_0$ shall be
\begin{eqnarray}
V_0 &\simeq& 10^{-21}\left(\frac{\Tilde{\alpha}}{10^{-8}}\right)\,m^4_{\rm Pl} \,,\\
H_{\rm inf} \simeq \sqrt{\frac{8\pi\,V_0}{3\,m^2_{\rm Pl}}} &\simeq& 
10^{-10}\left(\frac{\Tilde{\alpha}}{10^{-8}}\right)^{1/2}\,m_{\rm Pl}\,,
\end{eqnarray}
with $\Tilde{\alpha} < 10^{-8}$.
Further, recall that the tensor-to-scalar ratio for this class of models
is known to be $r \simeq 12\Tilde{\alpha}/N_k^2$ [cf. Eq.~\eqref{eq:r}]\,. Thus
for $\Tilde{\alpha} < 10^{-8}$, with $N_k=50$, we obtain
\begin{equation}
r < 4\times 10^{-11}\,.
\end{equation}
Needless to say, this value of $r$ predicted by this model is extremely small 
that it may not be detectable in near-future experiments.

At the end of (p)reheating, the energy of the inflaton must be converted into a relativistic thermal bath. But for the quartic $\alpha$-attractor models studied in this work, the total energy density already redshifts as radiation since the end of inflation. Therefore, we can easily estimate the maximum possible reheating temperature that can be attained, related to the total energy density as 
\be
T^{\rm max}_{\rm reh}  \simeq   \left(\frac{30}{\pi\,g_\ast}\rho_{\phi}\right)^{1/4} \simeq  \left(\frac{30}{\pi\,g_\ast}\right)^{1/4}
\left(\frac{V_0}{36\,\alpha^2}\phi^4_0(0)\right)^{1/4}
\simeq  10^{15}\,\Tilde{\alpha}^{1/4}\,{\rm GeV}\,,
\ee
where in the last equality we have used $g_\ast\simeq 10^2$, and the values of $V_0/\alpha$ and $\phi_0(0)$ given in Eqs. \eqref{eq:V0byalpha} and \eqref{eq:init_field_value} respectively. The bound on $\Tilde{\alpha} < 10^{-8}$ then suggests that
\begin{eqnarray}
T^{\rm max}_{\rm reh} &\simeq & 10^{13}\,\left(\frac{\Tilde{\alpha}}{10^{-8}}\right)^{1/4}\,c^{2/5}\,{\rm GeV}\,,
\end{eqnarray}
where with $c=1$, we infer that $T_{\rm reh} < 10^{13}\,{\rm GeV}$.
We should add that the above expression for upper bound helps us obtain the trend
mainly for the range $\tilde\alpha < 10^{-8}$.
We may contrast this bound of $T_{\rm reh} < 10^{13}$ GeV against an earlier
bound of $T_{\rm reh} < 10^{7}\,{\rm GeV}$ obtained from a similar argument of 
moduli production in the case of chaotic inflation~\cite{Giudice:2001ep}.
Our bound is higher by ${\cal O}(10^6)$ due to the fact that the production in
our class of models is efficiently suppressed over small values of $\tilde\alpha$ compared to generic chaotic inflation.
%%%%%%%%%%%%%%%%%%%%%%%%%%%%%%%%%%%%%%%%%%%%%%%%%%%%%%%%%%%%%%%%%%%%%%%%%%%%%%%

\section{Conclusion and discussions}

This numerical exercise, supplemented with the semi-analytical estimate, shows 
that the resonant production of moduli particles is suppressed in $\alpha$-attractor models with a decrease in the value of $\alpha$. Moduli production is triggered by the self-resonance of the inflaton, similar to what happens in a standard quartic chaotic model. Indeed, for small field values of
$\phi$, our potential in Eq.~\eqref{eq:Vphi} reduces to $V(\phi) \simeq \lambda \phi^4/4$, with $\lambda=V_0/(9\alpha^2)$. However the quartic chaotic and $\alpha$ attractor models differ in their behavior during inflation, and while Planck constraints on the primordial spectrum sets the value of $\lambda$ in a quartic model, it does only set the value of the ratio $V_0/\alpha$ in Planck units for the $\alpha$-attractor model. This means that the inflationary scale, and the post-inflationary physics, are not completely fixed by observations, and we may have viable inflationary models at different energy scales by varying the parameter $\tilde \alpha=8 \pi\alpha/m^2_{Pl}$. This is the key property that allows to control and suppress particle production during preheating by lowering the scale of inflation. 

As in the quartic chaotic model, the moduli abundance is proportional to the variance of the inflaton field, and we extract the behavior of the inflaton variance with the parameter $\tilde \alpha$ from the numerical simulations. 
We find that the ratio of $n_\sigma/s$ goes like ${\tilde \alpha}^{1/4}$ for $\tilde \alpha \gtrsim 10^{-4}$, but it behaves as $\Tilde{\alpha}^{5/4}$ over small values of $\Tilde{\alpha}$ and can be suppressed below the observational bound for 
$\Tilde{\alpha} < 10^{-8}$. We remark again that this conclusion is for the specific power of $p=4$ in the class of $\alpha$-attractor models i.e., for potential behaving as $V(\phi)\sim\phi^4$ close to the minimum. This kind of model have (a) a self-resonant behavior of inflaton fluctuations during preheating, and (b) the total energy density redshifts as radiation since the end of inflation. Although the determination of the reheating temperature $T_{\rm reh}$ requires knowing when the inflaton energy density converts into thermal relativistic degrees of freedom, we can set an upper limit on its value, from the upper limit on the inflationary scale, of the order $T_{\rm reh} <10^{13}$ GeV for $\tilde \alpha\simeq 10^{-8}$.

As to the future extensions of this work, we may expect a similar behavior or even stronger suppression in case of higher values of $p$ that determine the shape of the inflationary potential at the minimum. For our current work, we just focussed on $p =4$ case. More important possible future extensions include the development of the subtraction of the vacuum contribution for $\Tilde{\alpha} < 10^{-4}$. It is worthwhile to mention that our derivation of constraint on $\alpha$ crucially depends on the semi-analytical estimate for smaller values of $\tilde \alpha$.
Besides, throughout our analysis, we have not considered the mass term for the moduli potential. This limit is relevant for our constraints coming from the BBN that we use. Introduction and increase in the mass term may further suppress moduli production as heavier mass particles are in general difficult to produce. 

Moreover, we have focused only on the leading order term in the behavior of the inflaton potential around its minimum, such that the field oscillates always in the positive curvature region of the potential. Accounting for higher-order terms in the expansion of the complete potential may lend further insights into the dynamics of moduli production. 
One effect that we have not included in this first study is particle production due to the tachyonic instability in the potential when ending inflation, which is relevant for small values of $\alpha$. We have implicitly assumed that backreaction and re-scattering effects during the transition might be enough to shut down the instability before the complete fragmentation of the background field. We note that for the case of study in this work, with a linear coupling, moduli production is always triggered by inflaton fluctuations. Here we have focused on the induced production due to the inflaton self-resonant process. If non-vacuum inflaton fluctuations are already present before the first oscillation is complete, they may also source moduli fluctuations. Whether this will contribute to relaxing the bound on $\tilde \alpha$ or not, i.e., will impact moduli production, is an important issue to be studied in the future.

Further, it would be interesting to explore moduli production in the same class of $\alpha$-attractor models with different functional forms of coupling other than the linear form studied here. Establishing the trend of suppression with a decrease in $\alpha$ using a variety of couplings shall help us infer more about the significance of this parameter and hence the viability of this class of models. Also, another interesting avenue is to study the production of gravitational waves during this epoch and how the dynamics of moduli affect the amplitude and peak frequency of the gravitational waves. We relegate these examinations to future work. 

In summary, in this work, we have studied non-thermal light moduli production during preheating in $\alpha$-attractor inflation models, and we have found that to be consistent with observational bounds the parameter $\alpha$ needs to be $\alpha \lesssim 10^{-8}\,m^2_{\rm Pl}$.

%%%%%%%%%%%%%%%%%%%%%%%%%%%%%%%%%%%%%%%%%%%%%%%%%%%%%%%%%%%%%%%%%%%%%%%%%%%%%%%
\acknowledgments
MBG work has been partially supported by MICINN (PID2019-105943GB-I00/AEI/10.13039/501100011033), ``Junta de Andaluc\'ia" grant  P18-FR-4314 and FCT grant No.~CERN/FIS-PAR/0027/2021. 
HVR acknowledges support from Raman Research Institute through a postdoctoral fellowship. The work of K.D. is partially supported by MTR/2019/000395, and Core Research Grant CRG/2020/004347 funded by SERB, DST, Government of India. 
We thank the anonymous referee for critical comments on the draft, and in the process of addressing those points, the clarity of the presentation has improved.

%%%%%%%%%%%%%%%%%%%%%%%%%%%%%%%%%%%%%%%%%%%%%%%%%%%%%%%%%%%%%%%%%%%%%%%%%%%%%%%

\appendix 

%%%%%%%%%%%%%%%%%%%%%%%%%%%%%%%%%%%%%%%%%%%%%%%%%%%%%%%%%%%%%%%%%%%%%%%%%%%%%%%

\section{Details of conversion between physical and program variables of \texttt{Latticeeasy}}\label{app:prog_var}
In this appendix, we shall discuss the details of our usage of the numerical
package of \texttt{Latticeeasy}, a lattice simulation that tracks the evolution 
of the homogeneous background as well as the fluctuations in a typical FLRW 
spacetime, as governed by the potential provided.
It provides as output, the quantities that capture the behaviors of the fields, 
such as their means, variances, spectrum of number density and energy density of 
particles associated with the fluctuations of each field.
Let us briefly discuss the setup of $\texttt{Latticeeasy}$ and the modifications
we made as per the model of interest.

To evolve the system of $\phi$ and $\sigma$ through preheating and estimate the 
quantities such as their variances and energy densities of the fluctuations, 
the variables of the system such as a given field $f$, coordinates ${\bm x}, t$ and 
momentum ${\bm k}$ are rescaled to dimensionless variables using the model 
parameters namely, $V_0,\,\alpha$, the scale factor $a$ and the value of inflaton 
at the end of inflation $\phi_0(0)$. These rescaled variables effectively capture the 
dynamics of resonance and are called program variables.
The relations between physical and program variables are given by~\cite{Khlebnikov:1996mc,Felder:2000hq}
\begin{subequations}
\begin{eqnarray}
f_{\rm pr}&=&Aa^rf\,, \\
dt_{\rm pr}&=&Ba^{s}dt\,,\\ 
{\bm x}_{\rm pr}&=&B\,\bm x\,,\\
\bm k_{\rm pr}&=&\frac{\bm k}{B}\,,
\end{eqnarray}
\end{subequations}
where the rescaling factors are expressed in terms of model parameters as
$A=1/\phi_{0}(0)$, $B=\phi_{0}(0)\sqrt{V_{0}/9\alpha^2}$, 
$s=-1$ and $r=1$ for our potential of interest with $V \propto \phi^4$.
These relations provide the explicit expressions of program variables in terms
of physical variables as
\begin{subequations}
\begin{eqnarray}
f_{\rm pr}&=&\frac{a}{\phi_{0}(0)}f\,,\\
{\rm d}t_{\rm pr}&=&\frac{1}{3}\sqrt{\frac{V_{0}}{\alpha}}\left(\frac{\phi_{0}(0)}{\sqrt{\alpha}}\right)
\frac{{\rm d}t}{a}\,,\\
\bm x_{\rm pr}&=&\frac{1}{3}\sqrt{\frac{V_{0}}{\alpha}}\left(\frac{\phi_{0}(0)}{\sqrt{\alpha}}\right)\,\bm x\,,
\\
\bm k_{\rm pr}&=&3\sqrt{\frac{\alpha}{V_{0}}}\left(\frac{\sqrt{\alpha}}{\phi_0(0)}\right)
\bm k\,,
\end{eqnarray}
\label{eq:rescaling}
\end{subequations}
where $\phi_0(0)$, as mentioned earlier, is the value of the inflaton field when 
inflation ends and the preheating dynamics is expected to begin.
Recall that in our analysis, we maintain the ratio of $V_0/\alpha$ constant 
so as to be consistent with CMB constraint on the scalar power over large scales.

It is worth noting a few points about these definitions. The purpose of these
rescalings is to essentially evolve the system in its characteristic time and 
length scales. For instance, in the case of our potential for inflaton, the 
effective mass is [cf.~Eq.~\eqref{eq:effmass}]
\begin{equation}
m_{\phi} \equiv \sqrt{\frac{\partial^2 V(\phi)}{\partial \phi^2}} 
= \sqrt{\frac{V_0}{3\alpha}}\frac{\phi_0(0)}{\sqrt{\alpha}}\,.
\label{eq:char-mass}
\end{equation}
Therefore, the rescaling of $\bm k$ is such that $k_{\rm pr}\simeq k/m_\phi$,
by which we may capture the production of $m_\phi$ at $k_{\rm pr} \simeq 1$\,.
The mass term further implies that the characteristic frequency where we 
expect resonance is around $\omega_{k} \simeq m_\phi$\,. Hence the rescaling of 
time variable is such that ${\rm d}t_{\rm pr}\simeq{\rm d}t/(m_\phi\,a)$, by 
which we capture the time scale of resonance as well as account for the 
background expansion using time steps of ${\rm d}t_{\rm pr} \simeq 1$.
An interesting point to note is that $m_\phi$ at $t=0$ does not depend on 
$\tilde\alpha$. This is because, $V_0/\tilde\alpha$ being constant and $\phi_0\propto\sqrt{\tilde\alpha}$ [cf.~Eq.~\eqref{eq:init_field_value}]. However, as time progresses
this characteristic mass picks up an indirect $\tilde\alpha$ dependence due to $\phi(t)\propto a^{-1}(t)$ and $a(t)$ has a mild dependence on $\tilde\alpha$ through Friedmann equations.

\begin{figure}
\centering
\includegraphics[width=3.5in]{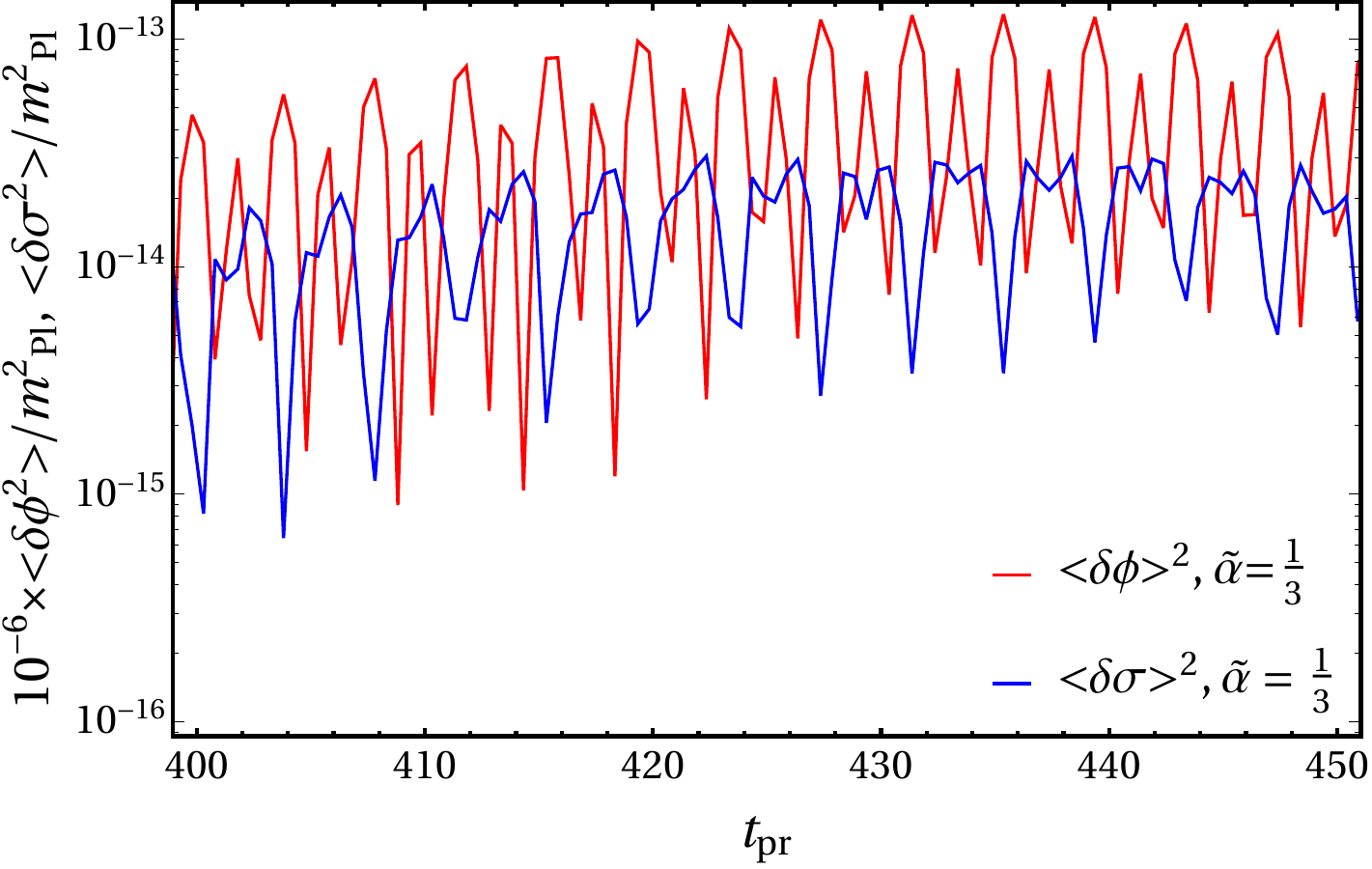}
\vskip -0.1in
\caption{The figure shows a detail of the time evolution of variances of the inflaton (solid red line) and the moduli (solid blue line) field for a fixed value of $\Tilde{\alpha}$.}
\label{fig:variance_1}
\end{figure}

We note that the last expression of Eq.~\eqref{eq:rescaling} above also shows the fact that the inflaton and the modulus field oscillates with the same frequency $\omega_{\rm k_{r}}=\left(\frac{V_{0}}{9\alpha^2}\right)\varphi^2_{0}(t)$ during resonance. This can be seen in Fig.~\ref{fig:variance_1} where we have plotted the variances of the fields during resonance. We see that both fields have the same frequency. Note that they are in opposite phases due to a relative sign in the source term governing their dynamics. 

%%%%%%%%%%%%%%%%%%%%%%%%%%%%%%%%%%%%%%%%%%%%%%%%%%%%%%%%%%%%%%%%%%%%%%%%%%%%%%%

\section{Calculating the Floquet exponent }\label{app:floq}
\begin{figure}[!t]
\centering
\includegraphics[width=3.47in]{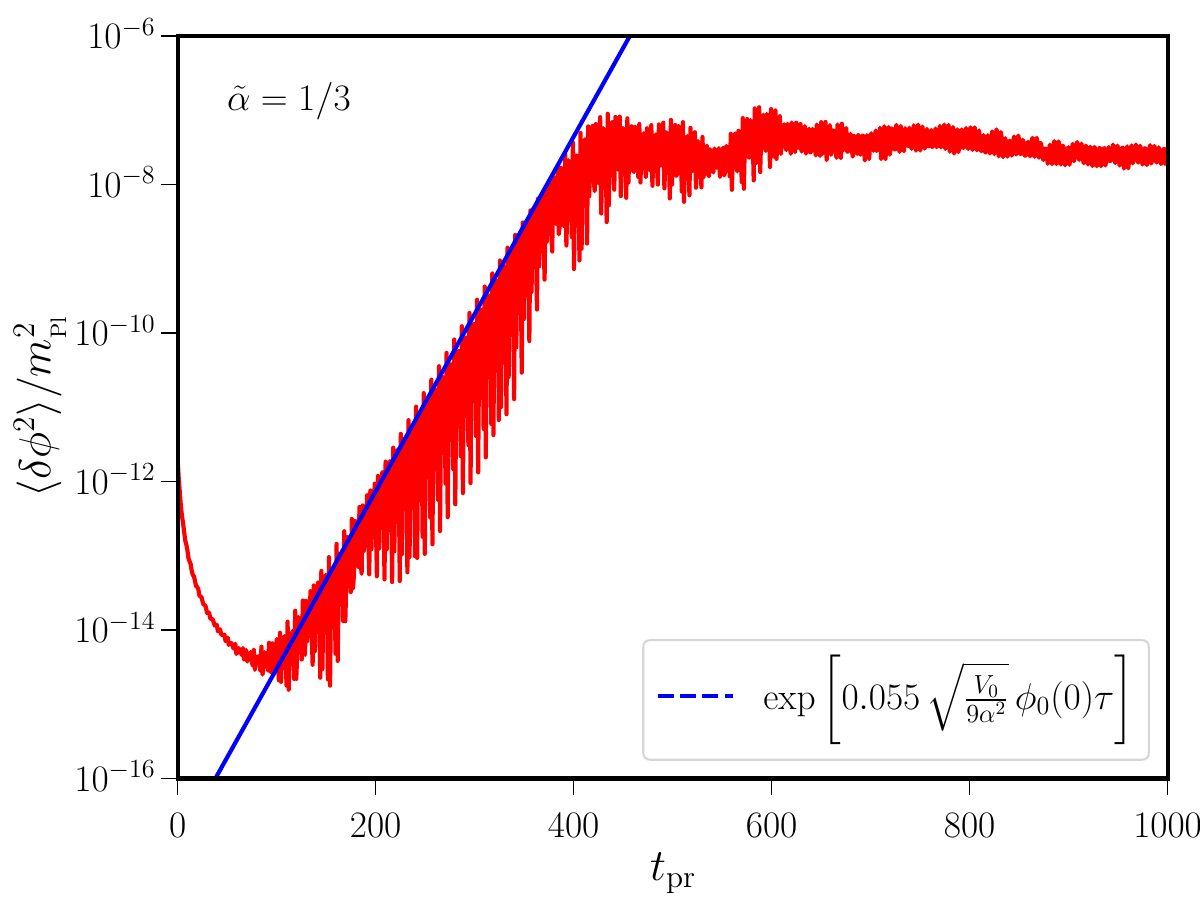}
\includegraphics[width=3.47in]{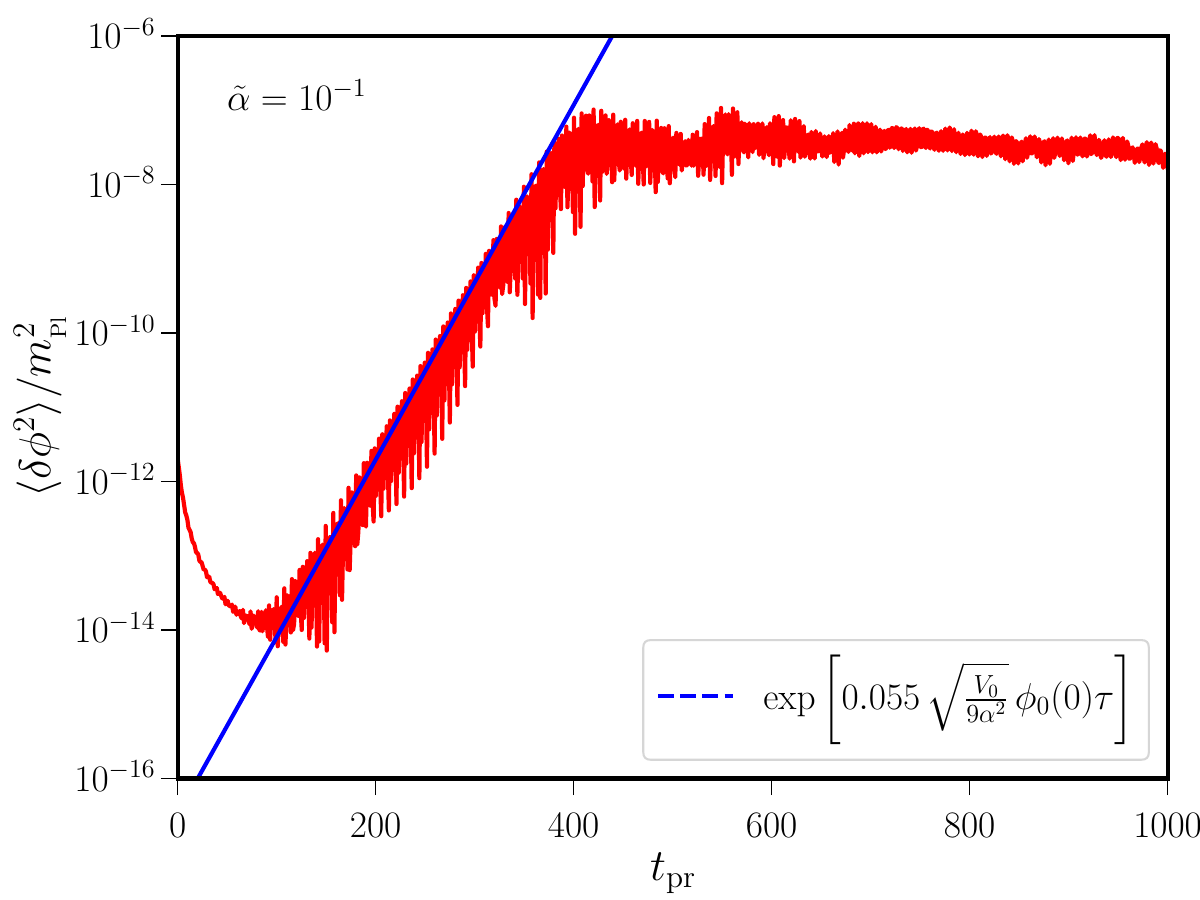}\\
\includegraphics[width=3.47in]{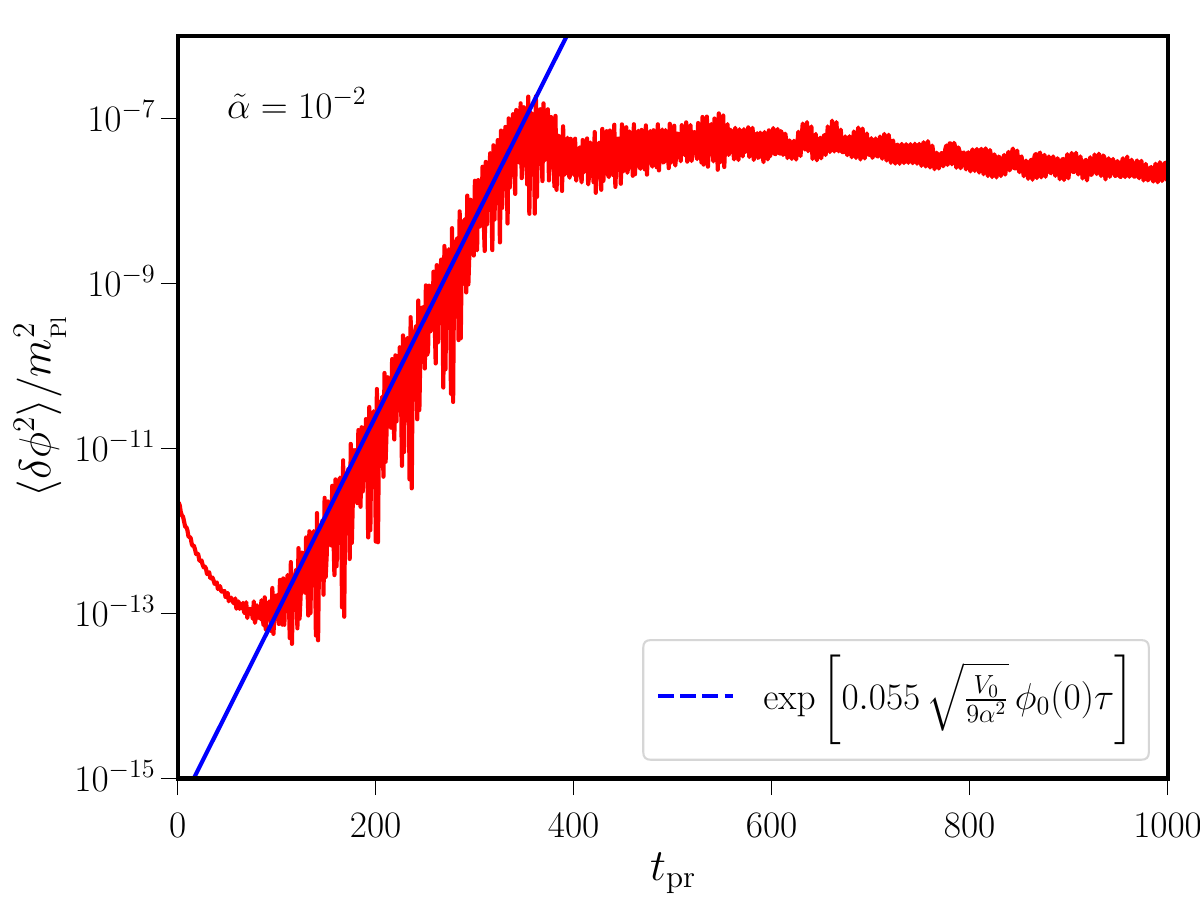}
\includegraphics[width=3.47in]{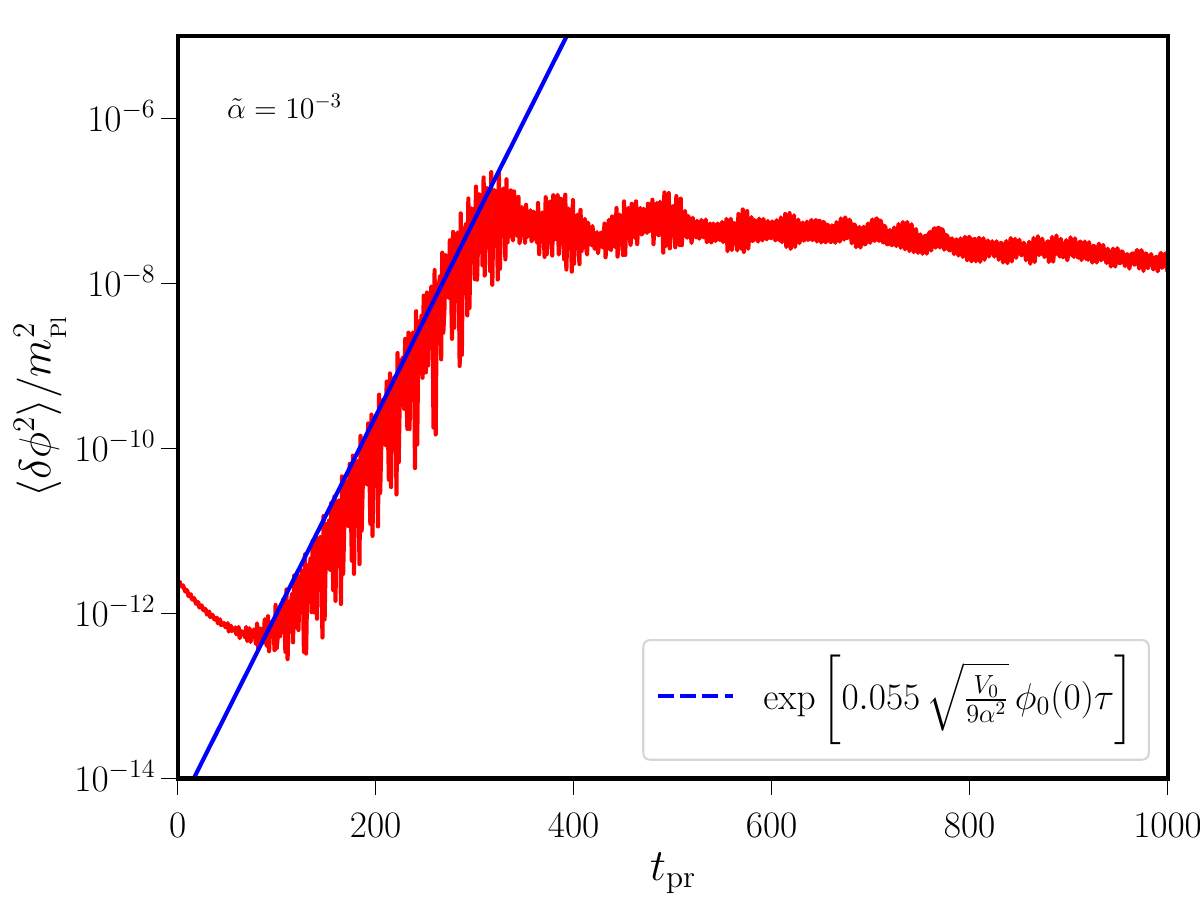}
\vskip -0.1in
\caption{These are the plots for finding the value of the Floquet exponent ($\mu$). The red lines of these plots are the inflaton field's variance, and the blue dashed line is the exponential fitting function ($\sim e^{f\tau}$) at the resonance time. The different plots correspond to the different values of $\Tilde{\alpha}$. From the fitting function, we can say $2\mu\simeq f$, and we can see the value of $2\mu\simeq 0.06\sqrt{\frac{V_{0}}{9\alpha^2}}\phi_{0}(0)$ for all values of $\Tilde{\alpha}$.}
\label{fig:mu_fit}
\end{figure}

In this appendix, we will estimate the Floquet exponent $\mu$ from the
numerical output of variance obtained from ${\tt Latticeeasy}$. From the mode equation of the inflaton field in Eq.~\eqref{inflaton_mode_equ}, we can say that the inflaton field can show the parametric resonance on its own, and its solution behaves like $\delta\phi_{k}(\tau)\sim e^{\mu\tau}$. Now we will explain how to find the value of $\mu$ numerically. From the Eq.~\eqref{variance}, we can say that the variance of the inflaton field $\langle\delta\phi^2(\tau)\rangle$ is related to the square of the absolute value of inflaton mode function $\vert\delta\varphi_{k}(\tau)\vert^2$. We also know that the maximum contribution to the variance comes from the resonance mode($k_{r}$). So the variance of the inflaton field will also behave as $\langle\delta\phi^2(\tau)\rangle\sim e^{\mu\tau}$ where the value of $\mu$ is roughly the same in both cases. To find the value of $\mu$, we will fit an exponential function($\sim e^{f\tau}$) with the variance of the inflaton field, which is calculated numerically. For the fitting function, we can say $2\mu=f$. From Fig.~\ref{fig:mu_fit}, we see the value of the $2\mu \simeq 0.06\sqrt{\frac{V_{0}}{9\alpha^2}}\phi_{0}(0)$ and the value of the $\mu$ is same for all values of $\alpha$.  

%%%%%%%%%%%%%%%%%%%%%%%%%%%%%%%%%%%%%%%%%%%%%%%%%%%%%%%%%%%%%%%%%%%%%%%%%%%%%
%%%%%%%%%%%%%%%%%%%%%%%%%%%%%%%%%%%%%%%%%%%%%%%%%%%%%%%%%%%%%%%%%%%%%%%%%%%%%
\section{Scaling of numerical error with $\alpha$}\label{app:num_err}

In this appendix, we shall study the scaling of numerical error in the estimate 
of number density $n_\sigma$ and hence on the ratio $n_\sigma/s$, particularly 
over small values of $\alpha$.

The comoving number density of $\sigma$ particles produced due to resonance is 
calculated as 
\begin{eqnarray}
n_\sigma\,a^3 &=& \int {\rm d}^3k\, (n_k - n^0_k)\,, \\
&=& \left(\frac{V_0}{9\alpha}\right)^{3/2}\left(\frac{\phi_0(0)}{\sqrt\alpha}\right)^3
\int {\rm d}^3k_{\rm pr}\,(n_{k_{\rm pr}} - n^0_{k_{\rm pr}})\,.
\end{eqnarray}
Recall that $n^0_k$ is the spectrum of comoving number density corresponding to
the background fluctuations.
The prefactor in the above expression is evidently due to the conversion relation
between physical and program units [cf.~Eq.~\eqref{eq:rescaling}]. In this 
expression, the quantities $n_{k_{\rm pr}}$ and $n^0_{k_{\rm pr}}$ shall contain 
the associated numerical errors.
Over small values of $\Tilde{\alpha} \leq 10^{-4}$, we see that 
$n_{k_{\rm pr}} \simeq n^0_{k_{\rm pr}}$ and hence the difference between the two
shall be dominated by these numerical errors [cf. Fig.~\ref{fig:rel_nk_alpha}]. 
Let us denote such a numerical error which is independent of any model parameter 
as $\Delta_{\rm pr}$. The number density corresponding to such numerical error 
alone shall be
\begin{eqnarray}
n_\sigma\,a^3 &=& 4\pi\,\left(\frac{V_0}{9\alpha}\right)^{3/2}
\left(\frac{\phi_0(0)}{\sqrt\alpha}\right)^3
(k_{\rm pr, max}^3 - k^3_{\rm pr, min})\Delta_{\rm pr}\,m^3_{_{\rm Pl}}\,.
\end{eqnarray}
This estimate shall dominate the number density over the regime 
$\Tilde{\alpha} \leq 10^{-4}\,M^2_{_{\rm Pl}}$.
Note that the ratio of $V_0/\alpha$ is retained as a constant to ensure
proper COBE normalization of the scalar power spectrum. 
Also, the value of $\phi_0(0) \propto \sqrt{\alpha}$.
Hence there is no overall dependence of the above $n_\sigma$ on $\alpha$ i.e.
$n_\sigma \propto \alpha^0$.
This can be seen in the behavior of comoving number density over the range of
$\Tilde{\alpha} < 10^{-4}$ in Fig.~\ref{fig:n_rho_alpha}.
If we set $\Delta_{\rm pr}=10^{-7}$, which is $10\%$ of the 
precision used in the simulation, we obtain the comoving number density to be
$10^{-20}\,m^3_{_{\rm Pl}}$. This is close to the value at which the quantity
settles at in Fig.~\ref{fig:n_rho_alpha}.

Besides, recall that 
\begin{eqnarray}
\rho &\simeq& \frac{V_0}{9\alpha^2}\phi_0^4(0)
\simeq  \left(\frac{9V_0}{4\alpha}\right)\alpha\,.
\end{eqnarray}
Again, as $V_0/\alpha$ is retained as a constant, the scaling of entropy density
comes out to be $s \simeq \rho^{3/4} \propto \alpha^{3/4}$.
Hence, for the estimate of number density due to $\Delta_{\rm pr}$, the scaling 
of the ratio of interest $n_\sigma/s$ with respect to $\alpha$ 
turns out to be
\begin{eqnarray}
\frac{n_\sigma}{s} &\propto& \alpha^{-3/4}.
\end{eqnarray}
This is the scaling of the error in the ratio of interest and it evidently
takes over the actual value in a regime with $\Tilde{\alpha} \leq 10^{-4}$ 
as can be seen in Fig.~\ref{fig:nbys_alpha}.

%%%%%%%%%%%%%%%%%%%%%%%%%%%%%%%%%%%%%%%%%%%%%%%%%%%%%%%%%%%%%%%%%%%%%%%%%%%%%%%
\section{Role of the coupling strength $c$}

\begin{figure}
\includegraphics[width=3.5in]{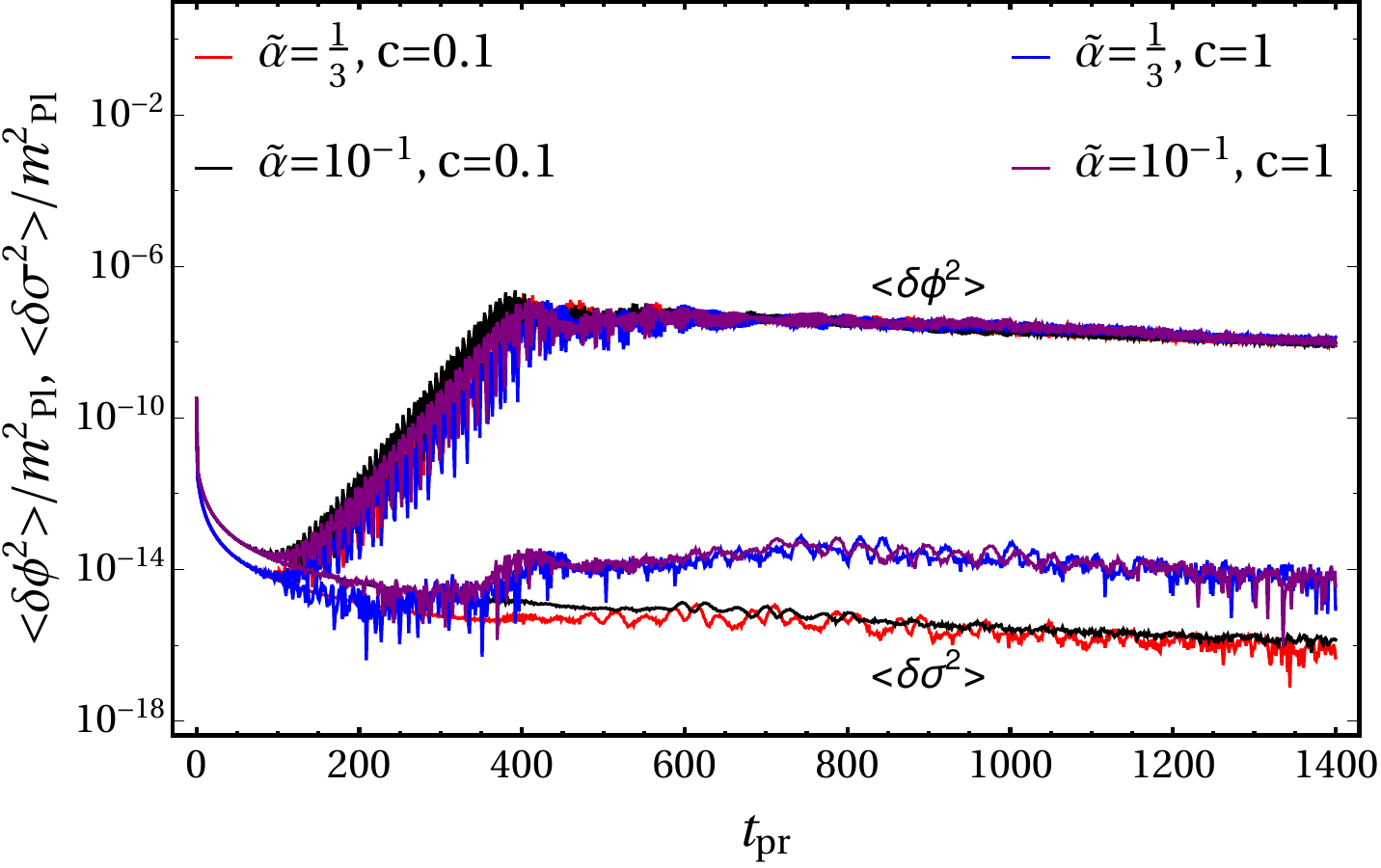}
\includegraphics[width=3.5in]{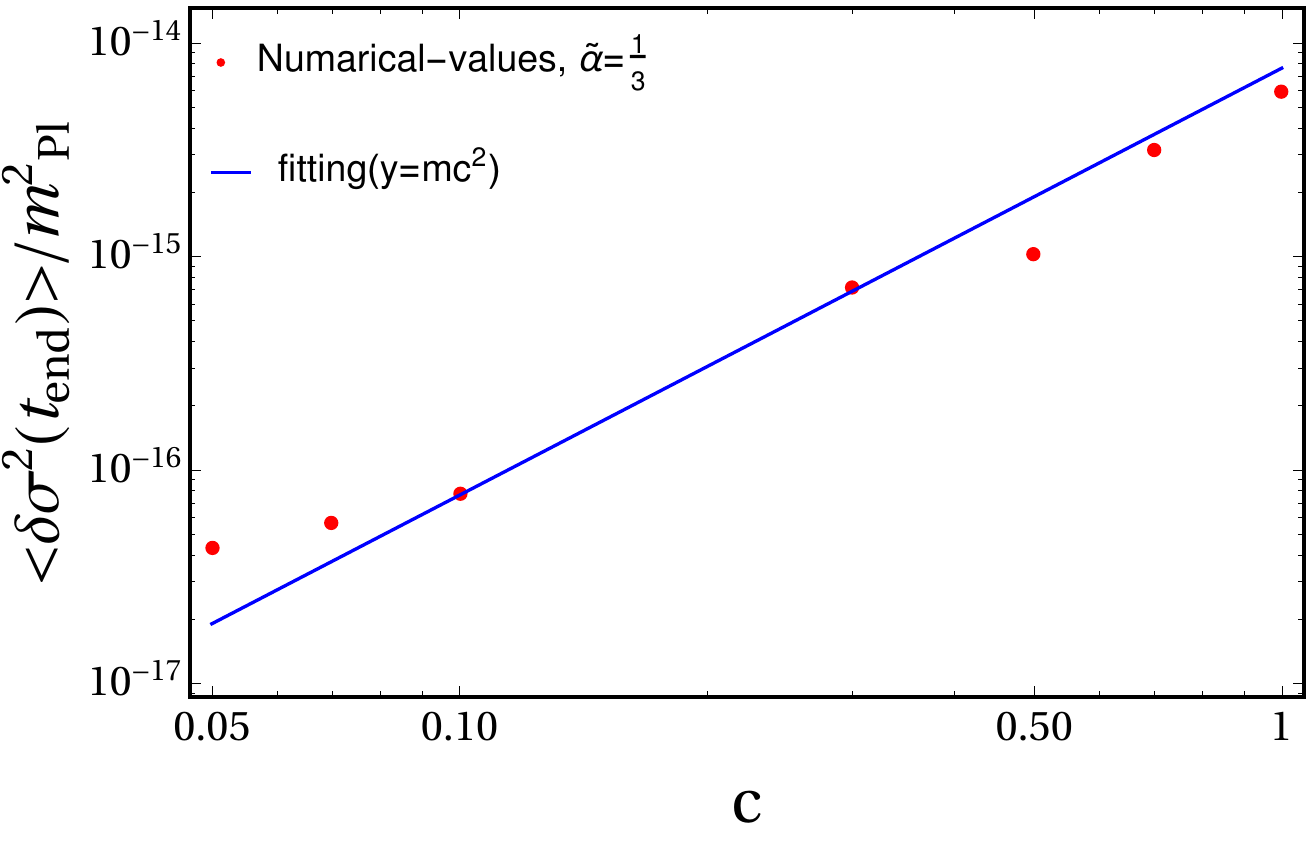}
\vskip -0.1in
\caption{In the left panel, we show variances of the inflaton and the moduli fields for
two sets of values of $\tilde\alpha$ and $c$.
In the right panel, we show the dependence of asymptotic moduli variances on the parameter $c$.}
\label{fig:variance}
\end{figure}

In this section, we will briefly discuss the role of the coupling strength $c$ between the moduli and the inflaton field as in Eq.~(\ref{pot}) on the moduli abundance 
 $n_{\sigma}/s$. To see the effect of the coupling, we fix the value of $\Tilde{\alpha}$. From the semi-analytical calculations in Sec.~\ref{emi-analytical approach}, we can say $n_{\sigma }/s \propto \langle\delta\sigma^2\rangle(t_{\rm end}) \propto c^2\,\langle\delta\phi^2\rangle(t_{\rm end})$ for fixed value of $\Tilde{\alpha}$. Numerically, if $\langle\delta\sigma^2\rangle(t_{\rm end})\propto\,c^2\,\langle\delta\phi^2\rangle(t_{end})$, this automatically satisfies $n_{\sigma}/s$ relation with $c$. From the left panel of  Fig.~\ref{fig:variance}, we see that the variance of the inflaton field at the end of our simulation ($\langle\delta\phi^2\rangle(t_{\rm end})$) does not change with $c$, but the variance of the moduli field changes with $10^{-2}$ order of magnitude for two different values of $c$ ($c=1$ and $c=0.1$). So it satisfies $\langle\delta\sigma^2\rangle(t_{\rm end}) \propto c^2\,\langle\delta\phi^2\rangle(t_{\rm end})$ for $\Tilde{\alpha}=1/3$. Similarly, we have repeated the test with $\Tilde{\alpha}=10^{-1}$ and checked it for two different values of $c$ ($c=1$ and $c=0.1$). In the right panel of Fig.~\ref{fig:variance}, we have plotted the variance of the moduli field at the end of the simulation for different values of $c$ where we have fixed $\Tilde{\alpha}=1/3$ and have fitted a function $y=mc^2$ $\left(y\equiv\langle\delta\sigma^2\rangle(t_{\rm end}\right)$ and $x\equiv c$) in a log-log scale. From this fitting function, we can say that the variance of moduli, and hence $n_\sigma/s$, obtained numerically behave with respect to $c$ as expected from the semi-analytical relation.

%%%%%%%%%%%%%%%%%%%%%%%%%%%%%%%%%%%%%%%%%%%%%%%%%%%%%%%%%%%%%%%%%%%%%%%%%%%%%%%
\section{On the initial conditions for the fluctuations}\label{app:ini_con}

During the numerical simulations, we have imposed Bunch-Davies initial conditions for the modes of fluctuations $\delta \phi_k$ and $\delta \sigma_k$ at the end of inflation. In this appendix, we check and justify this choice of initial conditions.

We inspect the behaviour of a wide range of modes around the end of inflation. If there were non-trivial background dynamics, the modes could have been excited to different initial states. However, as we find, until the end of inflation, there are no non-trivial phenomena in the evolution of the background that affect the modes, especially when in the sub-Hubble regime.

\begin{figure}[ht]
\vspace{-1.0cm}
\includegraphics[width=3.47in,height=2.0in]{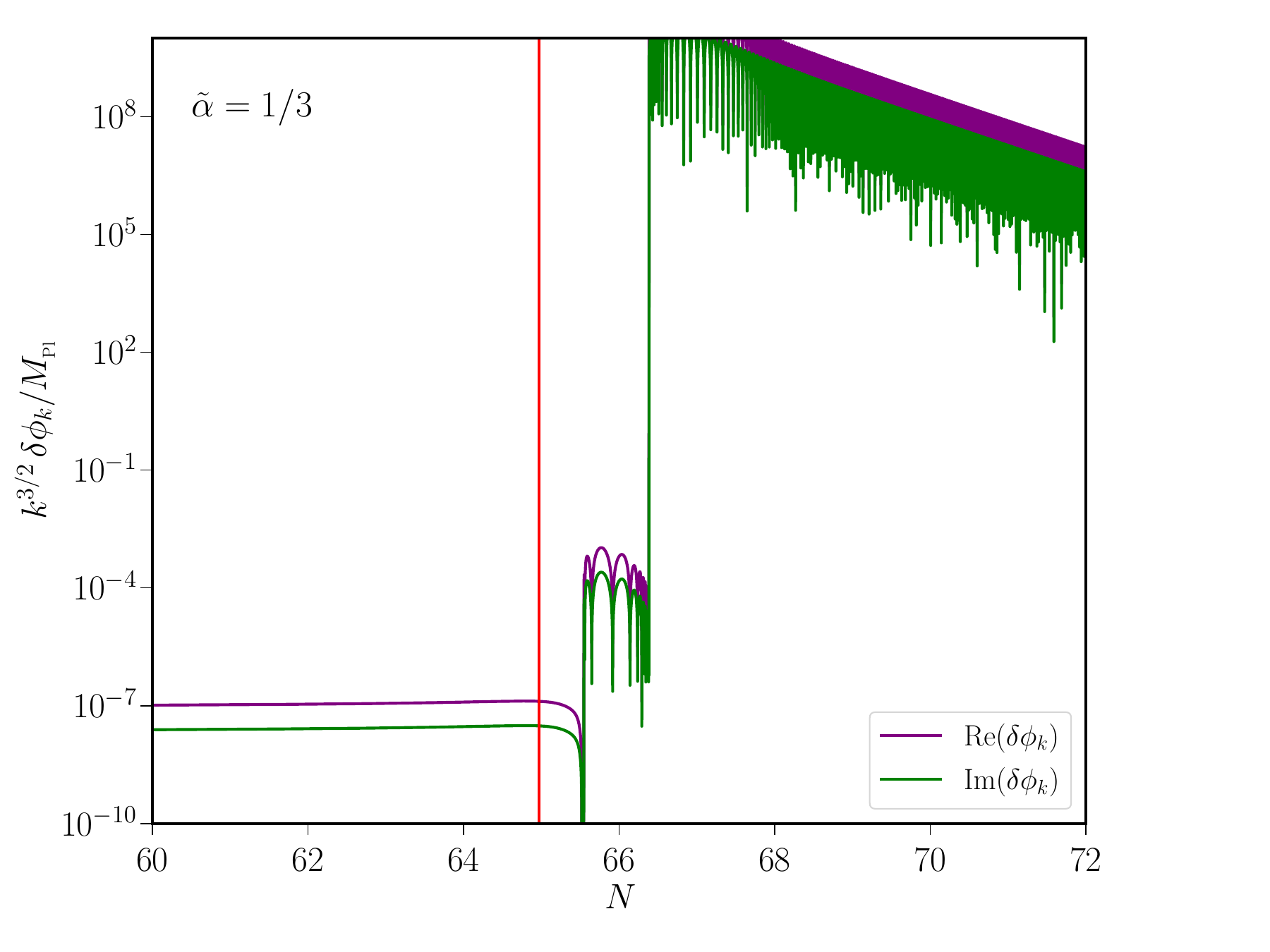}
\includegraphics[width=3.47in,height=2.0in]{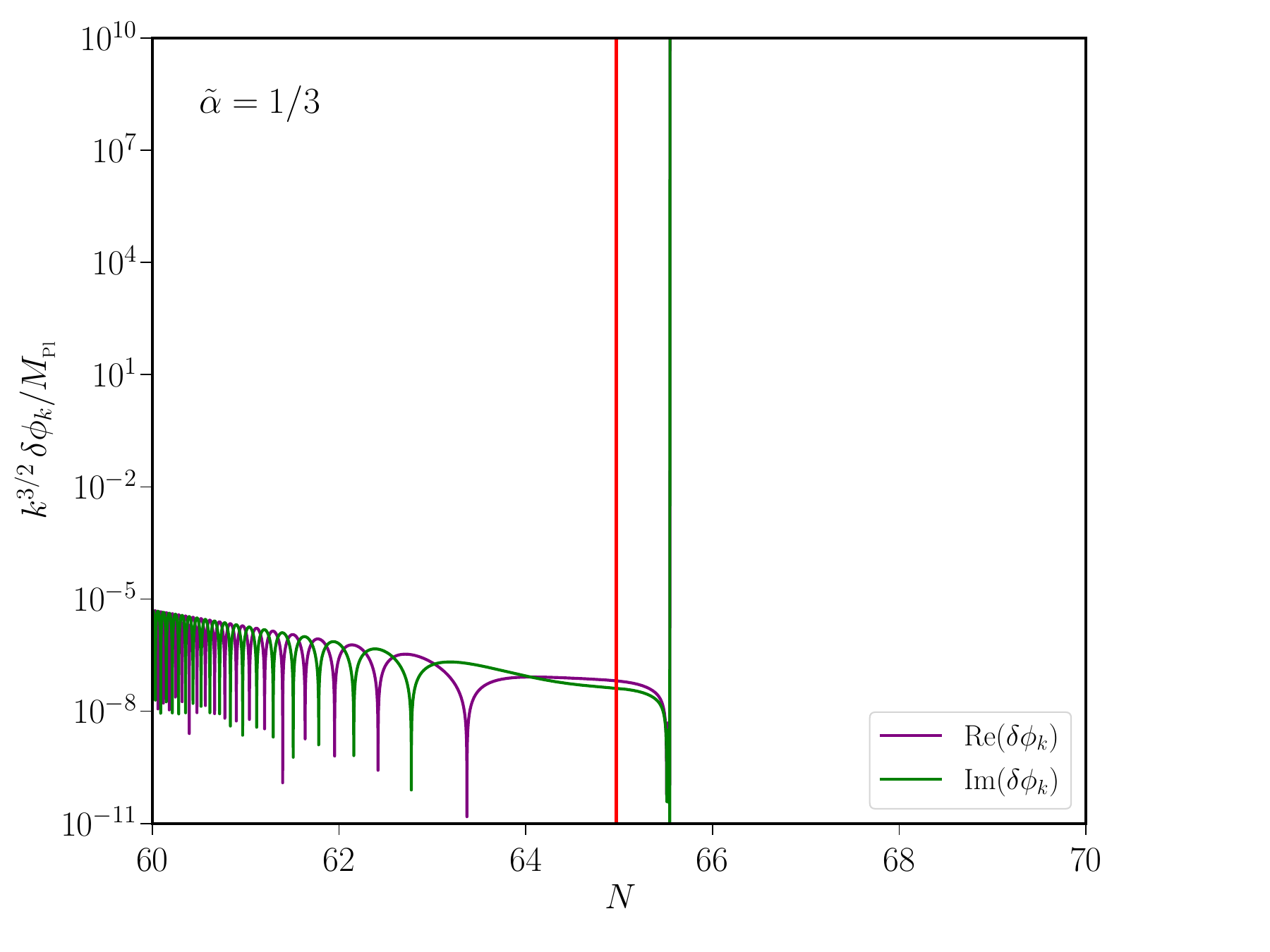}
\includegraphics[width=3.47in,height=2.0in]{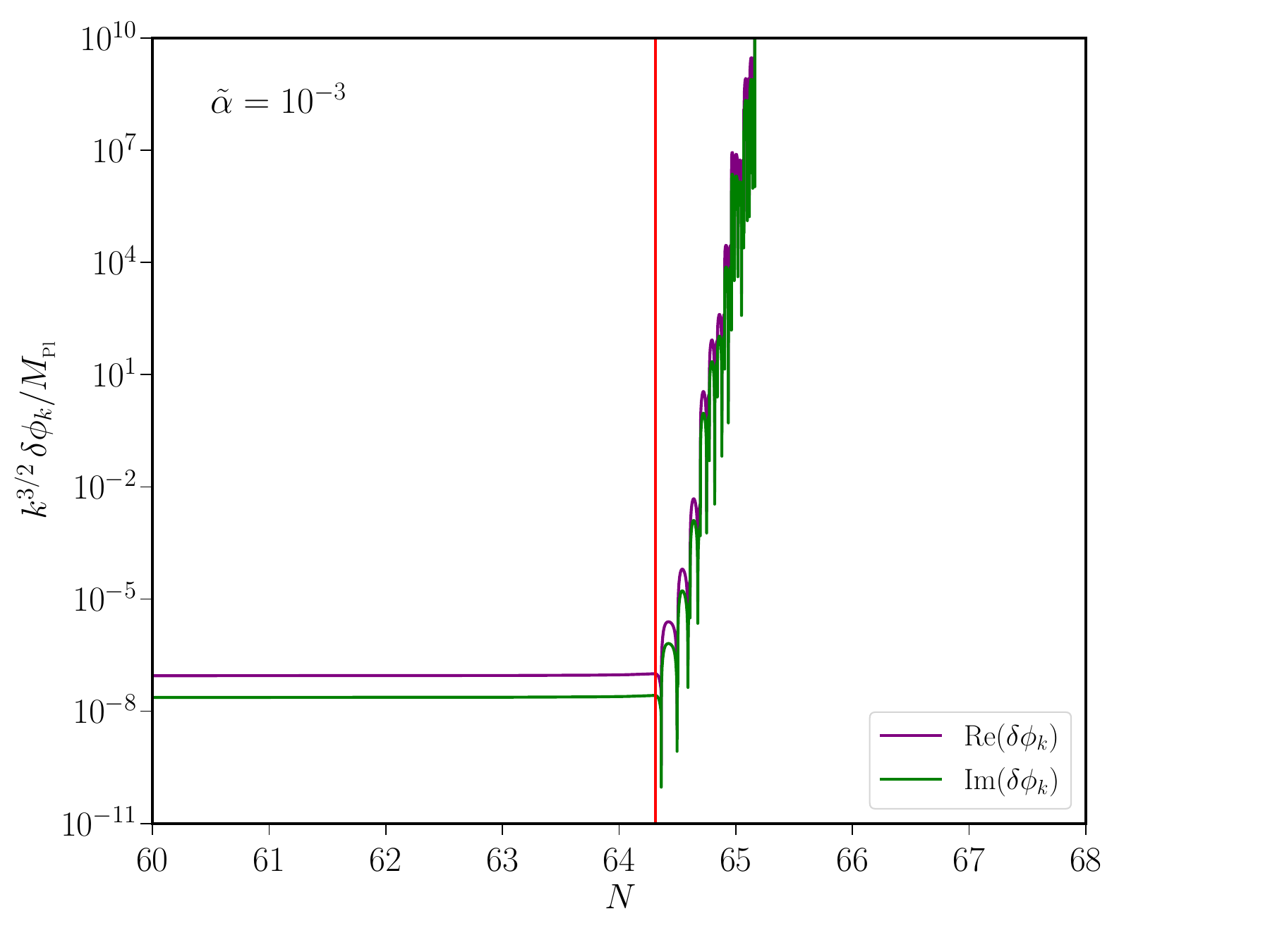}
\includegraphics[width=3.47in,height=2.0in]{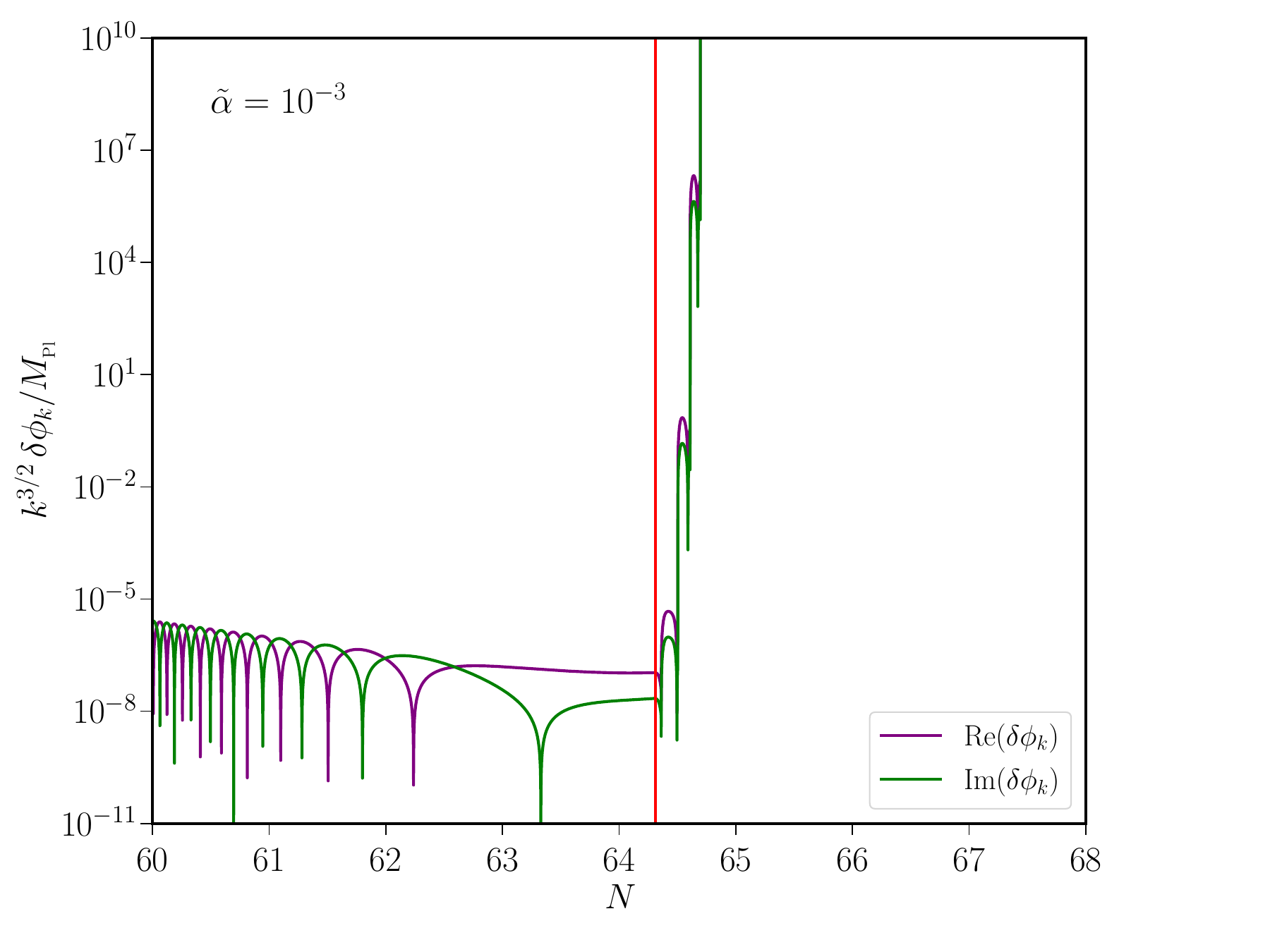}
\includegraphics[width=3.47in,height=2.0in]{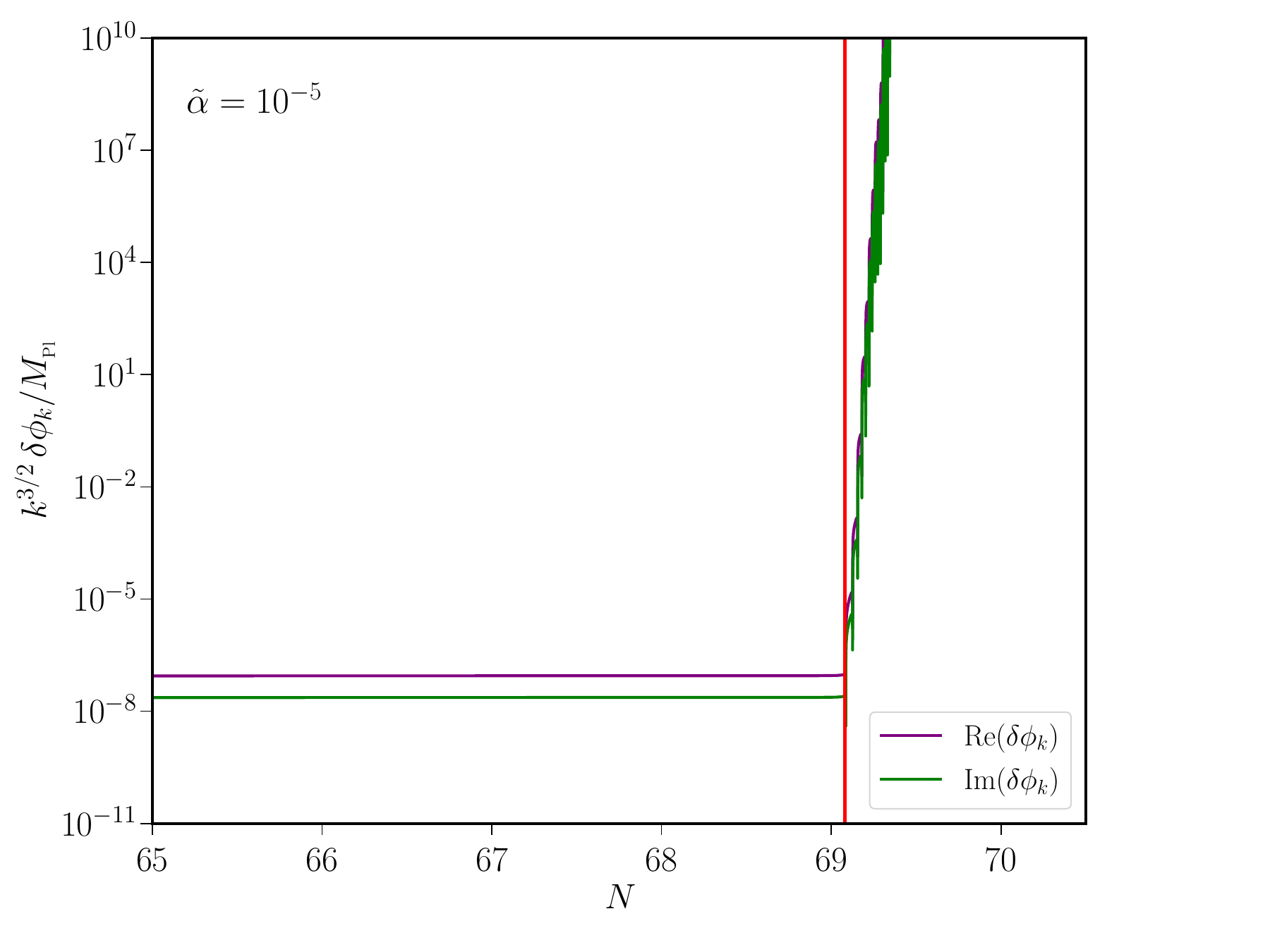}
\includegraphics[width=3.47in,height=2.0in]{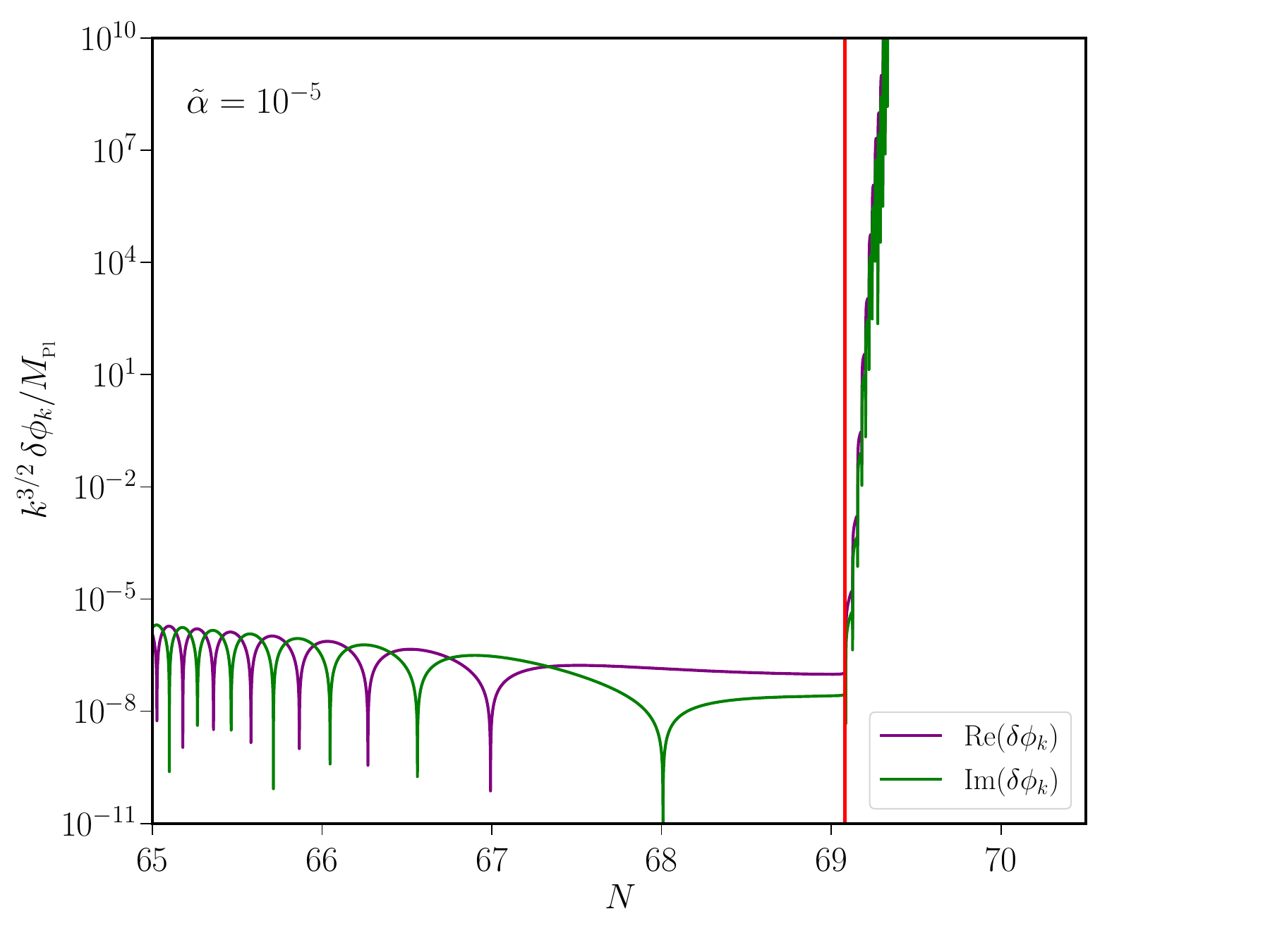}
\includegraphics[width=3.47in,height=2.0in]{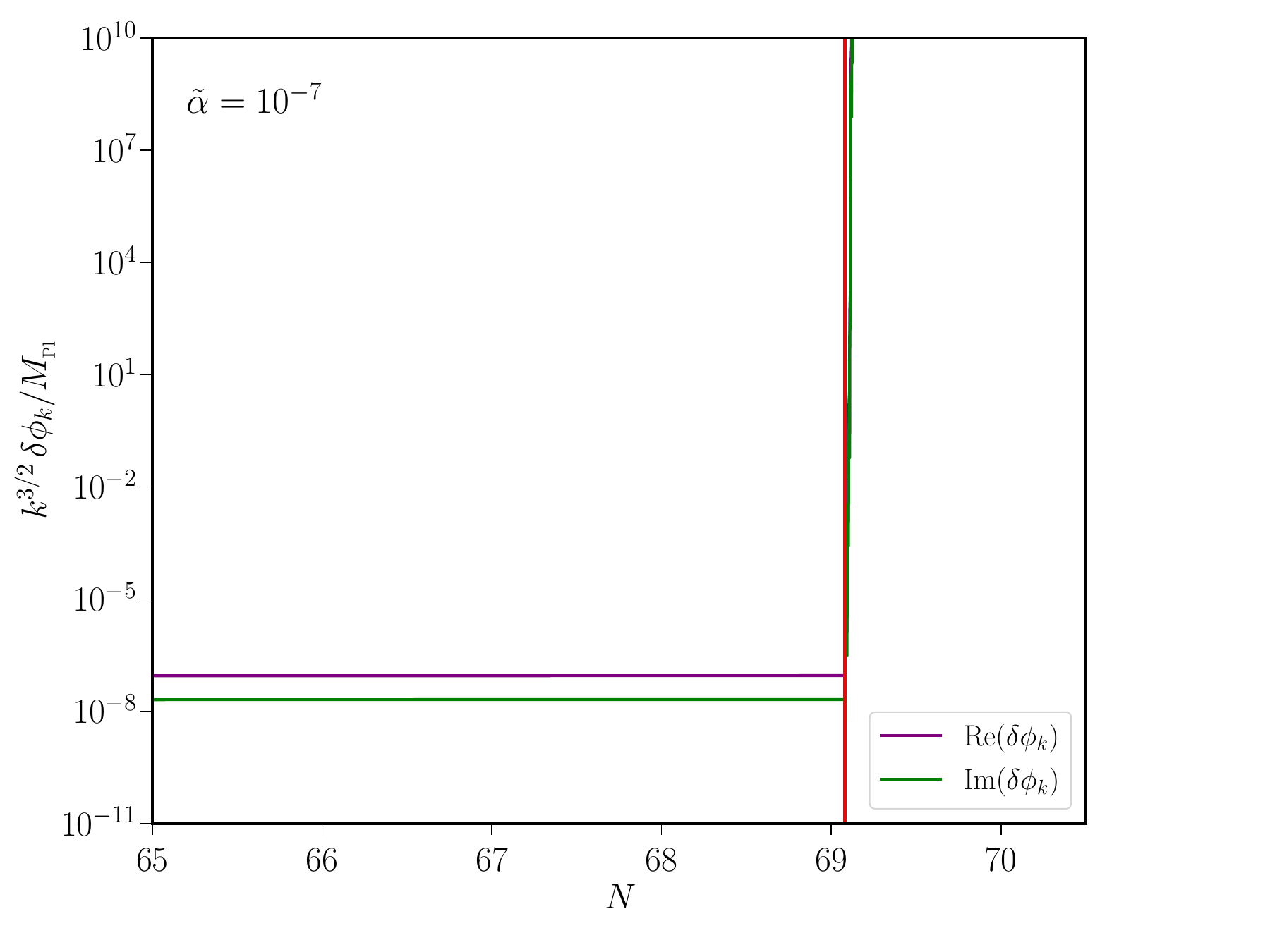}
\includegraphics[width=3.47in,height=2.0in]{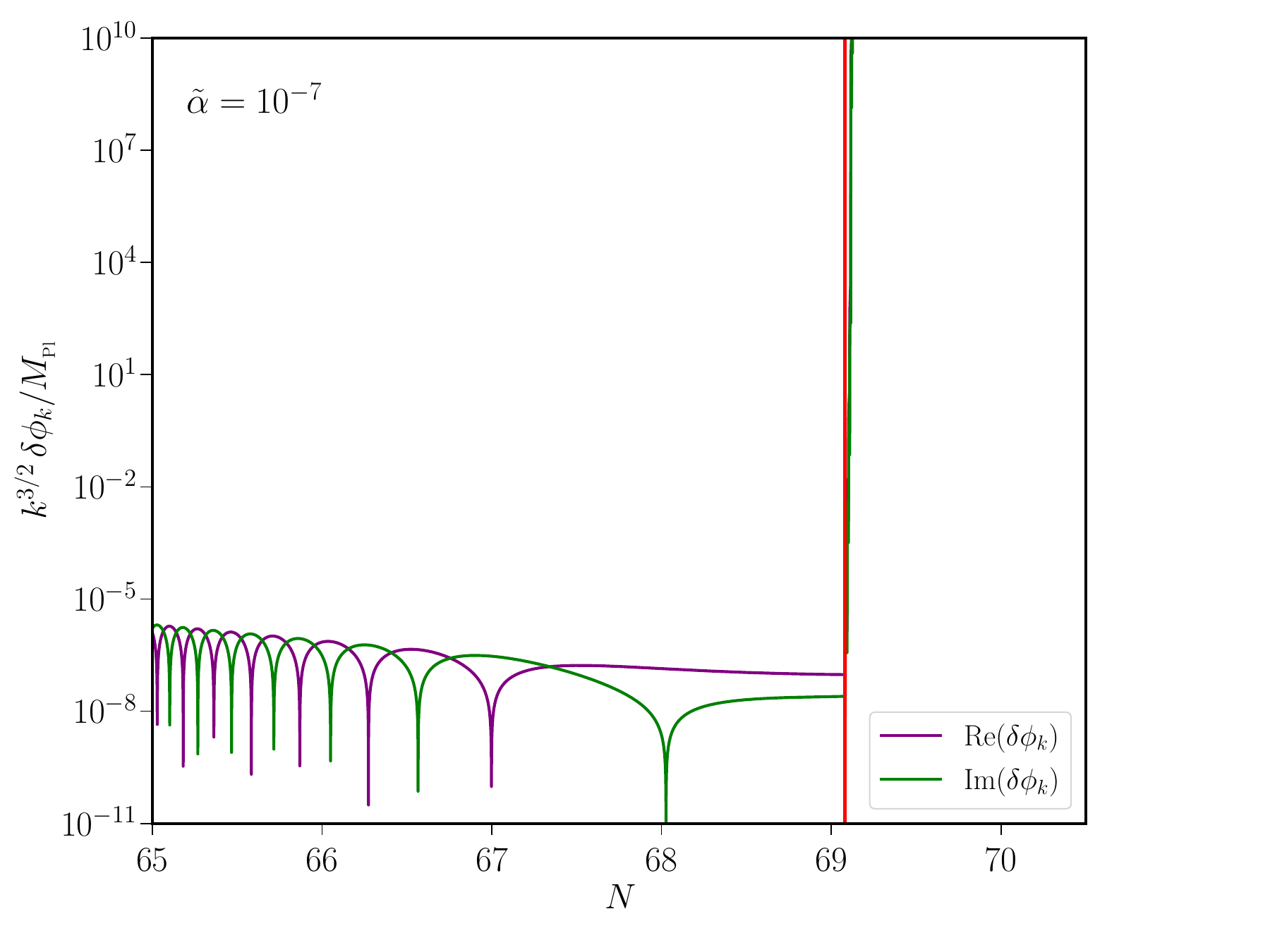}
\vskip -0.1in
\caption{We present the behavior of the real and imaginary parts of the quantity $k^{3/2}\,\delta \phi_k$, which is of dimension $m_{\rm Pl}$, around the end of inflation (marked in red) as a function of e-folds $N \sim \ln(a)$. We present them for two different scales, namely $k=0.05\,{\rm Mpc}^{-1}$ and $k = 10^{20}\,{\rm Mpc}^{-1}$ in left and right columns respectively. The behavior is plotted in each row for a specific value of $\tilde\alpha$ and it is varied as $\{1/3, 10^{-3}, 10^{-5}, 10^{-7}\}$ from top to bottom.}
\label{fig:deltaphi_at_end}
\end{figure}

We perform such an exercise by evolving the individual modes $\delta \phi_k$ using their equation of motion, for different wavenumbers, as presented in Fig.~\ref{fig:deltaphi_at_end}. Such an evolution is valid until the end of inflation, beyond which the non-linearities start playing a role, which then warrants a lattice simulation.
This numerical evolution of inflaton fluctuations, $\delta \phi_k$, as individual modes during inflation, is performed using a modified version of a publicly available package\footnote{\tt https://gitlab.com/ragavendrahv/pbs}. Hence the system is evolved in terms of e-folds $N \equiv \ln (a/a_i)$ instead of the program variable $t_{\rm pr}$, as used in the main text.

We consider two wavenumbers $k = (0.05,\,10^{20})$\,Mpc$^{-1}$, which are representative of large and smaller scale modes, that are in the super-Hubble and sub-Hubble regimes, respectively. We can see from the figure that the modes do not grow in amplitude or show any distinct deviation from their expected behaviors until the end of inflation. The super-hubble mode stays constant while the sub-Hubble mode continues oscillating during inflation. At the end of inflation, they start growing in amplitude with distinct step-like patterns indicating resonances. The overall evolution occurs over shorter time scales as we decrease $\tilde \alpha$, but the resonance does not occur before the end of inflation in any of the cases.
Beyond the end of inflation, such a mode-by-mode inspection becomes unreliable and we must resort to lattice simulation, as done in the main text. But the fact remains clear, that until the end of inflation the modes do not exhibit any deviation in their behavior. 
This is especially true for modes of smaller scales, i.e., $k > 10^{20}$\,Mpc$^{-1}$, which is the range of scales we have investigated in our analysis in the main text.
Thus, it is valid to impose Bunch-Davies initial conditions when these modes are in the sub-Hubble regime at the end of inflation, even for small values of $\tilde \alpha$.

%%%%%%%%%%%%%%%%%%%%%%%%%%%%%%%%%%%%%%%%%%%%%%%%%%%%%%%%%%%%%%%%%%%%%%%%%%%%%%%
%%%%%%%%%%%%%%%%%%%%%%%%%%%%%%%%%%%%%%%%%%%%%%%%%%%%%%%%%%%%%%%%%%%%%%%%%%%%%%%
\bibliographystyle{JHEP}
\bibliography{references}
%%%%%%%%%%%%%%%%%%%%%%%%%%%%%%%%%%%%%%%%%%%%%%%%%%%%%%%%%%%%%%%%%%%%%%%%%%%%%%%
\end{document}